\documentstyle[epsfig,12pt]{article}
\newcommand{\case}[2]{\mbox{\footnotesize $\displaystyle \frac{#1}{#2}$}}
\newcommand{\gsim}{\mathrel{\rlap{\lower4pt\hbox{\hskip0pt$\sim$}}
\raise2pt\hbox{$>$}}}
\newcommand{\qcdtm}{\mbox{QCD$^T_\mu$}}
\newcommand{\ucite}[1]{\mbox{$\,^{\ref{#1}}$}}
\begin{document}
{\Large\bf Nonperturbative effects in QCD at \\
Finite Temperature and Density}\\[2mm] 
{\large  C. D. Roberts}\\[1mm]
{\it Physics Division 203, Argonne National Laboratory, Argonne IL
60439-4843, USA}\\[2mm]
{\bf Abstract}\\ {\small These lecture notes illustrate the application of
Dyson-Schwinger equations in QCD.  The extensive body of work at zero
temperature and chemical potential is represented by a selection of
contemporary studies that focus on solving the Bethe-Salpeter equation,
deriving an exact mass formula in QCD that describes light and heavy
pseudoscalar mesons simultaneously, and the calculation of the
electromagnetic pion form factor and the vector meson electroproduction cross
sections.  These applications emphasise the qualitative importance of the
momentum-dependent dressing of elementary Schwinger functions in QCD, which
provides a unifying connection between disparate phenomena.  They provide a
solid foundation for an extension of the approach to nonzero temperature and
chemical potential.  The essential, formal elements of this application are
described and four contemporary studies employed to exemplify the method and
its efficacy.  They study the demarcation of the phase boundary for
deconfinement and chiral symmetry restoration, the calculation of bulk
thermodynamic properties of the quark-gluon plasma and the response of $\pi$-
and $\rho$-meson observables to $T$ and $\mu$.  Along the way a continuum
order parameter for deconfinement is introduced, an anticorrelation between
the response of masses and decay constants to $T$ and their response to $\mu$
elucidated, and a $(T,\mu)$-mirroring of the slow approach of bulk
thermodynamic quantities to their ultrarelativistic limit highlighted.  These
effects too are tied to the momentum-dependent dressing of the elementary
Schwinger functions.}\\[-1.2\baselineskip]
\begin{center}
\rule{50mm}{0.1mm}
\end{center}\vspace*{-2.3\baselineskip}

\tableofcontents
\vspace*{-1.6\baselineskip}

\begin{center}
\rule{50mm}{0.1mm}
\end{center}

\setcounter{section}{1}
{\large\bf \arabic{section}.~Introduction.}\\[0.7\baselineskip] 
\addcontentsline{toc}{section}{\arabic{section}.~Introduction}
In this article I describe the application of Dyson-Schwinger equations
(DSEs) to QCD at finite temperature, $T$, and quark chemical potential,
$\mu$.  It is not a pedagogical introduction, as this can be found in recent
reviews.\ucite{rw94}$^{,\!}$\ucite{iitap} The goal instead is to illustrate
how contemporary studies at $(T=0,\mu=0)$ can be used as a foundation and
springboard for the application of DSEs at finite $T$ and $\mu$, and to
describe some of these applications and their results.

The DSEs provide a nonperturbative, Poincar\'e invariant, continuum approach
to solving quantum field theories.  They are an infinite tower of coupled
integral equations, with the equation for a particular $n$-point function
involving at least one $m>n$-point function.  A tractable problem is only
obtained if one truncates the system, and historically this has provided an
impediment to the application of DSEs: {\it a priori} it can be difficult to
judge whether a particular truncation scheme will yield qualitatively or
quantitatively reliable results for the quantity sought.  As integral
equations, the analysis of observables using DSEs rapidly becomes a numerical
problem and hence a critical evaluation of truncation schemes often requires,
or is at least simplified, by easy access to high-speed
computers.\footnote[2]{The human and computational resources required are
still modest compared with those consumed in contemporary numerical
simulations of lattice-QCD.}  With such tools now commonplace, this
evaluation can be pursued fruitfully.

The development of efficacious truncation schemes is not a purely numerical
task, and neither is it always obviously systematic.  For some, this last
point diminishes the appeal of the approach.  However, with growing community
involvement and interest, the qualitatively robust results and intuitive
understanding that the DSEs can provide is becoming clear.  Indeed, someone
familiar with the application of DSEs in the late-70s and early-80s might be
surprised with the progress that has been made.  It is now
clear\ucite{bender96}$^{,\!}$\ucite{QC96} that truncations which preserve the
global symmetries of a theory; for example, chiral symmetry in QCD, are
relatively easy to define and implement and, while it is more difficult to
preserve local gauge symmetries, much progress has been made with Abelian
theories\ucite{ayse97} and more is being learnt about non-Abelian ones.

The simplest truncation scheme for the DSEs is the weak-coupling expansion.
Using this systematic procedure it is readily established that the DSEs {\it
contain} perturbation theory, in the sense that for any given theory the
weak-coupling expansion of the equations generates all the diagrams of
perturbation theory.  Hence, at the very least, the DSEs can be used as a
generating tool for perturbation theory, and in this application they are an
essential element in the proof of the renormalisability of a quantum field
theory.  This feature also places a constraint on other truncation schemes;
i.e., the scheme must ensure that perturbative results are recovered in that
domain on which a weak-coupling expansion is known to be valid.

The most important feature of the DSEs is the antithesis of this
weak-coupling expansion: the DSEs are intrinsically nonperturbative.  They
can be derived directly from the generating functional of a quantum field
theory and at no stage in this derivation is a DSE represented as a sum of
diagrams in perturbation theory.  Hence their solution contains information
that is {\it not} present in perturbation theory.  They are ideal for the
study of dynamical chiral symmetry breaking\footnote[2]{
Historically, the DSE for a fermion propagator has found widespread use in
the study of dynamical symmetry breaking; for example, it is the ``gap
equation'' that describes Cooper-pairing in an ordinary superconductor.
} (DCSB) and confinement in QCD, and of hadronic bound state structure and
properties.  In this application they provide a means of elucidating
identifiable signatures of the quark-gluon substructure of hadrons.

Their intrinsically nonperturbative nature also makes them well suited to
studying QCD at finite-$T$ and $\mu$, where the characteristics of the phase
transition to a quark-gluon plasma are a primary subject.  The order of the
transition, the critical exponents, and the response of bound states to
changes in these intensive variables: all must be elucidated.  The latter is
of particular importance because there lies the signals that will identify
the formation of the plasma and hence guide the current and future
experimental searches.

There is a significant overlap between contemporary DSE studies and numerical
simulations of lattice-QCD.  Of particular importance is that both admit the
simultaneous study of DCSB and confinement, the absence of which {\it
defines} the plasma.  The DSEs provide an adjunct to lattice simulations.
They are a means of checking them, and the simulations can provide input into
the development and constraint of DSE truncations.  A truncation that is
accurate on the common domain can be used to extrapolate into that domain
presently inaccessible to lattice-simulations, such as finite chemical
potential and the $T$- and $\mu$-dependence of hadron properties.


\addtocounter{section}{1}
\setcounter{subsection}{1}
{\large\bf \arabic{section}.~Essential Elements of the
DSEs.}\\[0.7\baselineskip]
\addcontentsline{toc}{section}{\arabic{section}.~Essential Elements of the
DSEs} 
In this section I summarise some of the results upon which much of the
successful DSE phenomenology is founded.  Before doing so it is important to
specify that I employ a Euclidean metric throughout.  For real $4$-vectors,
$a$, $b$:
\begin{equation}
a\cdot b := a_\mu\,b_\nu \delta_{\mu\nu} :=  \sum_{i=1}^4\, a_i \, b_i\,,
\end{equation}
and hence a spacelike vector, $Q_\mu$, has $Q^2>0$.  The Dirac matrices
satisfy
\begin{equation}
\gamma_\mu^\dagger = \gamma_\mu\,,\; 
\{\gamma_\mu,\gamma_\nu\} = 2 \, \delta_{\mu\nu}
\end{equation}
and $\gamma_5:= - \gamma_1\gamma_2\gamma_3\gamma_4$.

My point of view is that the Euclidean formulation is {\it primary}$\,$;
i.e., a field theory should be {\it defined} in Euclidean space, where the
propagators and vertices are properly called ``$n$-point Schwinger
functions''.  This is the perspective adopted in constructive field theory
and, at least as a pragmatic artifice, by practitioners of lattice-QCD.  If
the field theory is well-defined, it is completely specified once all its
Schwinger functions are known.  Analytic continuation in the Euclidean-time
variable yields the Wightman functions and, following appropriate
time-ordering, the Minkowski space propagators.  Additional details and
discussion can be found in Sec.~2.3 of Ref.~[\ref{rw94}].

It is important because the analytic structure of nonperturbatively dressed
Schwinger functions need not be the same as that of their free-particle
seeds.  Hence, {\it a priori} one cannot know the analytic properties of the
integrand in a DSE and any rotation of the integration contours, as in a
``Wick rotation'', is plagued by uncertainty: there may be poles or branch
cuts, etc., that cannot be anticipated from the free-particle form of the
Schwinger functions involved.  This is manifest in the fact that the {\it
transcription} {\it rules}:
\begin{center}
\parbox{126mm}{
\parbox{58mm}{Configuration Space
\begin{enumerate}
\item $\displaystyle \int^M\,d^4x^M \, \rightarrow \,-i \int^E\,d^4x^E$
\item $\slash\!\!\! \partial \,\rightarrow \, i\gamma^E\cdot \partial^E $
\item $\slash \!\!\!\! A \, \rightarrow\, -i\gamma^E\cdot A^E$
\item $A_\mu B^\mu\,\rightarrow\,-A^E\cdot B^E$
\end{enumerate}}\hspace*{10mm}
\parbox{58mm}{Momentum Space
\begin{enumerate}
\item $\displaystyle \int^M\,d^4k^M \, \rightarrow \,i \int^E\,d^4k^E$
\item $\slash\!\!\! k \,\rightarrow \, -i\gamma^E\cdot k^E $
\item $k_\mu q^\mu \, \rightarrow\, - k^E\cdot q^E$
\item $k_\mu x^\mu\,\rightarrow\,-k^E\cdot x^E$\,,
\end{enumerate}}}
\end{center}
are valid at every order in perturbation theory; i.e., the correct Minkowski
space integral for a given diagram in perturbation theory is obtained by
applying these transcription rules to the Euclidean integral.  However, for
skeleton diagrams; i.e., those in which each line and vertex represents a
fully dressed $n$-point function, this cannot be guaranteed.
\vspace*{0.5\baselineskip}

{\it \arabic{section}.\arabic{subsection})~Gluon
Propagator.}\\[0.3\baselineskip]
\addcontentsline{toc}{subsection}{\arabic{section}.\arabic{subsection})~Gluon
Propagator} \addtocounter{subsection}{1}
\begin{figure}[t] 
\hspace*{50mm}
Charge Screening  \hspace*{2.0cm} 
        Charge AntiScreening \\
\hspace*{62mm} $ \searrow$ \hspace*{43mm} $ \swarrow$ 

\centering{\ \epsfig{figure=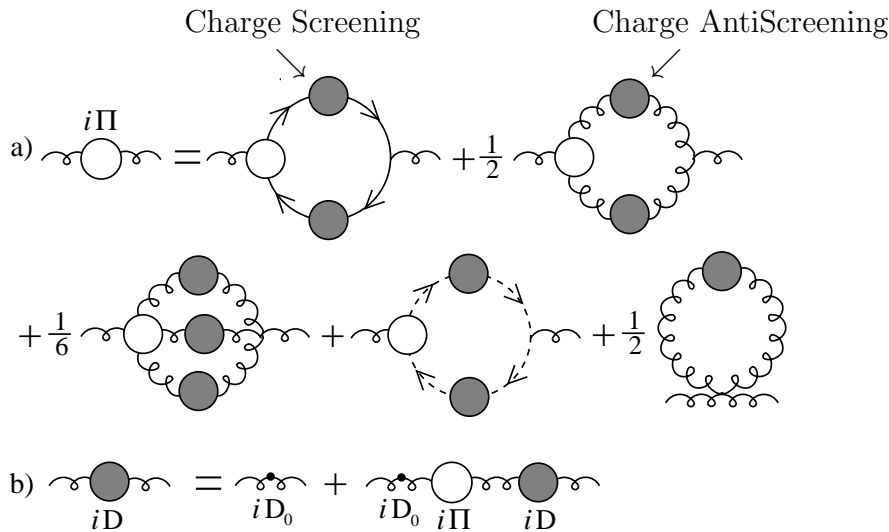,height=6.0cm} }
\caption{DSE for the gluon vacuum polarisation and propagator: solid line -
quark; spring - gluon; dotted-line - ghost.  The open circles are irreducible
vertices.  As indicated, the quark loop acts to screen the charge, as in QED,
while the gluon loop opposes this, Anti-screening the charge and enhancing the
interaction.
\label{gluondse}}
\end{figure}
In Landau gauge the two-point, dressed-gluon Schwinger function, or
dressed-gluon propagator, has the form
\begin{eqnarray}
g^2 D_{\mu\nu}(k)& = &
\left(\delta_{\mu\nu} - \frac{k_\mu k_\nu}{k^2}\right)
        \frac{{\cal G}(k^2)}{k^2}\,,\;
        {\cal G}(k^2):= \frac{g^2}{ 1+\Pi(k^2)} \,,
\end{eqnarray}
where $\Pi(k^2)$ is the gluon vacuum polarisation, which contains all the
dynamical information about gluon propagation.  This propagator satisfies the
DSE illustrated in Fig.~\ref{gluondse} (a nonlinear integral equation).
As already stated, a weak-coupling expansion of this DSE reproduces
perturbation theory.  Using this one sees directly that in the one-loop
expression for the running coupling constant:
\begin{equation}
\alpha_S(q^2) = \frac{12\pi}
        {\left( 11 N_c  -  2 N_f \right) 
        \ln\left(q^2/\Lambda^2_{\rm QCD}\right)}\,,
\end{equation}
the ``$11 N_c$'' comes from the charge-antiscreening gluon loop and the ``$2
N_f$'' from the charge-screening fermion loop, which illustrates how the
non-Abelian structure of QCD is responsible for asymptotic freedom and
suggests that confinement is related to the importance of gluon
self-interactions.

\begin{figure}[t]

\vspace*{30mm}

\hspace*{65mm}{\small Brown and Pennington}\vspace*{-3mm}

\hspace*{55mm} { $\swarrow$ } \vspace*{-1mm}

\hspace*{65mm}{\footnotesize\sf PRD 39 (1989) 2723}

\vspace*{-40mm}

\centering{\
\epsfig{figure=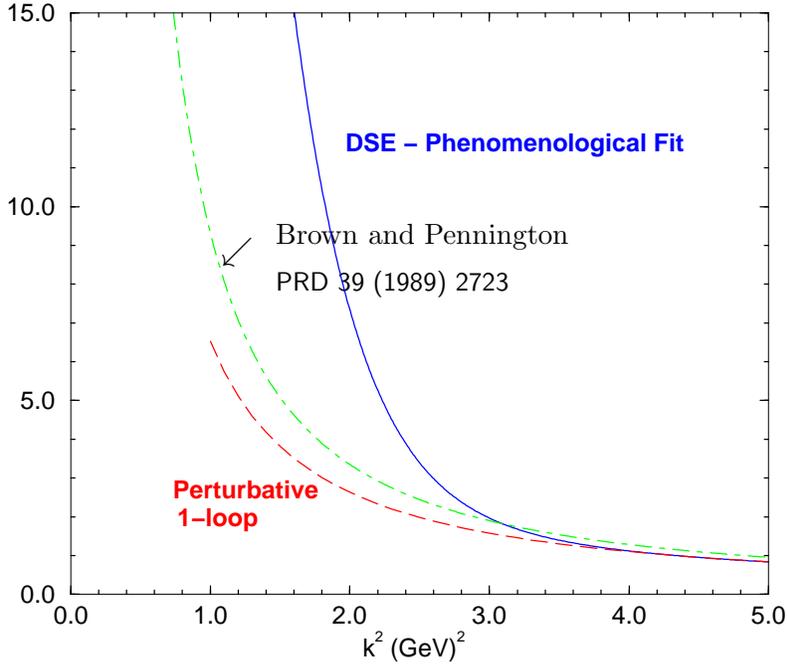,height=9cm}}
\caption{${\cal G}(k^2)/k^2$ from a solution$\,^{\protect\ref{bp89}}$ of the
gluon DSE (dash-dot line) compared with the one-loop perturbative result
(dashed line) and a fit (solid line) obtained following the method of
Ref.~[\protect\ref{mr97}]; i.e., by requiring that the gluon propagator lead,
via the quark DSE, to a good description of a range of hadron observables.
\label{gluonpic}}
\end{figure}

Studies of the gluon DSE have been reported by many authors\ucite{rw94} with
the conclusion that, if the ghost-loop is unimportant, then the
charge-antiscreening 3-gluon vertex dominates and, relative to the free gauge
boson propagator, the dressed gluon propagator is significantly enhanced in
the vicinity of $k^2=0$.  The enhancement persists to $k^2 \sim
1$-$2\,$GeV$^2$, where a perturbative analysis becomes quantitatively
reliable.  In the neighbourhood of $k^2=0$ the enhancement can be
represented\ucite{bp89} as a regularisation of $1/k^4$ as a distribution,
which is illustrated in Fig.~\ref{gluonpic}.
As I will elucidate, a dressed-gluon propagator with the illustrated
enhancement at $k^2\simeq 0$ generates confinement and DCSB {\it without}
fine-tuning.
\vspace*{0.5\baselineskip}

\begin{figure}[t] 
\centering{\ \epsfig{figure=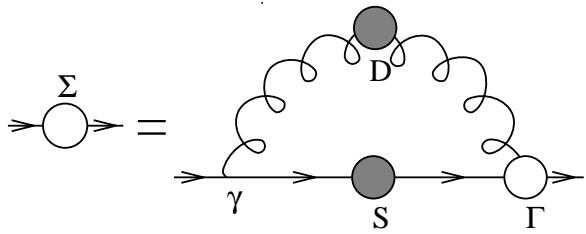,height=3.0cm} }
\caption{DSE for the dressed-quark self-energy.  The kernel of this equation
is constructed from the dressed-gluon propagator ($D$ - spring) and the
dressed-quark-gluon vertex ($\Gamma$ - open circle).  One of the vertices is
bare (labelled by $\gamma$) as required to avoid over-counting.
\label{quarkdse}}
\end{figure}

\parbox{160mm}{
{\it \arabic{section}.\arabic{subsection})~Quark
Propagator.}\\[0.3\baselineskip]
\addcontentsline{toc}{subsection}{\arabic{section}.\arabic{subsection})~Quark
Propagator} \addtocounter{subsection}{1}
In a covariant gauge the two-point, dressed-quark Schwinger function, or
dressed-quark propagator, can be written in a number of equivalent forms
\begin{eqnarray}
\label{Sp}
S(p) & := & \frac{1}{i\gamma\cdot p + \Sigma(p)} \\ 
& := & \frac{1}{i\gamma\cdot p\,A(p^2) + B(p^2)} 
        \equiv -i\gamma\cdot p \,\sigma_V(p^2) + \sigma_S(p^2) \,.
\end{eqnarray}}
$\Sigma(p)$ is the dressed-quark self-energy, which satisfies a nonlinear
integral equation: the quark DSE (depicted in Fig.~\ref{quarkdse})
\begin{eqnarray}
\label{gendse}
\Sigma(p) & = & ( Z_2 -1)\, i\gamma\cdot p + Z_4\,m_{\rm bm}
+\, Z_1\, \int^\Lambda_q \,
g^2 D_{\mu\nu}(p-q) \frac{\lambda^a}{2}\gamma_\mu S(q)
\Gamma^a_\nu(q,p) \,,
\end{eqnarray}
where $\Gamma^a_\nu(q;p)$ is the renormalised dressed-quark-gluon vertex,
$m_{\rm bm}$ is the $\Lambda$-dependent current-quark bare mass that appears
in the Lagrangian and $\int^\Lambda_q := \int^\Lambda d^4 q/(2\pi)^4$
represents mnemonically a {\em translationally-invariant} regularisation of
the integral, with $\Lambda$ the regularisation mass-scale.  The final stage
of any calculation is to remove the regularisation by taking the limit
$\Lambda \to \infty$.  The quark-gluon-vertex and quark wave function
renormalisation constants, $Z_1(\mu^2,\Lambda^2)$ and $Z_2(\mu^2,\Lambda^2)$,
depend on the renormalisation point, $\mu$, and the regularisation
mass-scale, as does the mass renormalisation constant $Z_m(\mu^2,\Lambda^2)
:= Z_2(\mu^2,\Lambda^2)^{-1} Z_4(\mu^2,\Lambda^2)$.

\begin{figure}[t]
\centering{\ 
\epsfig{figure=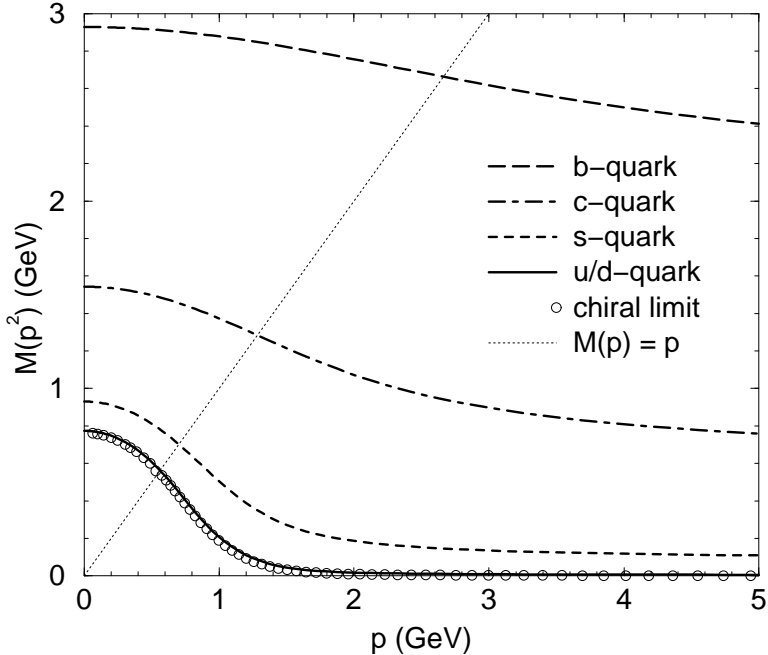,height=10.0cm}}\vspace*{-0.7cm}
\caption{Dressed-quark mass-function obtained in solving the quark DSE using
the dressed-gluon propagator of Ref.~[\protect\ref{mr97}].
\label{plotMpp}}
\end{figure}

One can define a quark mass-function: 
\begin{equation}
\label{Mpp}
M(p^2) := \frac{B(p^2)}{A(p^2)}
\end{equation}
and, as depicted in Fig.~\ref{plotMpp},
solving the quark DSE using a dressed-gluon propagator that behaves as
illustrated in Fig.~\ref{gluonpic} and a dressed-quark-gluon vertex,
$\Gamma_\mu(p,q)$, that does not exhibit particle-like singularities at
$(p-q)^2=0$,\footnote[2]{
A particle-like singularity is one of the form $(P^2)^{-\alpha}$, $\alpha \in
(0,1]$. In this case one can write a spectral decomposition for the vertex in
which the spectral densities are non-negative.  This is impossible if
$\alpha>1$. $\alpha=1$ is the ideal case of an isolated, $\delta$-function
singularity in the spectral densities and hence an isolated, free-particle
pole.  $\alpha \in (0,1)$ corresponds to an accumulation, at the particle
pole, of branch points associated with multiparticle production.
} yields a quark mass-function that mirrors the infrared enhancement of the
dressed-gluon propagator.  The results in Fig.~\ref{plotMpp} were
obtained\ucite{mr97} with the current-quark masses:
\begin{equation}
\label{baremasses}
\begin{array}{cccc}
m_{u/d}^\mu=3.7\, {\rm MeV},\; & 
m_s^\mu= 82\,{\rm MeV},\;  & 
m_c^\mu=0.59\,{\rm GeV},\; & 
m_b^\mu=2.0\,{\rm GeV},\; 
\end{array}
\end{equation}
at a renormalisation point of $\mu\simeq 20\,$GeV.  Applying the one-loop
evolution formula, Eq.~(\ref{masanom}), these masses correspond to:
\begin{equation}
\label{monegev}
\begin{array}{cccc}
m_{u/d}^{1\,{\rm GeV}}=5.5\, {\rm MeV},\, & 
m_s^{1\,{\rm GeV}}= 130\,{\rm MeV},\,  & 
m_c^{1\,{\rm GeV}}= 1.0\,{\rm GeV},\, & 
m_b^{1\,{\rm GeV}}= 3.4\,{\rm GeV} 
\end{array}
\end{equation}
and although it is obvious from Fig.~\ref{plotMpp} that the one-loop formula
does not describe correctly the momentum evolution of the mass-function down
to $p^2=1\,$GeV$^2$, the values in Eq.~(\ref{monegev}) provide a useful and
meaningful comparison with the values quoted conventionally.

The quark DSE was also solved in the chiral limit, which in QCD is obtained
by setting the Lagrangian current-quark bare mass to zero.\ucite{mr97} From
the figure one observes immediately that the mass-function is nonzero even in
this case.  That {\it is} DCSB: a momentum-dependent quark mass generated
dynamically in the absence of any term in the action that breaks chiral
symmetry explicitly.  This entails a nonzero value for the quark condensate
in the chiral limit.  The fact that $M(p^2)\neq 0$ in the chiral limit is
independent of the details of the dressed-gluon propagator in
Fig.~\ref{gluonpic}; they only affect the magnitude of $M(p^2)$.

Figure~\ref{plotMpp} illustrates that for light quarks ($u$, $d$ and $s$)
there are two distinct domains: perturbative and nonperturbative.  In the
perturbative domain the magnitude of the quark mass-function is governed by
the explicit chiral symmetry breaking mass-scale; i.e., the current-quark
mass.  For $p^2< 1\,$GeV$^2$ the mass-function rises sharply.  This is the
nonperturbative domain where the magnitude of $M(p^2)$ is determined by the
DCSB mechanism; i.e., the enhancement in the dressed-gluon propagator.  This
emphasises again that DCSB is more than just a nonzero value of the quark
condensate in the chiral limit!  The boundary, at $p^2 \sim 1\,$GeV$^2$, is
that point where the the enhancement in the dressed-gluon propagator becomes
significant.

The solution of $p^2=M^2(p^2)$ defines a Euclidean constituent-quark mass,
$M^E$.\footnote[2]{
In my Euclidean metric a true quark mass-pole exhibits itself as a real-$p^2$
solution of \mbox{$p^2+M^2(p^2)=0$}.  This is absent in the solutions of the
quark DSE illustrated in Fig.~\protect\ref{plotMpp}, which is a manifestation
of confinement, as discussed in Sec.~2.3).}
For a given quark flavour, the ratio $M^E_f/m_f^\mu$ is a single,
quantitative measure of the importance of the DCSB mechanism in modifying the
quark's propagation characteristics.  As illustrated in Eq.~(\ref{Mmratio}),
obtained using the dressed-gluon propagator in Ref.~[\ref{mr97}],
\begin{equation}
\label{Mmratio}
\begin{array}{l|c|c|c|c|c}
\mbox{\sf flavour} 
        &   u/d  &   s   &  c  &  b  &  t \\\hline
 \frac{M^E}{m_{\mu\sim 20\,{\rm GeV}}}
       &  150   &    10      &  2.3 &  1.4 & \to 1
\end{array}
\end{equation}
this ratio provides for a natural classification of quarks as either light or
heavy.  For light-quarks the ratio is characteristically $10$-$100$ while for
heavy quarks it is only $1$-$2$.\ucite{mr98} The values of this ratio signal
the existence of a characteristic mass-scale associated with DCSB, which I
will denote by $M_\chi$.  For $p^2>0$ the propagation characteristics of a
flavour with $m_f^\mu< M_\chi$ are altered significantly by the DCSB
mechanism, while for flavours with $m_f^\mu\gg M_\chi$ it is irrelevant, and
explicit chiral symmetry breaking dominates.  It is apparent from the figure
that $M_\chi \sim 0.2\,$GeV$\,\sim \Lambda_{\rm QCD}$.

The effect that the enhancement of the dressed-gluon propagator has on the
light-quark mass-function is fundamental in QCD and can be identified as the
source of many observable phenomena.  Further, that this enhancement little
affects heavy-quark propagation characteristics at spacelike-$p^2$ provides
for many simplifications in the study of heavy-meson observables.\ucite{misha}  
\vspace*{0.5\baselineskip}

\begin{figure}[t]
\centering{\ 
\epsfig{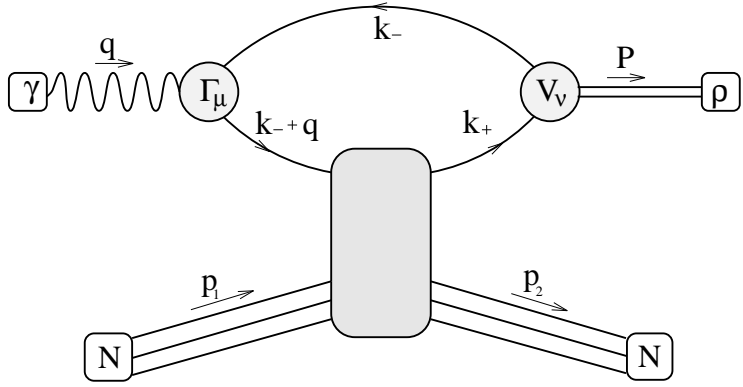}}
\caption{Illustration of the diffractive electroproduction of a vector meson:
$e^- N \to e^{-\prime} N \rho$ with the transition from photon to vector
meson proceeding via a quark loop.  The shaded region
represents$\,^{\protect\ref{pichowsky}}$ a Pomeron-exchange mechanism.
\label{eNeNrho}}
\end{figure}

{\it \arabic{section}.\arabic{subsection})~Confinement.}\\[0.3\baselineskip]
\addcontentsline{toc}{subsection}
{\arabic{section}.\arabic{subsection})~Confinement}
\addtocounter{subsection}{1}
One aspect of confinement is the absence of quark and gluon production
thresholds in colour-singlet-to-singlet ${\cal S}$-matrix amplitudes.  This
is manifest if, for example, the quark-loop illustrated in
Fig.~\ref{eNeNrho}, which describes\ucite{pichowsky} the diffractive,
Pomeron-induced $\gamma\to \rho$ transition, does not have pinch
singularities associated with poles at real-$p^2$ in the quark propagators.
This is ensured if the dressed-quark and -gluon propagators do not have a
Lehmann representation.

What is a Lehmann representation?  

Consider the 2-point free-scalar Schwinger function:
\begin{equation}
\Delta(k^2)=\frac{1}{k^2+m^2}\,.
\end{equation}
One can write 
\begin{equation}
\Delta(z)= \int_0^\infty\,d\sigma\, \frac{\rho(\sigma)}{z + \sigma}\,,
\end{equation}
where in this case the spectral density is 
\begin{equation}
\rho(x):= \frac{1}{2 \pi i} \lim_{\epsilon \to 0} 
\left[\Delta(-x - i\epsilon ) - \Delta(-x + i\epsilon )\right]
= \delta(m^2-x)\,,
\end{equation}
which is non-negative.  This is a Lehmann representation: each scalar function
necessary to completely specify the Schwinger function has a spectral
decomposition in which the spectral densities are non-negative.  Only those
functions whose poles or branch points lie at timelike, real-$k^2$ have a
Lehmann representation.  

The existence of a Lehmann representation for a dressed-particle propagator
is necessary if the construction of asymptotic ``in'' and ``out'' states for
the associated quanta is to proceed; i.e., it is necessary if these quanta
are to propagate to a ``detector''.  In its absence there are no asymptotic
states with the quantum numbers of the field whose propagation
characteristics are described by the Schwinger function.  Structurally, the
nonexistence of a Lehmann representation for the dressed-propagators of
elementary fields ensures the absence of pinch singularities in loops, such as
that illustrated in Fig.~\ref{eNeNrho}, and hence the absence of quark and
gluon production thresholds.

This mechanism can be generalised and applied to coloured bound states, such
as colour-antitriplet quark-quark composites (diquarks).  In this case a
study\ucite{bender96} of the 4-point quark-quark scattering matrix shows that
it does not have a spectral decomposition with non-negative spectral
densities and hence there are no diquark bound states.  The same argument
that demonstrates this absence of diquarks in the spectrum of $SU(N_c=3)$
also proves\ucite{QC96} that in $SU(N_c=2)$ the ``baryons'', which are
necessarily diquarks in this theory, are degenerate with the mesons.

The infrared-enhanced dressed-gluon propagators illustrated in
Fig.~\ref{gluonpic} do not have a Lehmann representation.  Using forms like
this in the kernel of the quark DSE yields automatically a dressed-quark
2-point function that does not have a Lehmann representation.  In this sense
confinement {\it breeds} confinement, without fine-tuning.
\vspace*{0.5\baselineskip}

{\it \arabic{section}.\arabic{subsection})~Hadrons: Bound
States.}\\[0.3\baselineskip]  
\addcontentsline{toc}{subsection}{\arabic{section}.\arabic{subsection})~Hadrons:
Bound States} 
\addtocounter{subsection}{1}
In QCD the observed hadrons are composites of the elementary quanta: mesons
of a quark and antiquark, and baryons of three quarks.  Their masses,
electromagnetic charge radii and other properties can be understood in terms
of their substructure by studying covariant bound state equations: the
Bethe-Salpeter equation (BSE) for mesons and the covariant Fadde'ev equation
for baryons.

\begin{figure}[t]
\centering{\ \epsfig{figure=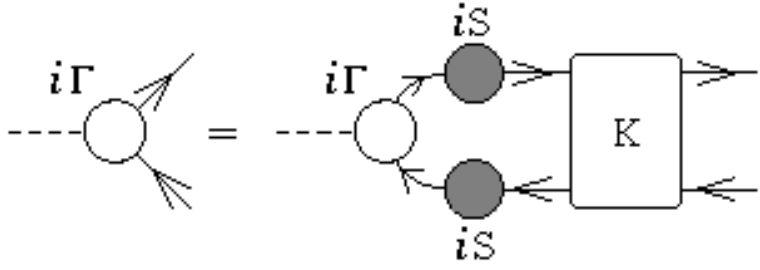,height=3.9cm} }
\caption{Homogeneous Bethe-Salpeter equation for a quark-antiquark bound
state: $\Gamma$ is the solution, the Bethe-Salpeter amplitude, $S$ is the
dressed-quark propagator and $K$ is the dressed-quark-antiquark scattering
kernel.
\label{bsepic}}
\end{figure}

As a two body problem, the mesons have been studied most extensively.  Their
internal structure is described by a Bethe-Salpeter amplitude, which is
obtained as a solution of the homogeneous BSE, illustrated in
Fig.~\ref{bsepic}.
For a pseudoscalar bound state the amplitude has the form
\begin{eqnarray}
\label{genpibsa}
\Gamma_H(k;P) & = &  T^H \gamma_5 \left[ i E_H(k;P) + 
\gamma\cdot P F_H(k;P) \rule{0mm}{5mm}\right. \\
\nonumber & & 
\left. \rule{0mm}{5mm}+ \gamma\cdot k \,k \cdot P\, G_H(k;P) 
+ \sigma_{\mu\nu}\,k_\mu P_\nu \,H_H(k;P) 
\right]\,,
\end{eqnarray}
where, if the constituents have equal current-quark masses, the scalar
functions $E$, $F$, $G$ and $H$ are even under $k\cdot P \to - k\cdot P$.  In
Eq.~(\ref{genpibsa}), $T^H$ is a flavour matrix that determines the mesonic
channel under consideration; e.g., $T^{K^+}:= (1/2)\left(\lambda^4 + i
\lambda^5\right)$, with $\{\lambda^j,j=1\ldots 8\}$ the Gell-Mann matrices.
The important new element in the BSE is $K$, the fully-amputated,
quark-antiquark scattering kernel: by definition it does not contain
quark-antiquark to single gauge-boson annihilation diagrams, such as would
describe the leptonic decay of the pion, nor diagrams that become
disconnected by cutting one quark and one antiquark line.

\begin{figure}[t]
  \centering{\ \hspace*{-2.5cm}\epsfig{figure=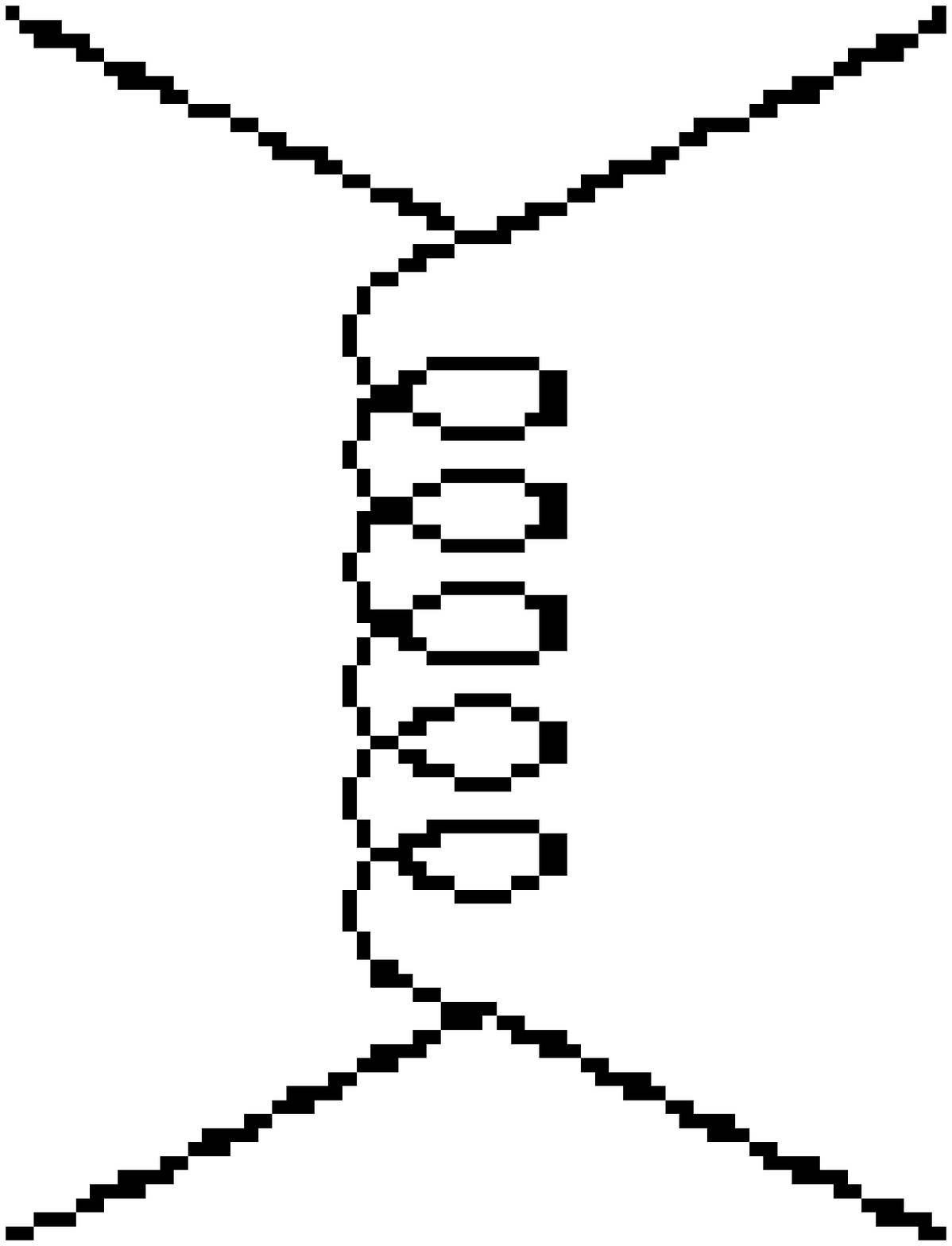,height=3.0cm} 
        \vspace*{-20mm} 

        \hspace*{15mm} $\longleftarrow$ { (1) -- Ladder}\vspace*{5mm} 

        \hspace*{65mm} { (2) -- Beyond Ladder}

        \hspace*{26mm}$\swarrow$

        \hspace*{0.5cm}\epsfig{figure=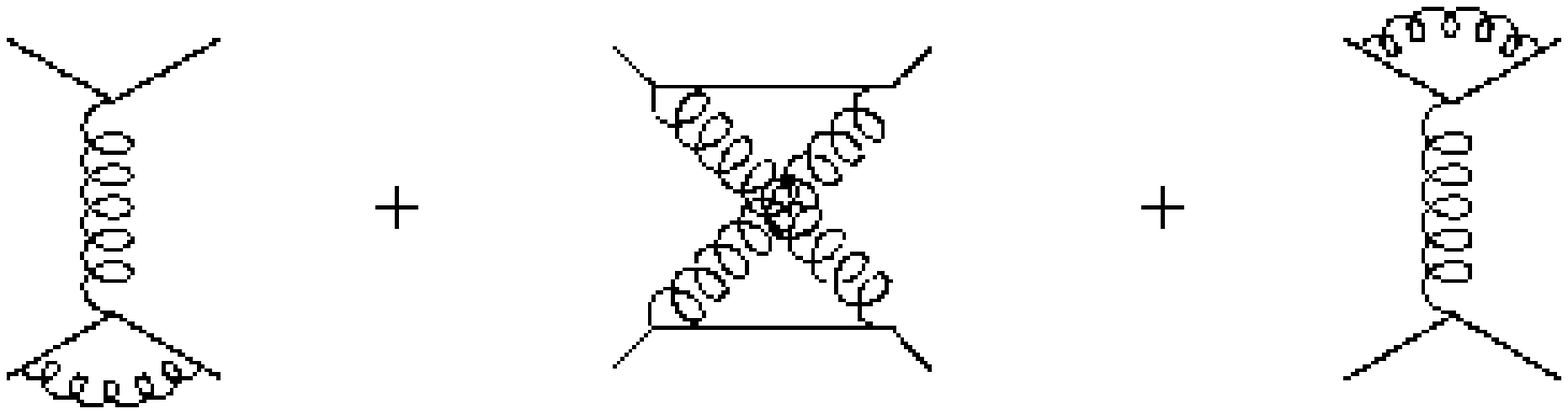,height=3.0cm} }
\caption{First two sets of contributions to a systematic expansion of the
quark-antiquark scattering kernel.  In this expansion, the propagators are
dressed but the vertices are bare.
\label{skeleton}}
\end{figure}

$K$ has a skeleton expansion in terms of the elementary, dressed-particle
Schwinger functions; e.g., the dressed-quark and -gluon propagators.  The
first two orders in one systematic expansion are depicted in
Fig.~\ref{skeleton}.
This particular expansion,\ucite{bender96} in concert with its analogue for
the kernel in the quark DSE, provides a means of constructing a kernel that,
order-by-order in the number of vertices, ensures the preservation of vector
and axial-vector Ward-Takahashi identities.  This is particularly important
in QCD where the Goldstone boson nature of the pion must be understood as a
{\it consequence} of its internal structure.

To proceed with a study of meson properties, one chooses a truncation for
$K$.  The homogeneous BSE is then fully specified as a linear integral
equation, which is straightforward to solve, yielding the bound state mass
and amplitude.  The ``ladder'' truncation of $K$ combined with the
``rainbow'' truncation of the quark DSE ($\Gamma_\mu \to \gamma_\mu$ in
Fig.~\ref{quarkdse}) is the simplest and most often used.  The expansion of
Fig.~\ref{skeleton} allows one to understand why this Ward-Takahashi identity
preserving truncation is accurate for flavour-nonsinglet pseudoscalar and
vector mesons: there are cancellations between the higher-order diagrams.
And also why it provides a poor approximation in the study of scalar mesons,
where the higher-order terms do not cancel, and for flavour-singlet mesons
where it omits timelike gluon exchange diagrams.

\pagebreak

\begin{figure}[t]
\hspace*{30mm}\epsfig{figure=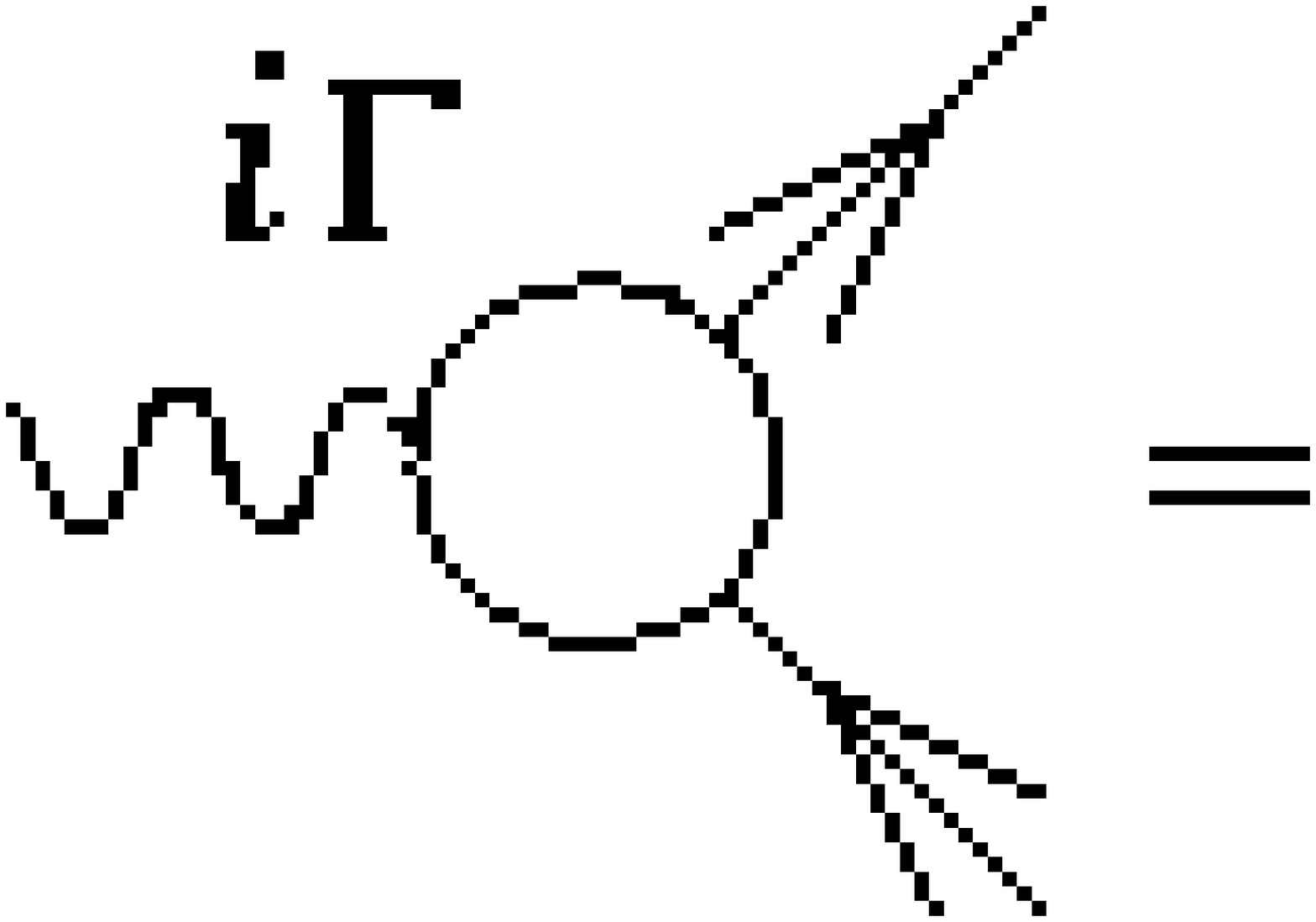,height=2.0cm}
\vspace*{-17.5mm} 

\hspace*{55mm}\epsfig{figure=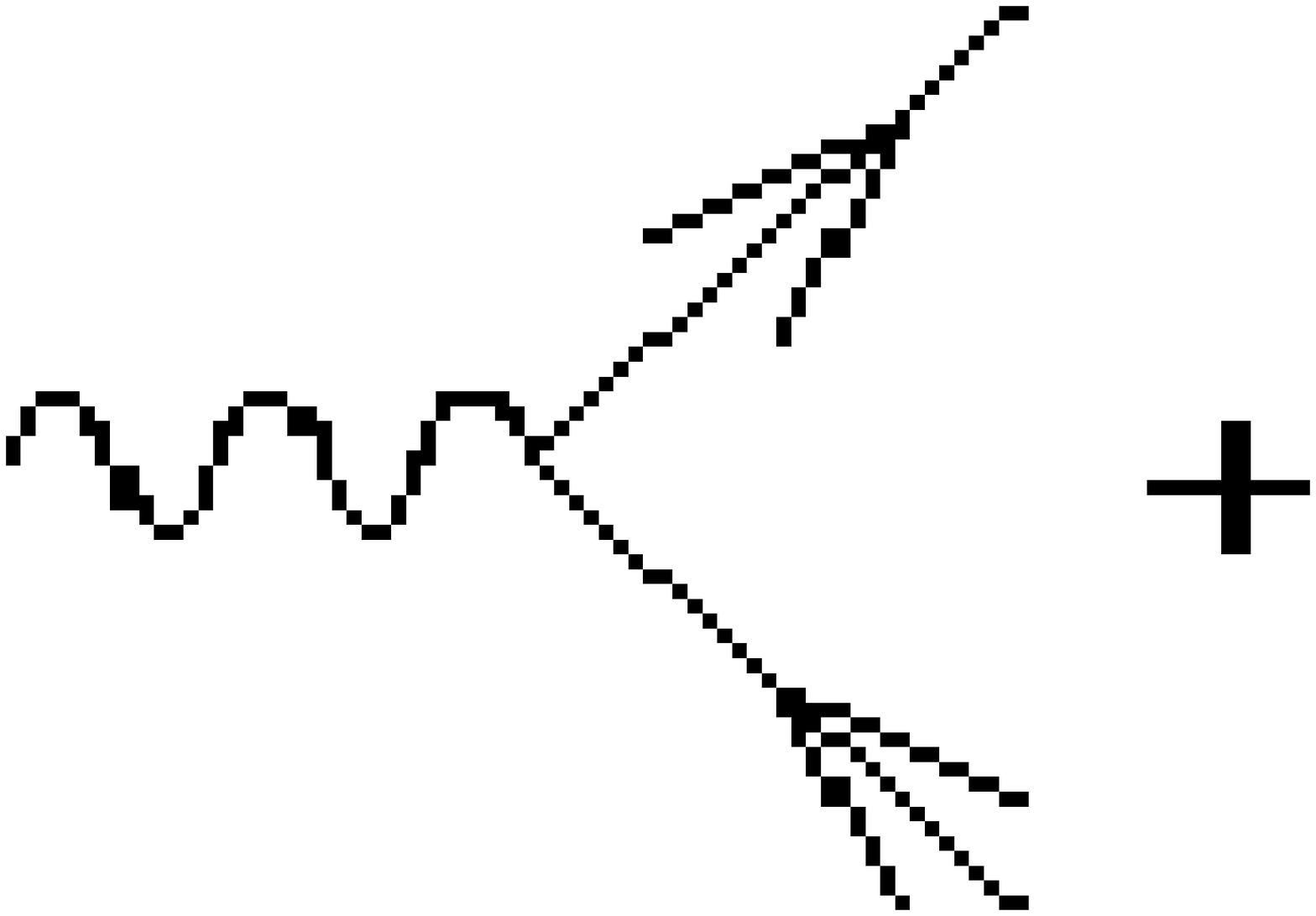,height=1.5cm} 
\vspace*{-21mm}

\hspace*{80mm}\epsfig{figure=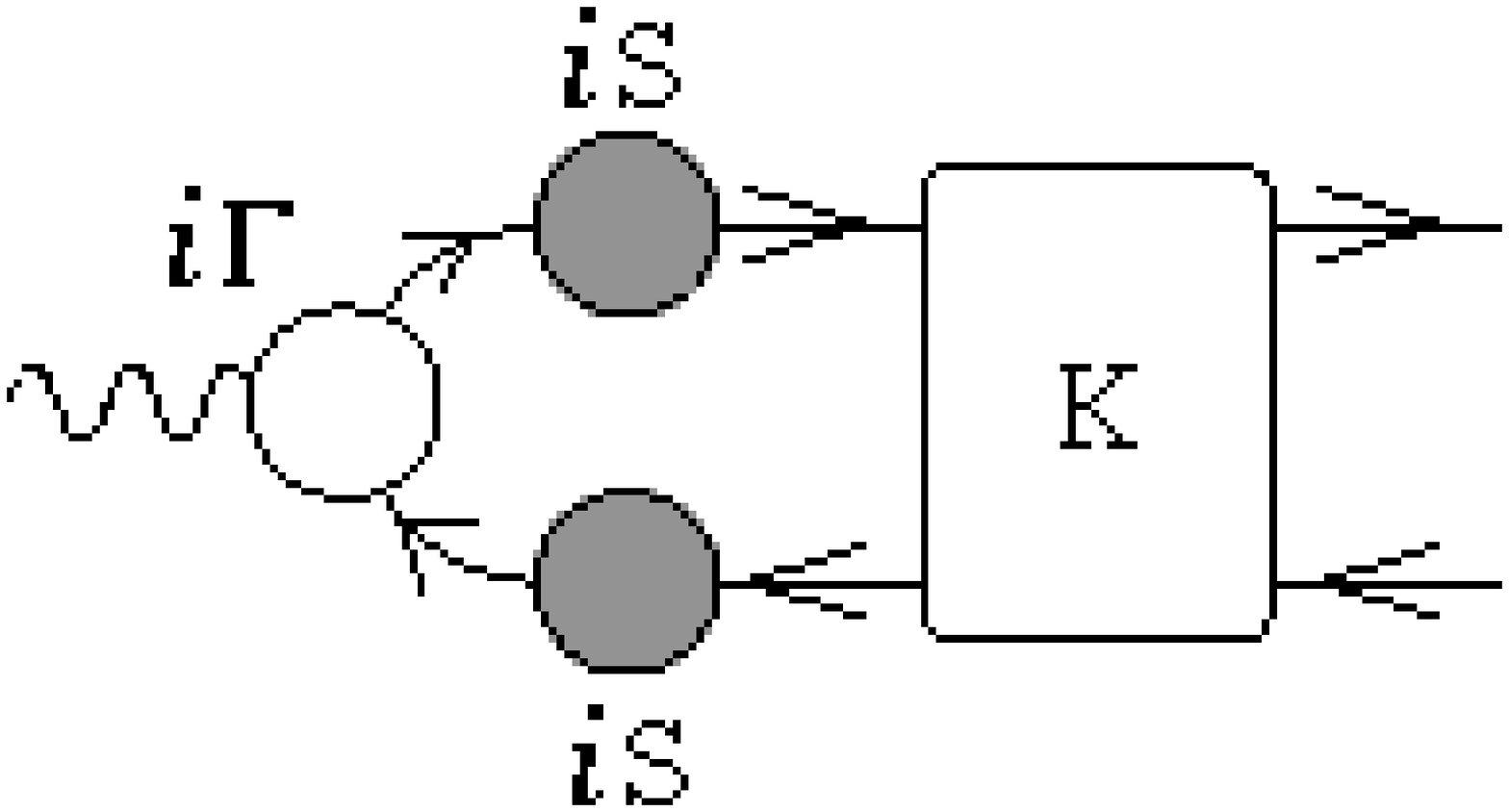,height=2.8cm}
\caption{DSE for the axial-vector vertex.  The driving term is the bare
vertex: $i\gamma_5\gamma_\mu$.
\label{avdse}}
\end{figure}

\addtocounter{section}{1}
\setcounter{subsection}{1}
{\large\bf \arabic{section}.~A QCD Mass Formula.}\\[0.7\baselineskip] 
\addcontentsline{toc}{section}{\arabic{section}.~A QCD Mass Formula}
The chiral-limit axial-vector Ward-Takahashi identity (AV-WTI)
\begin{equation}
\label{chavwti}
-i P_\mu \Gamma_{5\mu}^H(k;P) = {\cal S}^{-1}(k_+)\gamma_5\frac{T^H}{2}
+ \gamma_5\frac{T^H}{2} {\cal S}^{-1}(k_-)\,,
\end{equation}
where ${\cal S}:= {\rm diag}(S_u,S_d,\ldots)$, is the statement of
chiral-current conservation in massless QCD.  It relates the divergence of
the inhomogeneous axial-vector vertex, $\Gamma_{5\mu}^H(k;P)$, to a sum of
dressed-quark propagators.  The vertex is the solution of the DSE depicted in
Fig.~\ref{avdse}, which involves the quark-antiquark scattering kernel, $K$.
It is therefore not surprising that in order to preserve this identity when
truncating the DSEs the choice of $K$ and the vertex, $\Gamma_\mu$, in the
quark DSE, are tied together.  The divergence $P_\mu \Gamma_{5\mu}^H(k;P)$ is
a pseudoscalar and hence contains information about pseudoscalar mesons;
i.e., Goldstone modes.
\vspace*{0.5\baselineskip}

{\it \arabic{section}.\arabic{subsection})~Dynamical Chiral Symmetry Breaking
and Goldstone's 
Theorem.}\\[0.3\baselineskip]
\addcontentsline{toc}{subsection}{\arabic{section}.\arabic{subsection})~Dynamical
Chiral Symmetry Breaking and Goldstone's Theorem}
\addtocounter{subsection}{1}
In the chiral limit, the axial-vector vertex has the form\ucite{mr97}
\begin{eqnarray}
\label{truavv}
\Gamma_{5 \mu}^H(k;P) & = &
\frac{T^H}{2} \gamma_5 
\left[ \rule{0mm}{5mm}\gamma_\mu F_R(k;P) + \gamma\cdot k k_\mu G_R(k;P) 
- \sigma_{\mu\nu} \,k_\nu\, H_R(k;P) \right]\\
&+ & \nonumber
 \tilde\Gamma_{5\mu}^{H}(k;P) 
+\,f_H\,  \frac{P_\mu}{P^2 } \,\Gamma_H(k;P)\,,
\end{eqnarray}
where: $F_R$, $G_R$, $H_R$ and $\tilde\Gamma_{5\mu}^{H}$ are regular as
$P^2\to 0$; $P_\mu \tilde\Gamma_{5\mu}^{H}(k;P) \sim {\rm O }(P^2)$;
$\Gamma_H(k;P)$ is the pseudoscalar Bethe-Salpeter amplitude in
Eq.~(\ref{genpibsa}); and the residue of the pseudoscalar pole in the
axial-vector vertex is $f_H$, the leptonic decay constant:
\begin{eqnarray}
\label{caint}
f_H P_\mu = 
Z_2\int^\Lambda_q\,\case{1}{2}
{\rm tr}\left[\left(T^H\right)^{\rm t} \gamma_5 \gamma_\mu 
{\cal S}(q_+) \Gamma_H(q;P) {\cal S}(q_-)\right]\,,
\end{eqnarray}
with the trace over colour, Dirac and flavour indices.  This expression is
exact: the dependence of $Z_2$ on the renormalisation point, regularisation
mass-scale and gauge parameter is just that necessary to ensure that the
left-hand-side, $f_H$, is independent of all these things.

It now follows from the chiral-limit AV-WTI that
\begin{eqnarray}
\label{bwti} 
f_H E_H(k;0)  &= &  B(k^2)\,, \\
\label{fwti}
 F_R(k;0) +  2 \, f_H F_H(k;0)                 & = & A(k^2)\,, \\
\label{rgwti}
G_R(k;0) +  2 \,f_H G_H(k;0)    & = & 2 A^\prime(k^2)\,,\\
\label{gwti} 
H_R(k;0) +  2 \,f_H H_H(k;0)    & = & 0\,,
\end{eqnarray}
where $A(k^2)$ and $B(k^2)$ are the solutions of the quark DSE in the chiral
limit.  As emphasised in Sec.~2.2), the appearance of a $B(k^2)\neq 0$
solution of the quark DSE in the chiral limit signals DCSB.  It is an
intrinsically nonperturbative effect: in perturbation theory $B(k^2) \propto
\hat m$, the renormalisation-point independent current-quark mass, and hence
vanishes in the chiral limit.  Equations~(\ref{truavv}) and
(\ref{bwti})-(\ref{gwti}) show that when chiral symmetry is dynamically
broken: 1) the homogeneous, flavour-nonsinglet, pseudoscalar BSE has a
massless, $P^2=0$, solution; 2) the Bethe-Salpeter amplitude for the massless
bound state has a term proportional to $\gamma_5$ alone, with the
momentum-dependence of $E_H(k;0)$ completely determined by that of the scalar
part of the quark self energy, in addition to terms proportional to other
pseudoscalar Dirac structures, $F_H$, $G_H$ and $H_H$, that are nonzero in
general; and 3) the axial-vector vertex, $\Gamma_{5 \mu}^H(k;P)$, is
dominated by the pseudoscalar bound state pole for $P^2\simeq 0$.  The
converse is also true.

Hence, in the chiral limit, the pion is a massless composite of a quark and
an antiquark, each of which has an effective mass $M^E \sim 450\,$MeV.  With
a dressed-gluon propagator of the type depicted in Fig.~\ref{gluonpic}, this
occurs without fine-tuning.
\vspace*{0.5\baselineskip}

{\it \arabic{section}.\arabic{subsection})~Nonzero Quark Masses: A Mass
Formula.}\\[0.3\baselineskip] 
\addcontentsline{toc}{subsection}{\arabic{section}.\arabic{subsection})~Nonzero
Quark Masses: A Mass Formula} 
\addtocounter{subsection}{1}
When the current-quark masses are nonzero the AV-WTI is modified:
\begin{equation}
\label{avwti}
-i P_\mu \Gamma_{5\mu}^H(k;P)  = {\cal S}^{-1}(k_+)\gamma_5\frac{T^H}{2}
+  \gamma_5\frac{T^H}{2} {\cal S}^{-1}(k_-) 
- M_{(\mu)}\,\Gamma_5^H(k;P) - \Gamma_5^H(k;P)\,M_{(\mu)} \,,
\end{equation}
where: $M_{(\mu)}= {\rm diag}(m_u^\mu,m_d^\mu,m_s^\mu,\ldots)$, is the
current-quark mass matrix.  In this case both the axial-vector and the
pseudoscalar vertices have a pseudoscalar pole: i.e.,
\begin{eqnarray}
\label{truavvm}
\Gamma_{5 \mu}^H(k;P) & = &
\frac{T^H}{2} \gamma_5 
\left[ \gamma_\mu F_R^H(k;P) + \gamma\cdot k k_\mu G_R^H(k;P) 
- \sigma_{\mu\nu} \,k_\nu\, H_R^H(k;P) \right]\\
&+ & \nonumber
 \tilde\Gamma_{5\mu}^{H}(k;P) 
+\,f_H\,  \frac{P_\mu}{P^2 +m_H^2} \,\Gamma_H(k;P)\,,
\end{eqnarray}
and
\begin{eqnarray}
\label{trupvvm}
\Gamma_{5}^H(k;P) & = &
\frac{T^H}{2} \gamma_5 
\left[ i {\cal E}_R^H(k;P) + 
\gamma\cdot P\, {\cal F}_R^H(k;P) 
+ \gamma\cdot k\, k\cdot P\, {\cal G}_R^H(k;P) 
\right.\\
& & \nonumber
\left. + \sigma_{\mu\nu} \,k_\mu P_\nu\, {\cal H}_R^H(k;P) \right]
+\,r_H\,  \frac{1}{P^2 +m_H^2} \,\Gamma_H(k;P)\,,
\end{eqnarray}
with: ${\cal E}_R^H$, $F_R^H$, ${\cal F}_R^H$, $G_R^H$, ${\cal G}_R^H$,
$H_R^H$, ${\cal H}_R^H$ and $\tilde\Gamma_{5\mu}^{H}$ regular as $P^2\to
-m_H^2$ and $P_\mu\tilde\Gamma_{5\mu}^{H}(k;P) \sim {\rm O}(P^2)$.  The
AV-WTI entails\ucite{mr97} that
\begin{equation}
\label{gmora}
f_H\,m_H^2 = r_H \, {\cal M}_H\,,\;\;
{\cal M}_H := {\rm tr}_{\rm flavour}
\left[M_{(\mu)}\,\left\{T^H,\left(T^H\right)^{\rm t}\right\}\right]\,,
\end{equation}
where $f_H$ is given by Eq.~(\ref{caint}), with massive quark propagators in
this case, and the residue of the pole in the pseudoscalar vertex is
\begin{equation}
\label{rH}
i r_H = Z_4\int^\Lambda_q\,\case{1}{2}
{\rm tr}\left[\left(T^H\right)^{\rm t} \gamma_5 
{\cal S}(q_+) \Gamma_H(q;P) {\cal S}(q_-)\right]\,.
\end{equation}
The renormalisation constant $Z_4$ on the right-hand-side depends on the
gauge parameter, the regularisation mass-scale and the renormalisation point.
This dependence is exactly that required to ensure that: 1) $r_H$ is finite
in the limit $\Lambda\to \infty$; 2) $r_H$ is gauge-parameter independent;
and 3) the right-hand-side of Eq.~(\ref{gmora}) is renormalisation point {\it
independent}.  This is obvious at one-loop order, especially in Landau-gauge
where $Z_2\equiv 1$ and hence $Z_4 = Z_m$.

Equation~(\ref{gmora}) is a mass formula for flavour-octet pseudoscalar
mesons that is valid {\it independent} of the magnitude of the current-quark
masses of a meson's constituents.  For small current-quark masses, using
Eqs.~(\ref{genpibsa}) and (\ref{bwti})-(\ref{gwti}), Eq.~({\ref{rH}) yields
\begin{equation}
\label{cbqbq}
\begin{array}{lcr}
\displaystyle
r_H^0  =  -\,\frac{1}{f_H^0}\, \langle \bar q q \rangle_\mu^0 \,
, & & 
\displaystyle
\,-\,\langle \bar q q \rangle_\mu^0 :=  
Z_4(\mu^2,\Lambda^2)\, N_c \int^\Lambda_q\,{\rm tr}_{\rm Dirac}
        \left[ S_{\hat m =0}(q) \right]\,,
\end{array}
\end{equation}
where the superscript ``$0$'' denotes that the quantity is evaluated in the
chiral limit and $ \langle \bar q q \rangle_\mu^0 $, as defined here, is the
chiral limit {\it vacuum quark condensate}, which is renormalisation-point
dependent but independent of the gauge parameter and the regularisation
mass-scale.  Hence Eq.~(\ref{trupvvm}) is the statement that {\it in the
chiral limit the residue of the bound state pole in the flavour-nonsinglet
pseudoscalar vertex is} $(-\,\langle \bar q q \rangle_\mu^0)/f_H^0$.

Now one obtains immediately from Eqs.~(\ref{gmora}) and (\ref{cbqbq})
\begin{eqnarray}
\label{gmorepi}
f_{\pi}^2 m_{\pi}^2 & = &-\,\left[m_u^\mu + m_d^\mu\right]
       \langle \bar q q \rangle_\mu^0 + {\rm O}\left(\hat m_q^2\right)\,\\
\label{gmoreKp}
f_{K^+}^2 m_{K^+}^2 & = &-\,\left[m_u^\mu + m_s^\mu\right]
       \langle \bar q q \rangle_\mu^0 + {\rm O}\left(\hat m_q^2\right)\,,
\end{eqnarray}
which exemplify what is commonly known as the Gell-Mann--Oakes--Renner
relation.  

The primary result, Eq.~(\ref{gmora}), is valid {\it independent} of the
magnitude of $\hat m_q$, and can be rewritten in the form
\begin{equation}
\label{gmorqbqM}
f_H^2\, m_H^2 = \,-\,\langle \bar q q \rangle_\mu^H\,{\cal M}_H
\end{equation}
where I have introduced the {\it notation}
\begin{equation}
\label{qbqM}
 -\,\langle \bar q q \rangle_\mu^H := f_H\,r_H =f_H
Z_4\int^\Lambda_q\,\mbox{$\case{1}{2}$} {\rm tr}\left[\left(T^H\right)^{\rm
t} \gamma_5 {\cal S}(q_+) \Gamma_H(q;P) {\cal S}(q_-)\right] \,, 
\end{equation}
which defines an {\it in-meson} condensate.  This emphasises that, for
nonzero current-quark masses, Eq.~(\ref{gmora}) {\it does not}$\,$ involve a
difference of vacuum massive-quark condensates; a phenomenological assumption
often employed.

As elucidated elsewhere,\ucite{cssm} Eq.~(\ref{gmora}) has another important
corollary: it predicts that the mass of a heavy pseudoscalar meson rises
linearly with the current-quark mass of its heavy constituent(s).  Model
calculations\ucite{mr98} show that this linear evolution is dominant at
$\approx 2 \hat m_s$, in agreement with experiment where the mass of the $K$,
$D$ and $B$ mesons lie precisely on the same straight line.

In Eq.~(\ref{gmora}) one therefore has a single mass formula, exact in QCD,
that provides a unified understanding of light- and heavy-meson masses.

\vspace*{\baselineskip}

\addtocounter{section}{1}
\setcounter{subsection}{1}
{\large\bf \arabic{section}.~An Illustrative Model.}\\[0.7\baselineskip] 
\addcontentsline{toc}{section}{\arabic{section}.~An Illustrative Model}
I have already made use of a model\ucite{mr97} in illustrating some of the
robust results of DSE studies.  To further elucidate the method I will
describe that model in more detail.  For the kernel of the quark DSE it
employs the analogue of the lowest-order BSE kernel in Fig.~\ref{skeleton}:
\begin{equation}
\label{ouransatz}
Z_1\, \int^\Lambda_q \,
g^2 D_{\mu\nu}(p-q) \frac{\lambda^a}{2}\gamma_\mu S(q)
\Gamma^a_\nu(q,p)
\to
\int^\Lambda_q \,
{\cal G}((p-q)^2)\, D_{\mu\nu}^{\rm free}(p-q)
 \frac{\lambda^a}{2}\gamma_\mu S(q)
\frac{\lambda^a}{2}\gamma_\nu \,.
\end{equation}
This is the ``rainbow'' approximation, in which the specification of the
model is complete once a form is chosen for the ``effective coupling'' ${\cal
G}(k^2)$.

A choice for ${\cal G}(k^2)$ can be motivated by observing that at
large-$Q^2:= (p-q)^2$ in an asymptotically free theory the quark-antiquark
scattering kernel satisfies
\begin{eqnarray}
\label{uvkernel}
\lefteqn{g^2(\mu^2)\, D_{\mu\nu}(p-q) \,
\left[\rule{0mm}{0.7\baselineskip} \Gamma^a_\mu(p_+,q_+)S\,(q_+) \right] 
\times 
\left[ \rule{0mm}{0.7\baselineskip}S(q_-)\,\Gamma^a_\nu(q_-,p_-) \right] }\\
&& \nonumber  
= 4\pi\, \alpha(Q^2)\, D_{\mu\nu}^{\rm free}(p-q)\,
\left[\rule{0mm}{0.7\baselineskip}
        \frac{\lambda^a}{2}\gamma_\mu \,S^{\rm free}(q_+)\right]\times
\left[\rule{0mm}{0.7\baselineskip}S^{\rm free}(q_-)\,
        \frac{\lambda^a}{2}\gamma_\nu\right]\,,
\end{eqnarray}
where $P$ is the total momentum of the quark-antiquark pair, $p_+ := p +
\eta_P P$ and $p_- := p - (1-\eta_P) P$ with $0\leq\eta\leq 1$.  Choosing a
truncation of $K$ in which this right-hand-side is identified with the lowest
order contribution in Fig.~\ref{skeleton} then consistency with the AV-WTI
requires
\begin{equation}
\label{abapprox}
{\cal G}(Q^2):= 4\pi\,\alpha(Q^2)\,.
\end{equation}
Thus the form of ${\cal G}(Q^2)$ at large-$Q^2$ is fixed by that of the
running coupling constant.  This {\it Ansatz} is often described as the
``Abelian approximation'' because the left- and right-hand-sides are {\it
equal} in QED.  In QCD, equality between the two sides of
Eq.~(\ref{abapprox}) cannot be obtained easily by a selective resummation of
diagrams.  As reviewed in Ref.~[\ref{rw94}], Eqs.~(5.1) to (5.8), it can only
be achieved by enforcing equality between the renormalisation constants for
the ghost-gluon vertex and ghost wave function: $\tilde Z_1=\tilde Z_3$.

The explicit form of the {\it Ansatz} employed in Ref.~[\ref{mr97}] is
\begin{equation}
\label{gk2}
\frac{{\cal G}(k^2)}{k^2} =
8\pi^4 D \delta^4(k) + \frac{4\pi^2}{\omega^6} D k^2 {\rm e}^{-k^2/\omega^2}
+ 4\pi\,\frac{ \gamma_m \pi}
        {\case{1}{2}
        \ln\left[\tau + \left(1 + k^2/\Lambda_{\rm QCD}^2\right)^2\right]}\,
{\cal F}(k^2) \,,
\end{equation}
with ${\cal F}(k^2)= [1 - \exp(-k^2/[4 m_t^2])]/k^2$ and $\tau={\rm
e}^2-1$,$N_f=4$ and $\Lambda_{\rm QCD}^{N_f=4}= 0.234\,{\rm GeV}$. 

The qualitative features of Eq.~(\ref{gk2}) are clear.  The first term is an
integrable infrared singularity\ucite{mn83} and the second is a finite-width
approximation to $\delta^4(k)$, normalised such that it has the same $\int
d^4k$ as the first term.  In this way the infrared singularity is split into
the sum of a zero-width and a finite-width piece.  The last term in
Eq.~(\ref{gk2}) is proportional to $\alpha(k^2)/k^2$ at large spacelike-$k^2$
and has no singularity on the real-$k^2$ axis.  Gluon confinement is manifest
since ${\cal G}(k^2)/k^2$ doesn't have a Lehmann representation.
\vspace*{0.5\baselineskip}

{\it \arabic{section}.\arabic{subsection})~Solving the Quark
DSE.}\\[0.3\baselineskip] 
\addcontentsline{toc}{subsection}{\arabic{section}.\arabic{subsection})~Solving
the Quark DSE} 
\addtocounter{subsection}{1}
There are ostensibly three parameters in Eq.~(\ref{gk2}): $D$, $\omega$ and
$m_t$.  However, in the numerical studies the values
$\omega=0.3\,$GeV$(=1/[.66\,{\rm fm}])$ and $m_t=0.5\,$GeV$(=1/[.39\,{\rm
fm}])$ were fixed, and only $D$ and the renormalised $u/d$- and
$s$-current-quark masses varied in order to satisfy the goal of a good
description of low-energy $\pi$- and $K$-meson properties.  This was achieved
with
\begin{equation}
\label{params}
\begin{array}{ccc}
D= 0.781\,{\rm GeV}^2\,,\; &
m_{u/d}^\mu = 3.74\,{\rm MeV}\,,\; &
m_s^\mu = 82.5\,{\rm MeV}
\end{array}\,,
\end{equation}
at $\mu\approx 20\,$GeV, which is large enough to be in the perturbative
domain.  The effective coupling obtained is depicted in Fig.~\ref{Gmodel}.
\begin{figure}[t]
\centering{\
\epsfig{figure=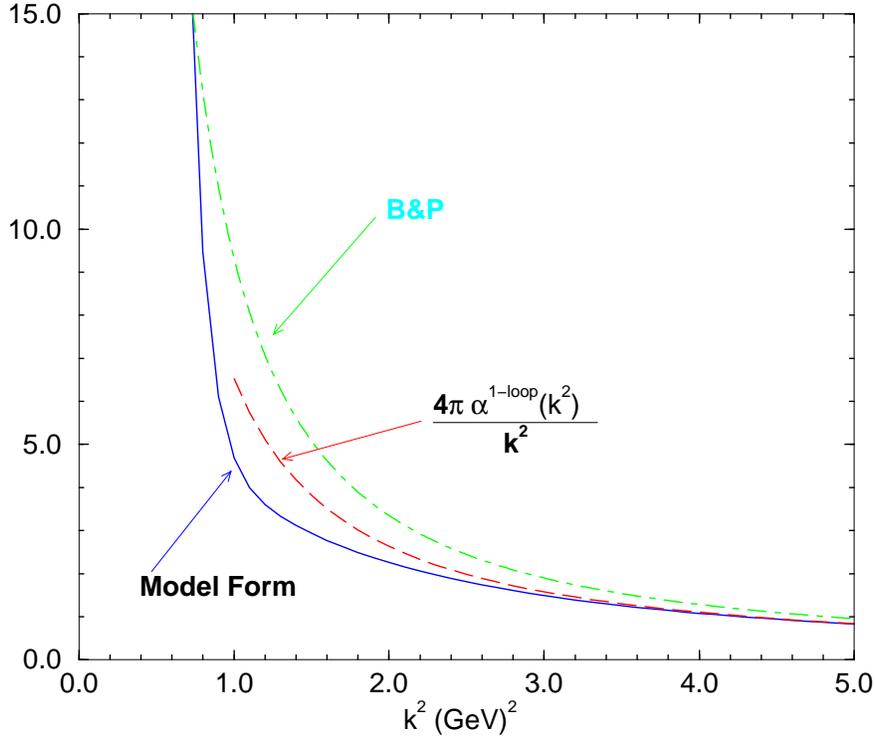,height=10.0cm}}\vspace*{-1.0\baselineskip}
\caption{{\it Ansatz}$\,$ for ${\cal G}(k^2)/k^2$ employed in
Ref.~[\protect\ref{mr97}].  ``B\&P'' labels a
solution$\,^{\protect\ref{bp89}}$ of the gluon DSE, which is presented for
comparison, as is the one-loop running coupling in QCD.
\label{Gmodel}}
\end{figure}

Using Eqs.~(\ref{ouransatz}), (\ref{gk2}), (\ref{params}), and the
renormalisation boundary condition
\begin{equation}
\label{rencond}
\left.S(p)^{-1}\right|_{p^2=\mu^2}  = i\gamma\cdot p + m^\mu\,,
\end{equation}
the quark DSE, Eq.~(\ref{gendse}), is completely specified and can be solved
by iteration.

The chiral limit in QCD is unambiguously defined by $\hat m = 0$.  In this
case there is no perturbative contribution to the scalar piece of the quark
self energy, $B(p^2,\mu^2)$: in fact, there is no scalar, mass-like
divergence in the perturbative calculation of the self energy.  It follows
that $Z_2(\mu^2,\Lambda^2) \,m_{\rm bm}(\Lambda^2)=0\,,\forall \Lambda$ and,
from Eq.~(\ref{rencond}), that there is no subtraction in the equation for
$B(p^2,\mu^2)$.  In terms of the renormalised current-quark mass the
existence of DCSB means that, in the chiral limit, $M(\mu^2) \sim {\rm
O}(1/\mu^2)$, up to $\ln\mu^2$-corrections.\footnote[2]{This is a
model-independent statement; i.e., it is true in any study that preserves the
one-loop renormalisation group behaviour of QCD.}

Figure~\ref{spplot} depicts the renormalised dressed-quark mass function,
$M(p^2)$, obtained by solving the quark DSE using the parameters in
Eq.~(\ref{params}), and in the chiral limit.
\begin{figure}[t]
\centering{\
\epsfig{figure=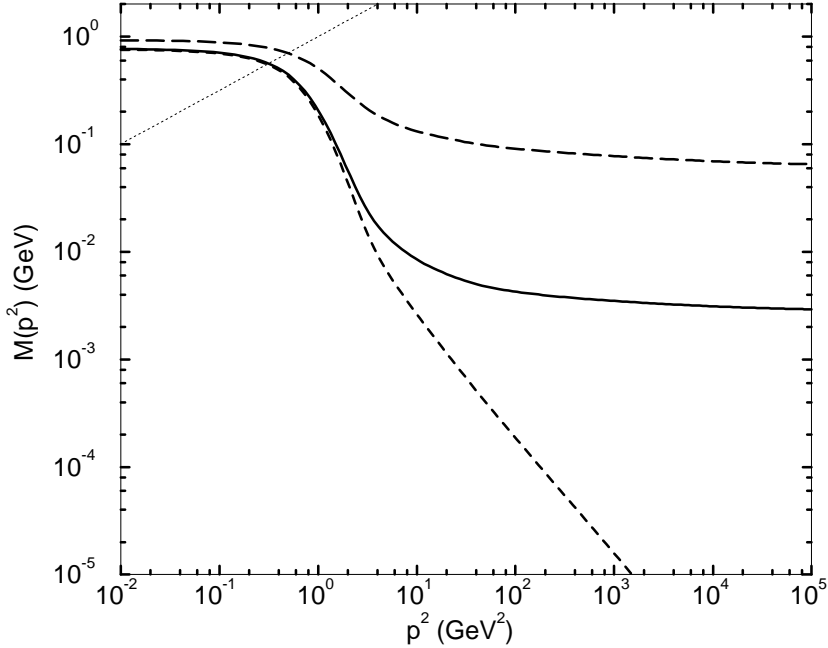,height=10.0cm}}\vspace*{-1.5\baselineskip}
\caption{The renormalised dressed-quark mass function, $M(p^2)$, obtained in
solving the quark DSE using the parameters in Eq.~(\protect\ref{params}):
$u/d$-quark (solid line); $s$-quark (long-dashed line); and chiral limit
(dashed line).  The renormalisation point is $\mu\approx 20\,$GeV.  The
intersection of the line $M^2(p^2)=p^2$ (dotted line) with each curve defines
the Euclidean constituent-quark mass, $M^E$.
\label{spplot}}
\end{figure}
It is a complement to Fig.~\ref{plotMpp} because it highlights the
qualitative difference between the behaviour of $M(p^2)$ in the chiral limit
and in the presence of explicit chiral symmetry breaking.  In the latter case
\begin{equation}
\label{masanom}
 M(p^2) \stackrel{{\rm large}-p^2}{=} \frac{ \hat m}
{\left(\mbox{\footnotesize$\frac{1}{2}$}\ln\left[\frac{p^2}{\Lambda_{\rm QCD}^2}
\right]\right)^{\gamma_m}} \left\{ 1 + {\it two~loop}\right\}\,, \;
\gamma_m=\frac{12}{33- 2 N_f}\,.
\end{equation}
However, in the chiral limit the ultraviolet behaviour is given by
\begin{equation}
\label{Mchiral}
M(p^2) \stackrel{{\rm large}-p^2}{=}\,
\frac{2\pi^2\gamma_m}{3}\,\frac{\left(-\,\langle \bar q q \rangle^0\right)}
           {p^2
        \left(\case{1}{2}\ln\left[\frac{p^2}{\Lambda_{\rm QCD}^2}\right]
        \right)^{1-\gamma_m}}\,,
\end{equation}
where $\langle \bar q q \rangle^0$ is the renormalisation-point-independent
vacuum quark condensate. Analysing the chiral limit solution yields
\begin{equation}
\label{qbqM0}
-\,\langle \bar q q \rangle^0 = (0.227\,{\rm GeV})^3\,,
\end{equation}
which is a reliable means of determining $\langle \bar q q \rangle^0$ because
corrections to Eq.~(\ref{Mchiral}) are suppressed by powers of $\Lambda_{\rm
QCD}^2/\mu^2$.

Equation~(\ref{cbqbq}) defines the renormalisation-point-dependent vacuum
quark condensate
\begin{equation}
\label{qbq19}
\left.-\,\langle \bar q q \rangle_\mu^0 \right|_{\mu=19\,{\rm GeV}}:=  
\left(\lim_{\Lambda\to\infty}
\left. Z_4(\mu,\Lambda)\, N_c \int^\Lambda_q\,{\rm tr}_{\rm Dirac}
        \left[ S_{\hat m =0}(q) \right]\right)\right|_{\mu=19\,{\rm GeV}}\,
= (0.275\,{\rm GeV})^3\,,
\end{equation}
the calculated result.  It is straightforward to establish explicitly that $
m^\mu\,\langle \bar q q \rangle_\mu^0 =\,$constant, independent of $\mu$, and
hence
\begin{equation}
\label{rgiprod}
m^\mu\,\langle \bar q q \rangle_\mu^0 :=  \hat m \,\langle \bar q q
\rangle^0\,,
\end{equation}
which unambiguously defines the renormalisation-point-independent
current-quark masses.  From this and Eqs.~(\ref{params}), (\ref{qbqM0}) and
(\ref{qbq19}) one obtains the values of these masses appropriate to this
model
\begin{equation}
\label{rgimass}
\hat m_{u/d} = 6.60 \, {\rm MeV} \,,\;
\hat m_s = 147\, {\rm MeV}\,.
\end{equation}
Using the one-loop evolution in Eq.~(\ref{masanom}) these values yield
$m_{u/d}^\mu= 3.2\,$MeV and $m_s^\mu= 72\,$MeV, which are within $\sim 10$\%
of the actual values in Eq.~(\ref{params}).  This indicates that higher-loop
corrections to the one-loop formulae, which are present in the solution of
the integral equation as made evident by $A(p^2,\mu^2)\not\equiv 1$ in Landau
gauge, provide contributions of $<10$\% at $p^2 = \mu^2$.  The higher-loop
contributions decrease with increasing $p^2$.

From the renormalisation-point-invariant product in Eq.~(\ref{rgiprod}) one
obtains
\begin{equation}
\label{qbq1}
\left.-\,\langle \bar q q
\rangle_\mu^0\right|_{\mu=1\,{\rm GeV}}
:= \left(\ln\left[1/\Lambda_{\rm QCD}\right]\right)^{\gamma_m}
\, \langle \bar q q \rangle^0
= (0.241\,{\rm GeV})^3\,.
\end{equation}
This result can be compared directly with the value of the quark condensate
employed in contemporary phenomenological studies:\ucite{derek} $ (0.236\pm
0.008\,{\rm GeV})^3$.  Increasing $\omega \to 1.5\,\omega$ in ${\cal G}(k^2)$
raises the calculated value in Eq.~(\ref{qbq1}) by $\sim 10$\%, a weak
sensitivity.

After this discussion of the vacuum quark condensate it is straightforward to
determine the accuracy of Eqs.~(\ref{gmorepi}) and (\ref{gmoreKp}).  Using
experimental values on the left-hand-side, one finds:
\begin{eqnarray}
\label{picf}
(0.0924 \times 0.1385)^2 = (0.113\,{\rm GeV})^4 & \;{\rm cf.} \;& 
(0.111\,{\rm GeV})^4 = 2\times 0.0055 \times 0.24^3 \\
\label{Kcf}
(0.113 \times 0.495)^2 = (0.237\,{\rm GeV})^4& \;{\rm cf.} \;& 
(0.206\,{\rm GeV})^4 = (0.0055 + 0.13)\times 0.24^3\,,
\end{eqnarray}
which indicates that O$(\hat m^2)$-corrections begin to become important at
current-quark masses near that of the $s$-quark, as demonstrated further in
Ref.~[\ref{mr98}].
\vspace*{0.5\baselineskip}

{\it \arabic{section}.\arabic{subsection})~Solving the Pseudoscalar Meson
BSE.}\\[0.3\baselineskip] 
\addcontentsline{toc}{subsection}{\arabic{section}.\arabic{subsection})~Solving
the Pseudoscalar Meson BSE} 
\addtocounter{subsection}{1}
The model quark DSE described above employs the rainbow truncation.
Following Fig.~\ref{skeleton} the consistent Ward-Takahashi identity
preserving truncation of the quark-antiquark scattering kernel is the ladder
approximation:
\begin{equation}
K^{rs}_{tu}(q,k;P) = - \,{\cal G}((k-q)^2)\,D^{\rm free}_{\mu\nu}(k-q)\,
        \left(\gamma_\mu\frac{\lambda^a}{2}\right)_{tr}\,
        \left(\gamma_\nu\frac{\lambda^a}{2}\right)_{su}\,,
\end{equation}
in which case the explicit form of Fig.~(\ref{bsepic}) is
\begin{equation}
\label{bsemod}
\Gamma_H(k;P) + \int^\Lambda_q
{\cal G}((k-q)^2)\, D_{\mu\nu}^{\rm free}(k-q)
 \frac{\lambda^a}{2}\gamma_\mu {\cal S}(q_+)\Gamma_H(q;P){\cal S}(q_-)
\frac{\lambda^a}{2}\gamma_\nu = 0\,.
\end{equation}
Having an {\it Ansatz} for ${\cal G}(k^2)$, $S(q)$ in Eq.~(\ref{bsemod})
follows by solving the quark DSE.  The kernel of the BSE is then completely
specified and solving the equation for $\Gamma_H(k;P)$ and the bound state
mass is a straightforward numerical problem.  Then, with $D_{\mu\nu}(k)$,
$S(p)$ and $\Gamma_H(k;P)$, the calculation of other observables such as: the
leptonic decay constant, $f_H$; meson charge radius, $r_H$; and
electromagnetic form factor, $F_H(Q^2)$; etc., is possible.

The general form of the solution of Eq.~(\ref{bsemod}) is given in
Eq.~(\ref{genpibsa}), where the scalar functions depend on the variables
$k^2$ and $k\cdot P$ and are labelled by the eigenvalue $P^2$.  From this it
is clear that the integrand in Eq.~(\ref{bsemod}) depends on the scalars:
$k^2$, $k\cdot q$, $q^2$, $q\cdot P$ and $P^2$, which takes a fixed-value at
the solution; i.e., at each value of $P^2$ the kernel is a function of four,
independent variables.  Solving Eq.~(\ref{bsemod}) can therefore require
large-scale computing resources, especially since there are four, independent
scalar functions in the general form of the solution.

Two different solution techniques can be employed.  In one procedure, which I
will label: (A), the scalar functions are treated directly as dependent on
two, independent variables: $E(k^2,k\cdot P; P^2)$, etc.  This requires
straightforward, multidimensional integration at every iteration.  Storing
the multidimensional kernel requires a large amount of computer memory but
the iteration proceeds quickly.

An adjunct, which I will label: (B), employs a Chebyshev decomposition
procedure.  It is implemented by writing
\begin{equation}
\label{chebexp}
E(k^2,k\cdot P;P^2) \approx \sum_{i=0}^{N_{\rm max}}
 \,^i\!E(k^2;P^2)\,U_i(\cos\beta)\,,
\end{equation}
with similar expansions for $F$, \underline{$\hat G:= k\cdot P\,G$} and $H$,
where $k\cdot P := \cos\beta\sqrt{k^2 P^2}$ and
$\{U_i(x);i=0,\ldots,\infty\}$ are Chebyshev polynomials of the second kind,
orthonormalised according to:
\begin{equation}
\frac{2}{\pi}\int_{-1}^1\,dx\,\sqrt{1-x^2}\, U_i(x) U_j(x) = \delta_{ij}\,.
\end{equation}
This procedure requires a large amount of time to set up the kernel but does
not require large amounts of computer memory.

\begin{table}[t]
\begin{center}
\begin{tabular}{l|ccc|ccc|ccc}
All amplitudes & 
\multicolumn{3}{c|}{$\pi$} & 
\multicolumn{3}{c|}{chiral limit} & 
\multicolumn{3}{c}{$s\bar s$}\\
& $m_\pi$ & $f_\pi$ & ${\cal R}_H$ &
  $m_0$   & $f^0$   & ${\cal R}_H$ & 
$m_{s\bar s}$ & $f_{s\bar s}$ & ${\cal R}_H$ \\\hline
Method (A) &
0.1385 & 0.0924 & 1.01 & 0.0 & 0.0898 & 1.00 & 0.685 & 0.129 & 1.00 \\
$U_0$ only & 
0.136 & 0.0999 & 0.95 & 0.0 & 0.0972 & 0.94 & 0.675 & 0.137 & 0.95 \\
$U_0$ and $U_1$ &
0.1385 & 0.0925 & 1.00 & 0.0 & 0.0898 & 1.00 & 0.685 & 0.129 & 1.00  
\\ \hline\hline
$E$ only & & & & & & & & & \\\hline
Method (A)
& 0.105 & 0.0667 & 1.82  & 0.0 & 0.0649 & 1.81  & 0.512 & 0.092 & 1.68\\
$U_0$ only & 
0.105 & 0.0667 &1.82 & 0.0 & 0.0649 & 1.81 & 0.513 & 0.092 & 1.69  
\\\hline\hline
$E$, $F$ & & & & & & & & & \\\hline
Method (A)
& 0.136 & 0.0992 & 0.95  & 0.0 & 0.0965 & 0.95  & 0.677 & 0.137 & 0.95\\
$U_0$ only & 
0.136 & 0.0992 & 0.95 & 0.0 & 0.0965 & 0.95 & 0.678 & 0.138 & 0.95 
\\\hline\hline
$E$, $F$, $\rule{0mm}{4mm}\hat G$ & & & & & & & & & \\\hline
Method (A)
& 0.140 & 0.0917 & 1.01  & 0.0 & 0.0891 & 1.00  & 0.688 & 0.128 & 1.01\\
$U_0$ only & 
0.136 & 0.0992 & 0.95 & 0.0 & 0.0965 & 0.95 & 0.678 & 0.138 & 0.95 \\
$U_0$ and $U_1$ & 
0.140 & 0.0917 & 1.01 & 0.0 & 0.0891 & 1.00 & 0.689 & 0.128 & 1.01  
\end{tabular}
\end{center}
\caption{Calculated values of the properties of light, pseudoscalar mesons
composed of a quark and antiquark of equal-mass.  The mass $(m_\pi^{\rm
exp}=0.1385)$ and decay constant ($f_\pi^{\rm exp}=0.0924$) are in GeV,
${\cal R}_H$ is dimensionless.  With the exception of the calculations that
retain only the zeroth Chebyshev moment, labelled by ``$U_0$ only'', the
results are independent of the momentum partitioning parameter, $\eta_P$, in
Eq.~(\protect\ref{uvkernel}).
\label{respi}}
\end{table}
In Tables~\ref{respi} and \ref{resK} I list values of the dimensionless
ratio
\begin{equation}
{\cal R}_H := \,-\,\frac{\langle \bar q q \rangle_\mu^H {\cal M}_H}
        {f_H^2 m_H^2}\,.
\end{equation}
A value of ${\cal R}_H=1$ means that Eq.~(\ref{gmora}) is satisfied and hence
so is the AV-WTI.\footnote[2]{It illustrates that the pseudoscalar-meson pole in
the axial-vector vertex is related to the pseudoscalar-meson pole in the
pseudoscalar vertex in the manner elucidated above.  A finite value in the
chiral limit emphasises that $m_H^2 \propto {\cal M}_H$ as ${\cal M}_H\to
0$.}  Looking at the tabulated values of ${\cal R}_H$ it is clear that the
scalar function $H$ is not quantitatively important, with the AV-WTI being
satisfied numerically with the retention of $E$, $F$ and $G$ in the
pseudoscalar meson Bethe-Salpeter amplitude.  The values of ${\cal R}_H$, and
the other tabulated quantities, highlight the importance of $F$ and $\hat G$:
$F$ is the most important of these functions but $\hat G$ nevertheless
provides a significant contribution, particularly for bound states of
unequal-mass constituents.
\begin{table}[t]
\begin{center}
\begin{tabular}{l|ccc|ccc|ccc}
All amplitudes & 
\multicolumn{3}{c|}{$\eta_P=0.50$} & 
\multicolumn{3}{c|}{$\eta_P=0.25$} & 
\multicolumn{3}{c}{$\eta_P=0.00$}\\
& $m_K$ & $f_K$ & ${\cal R}_K$ &
  $m_K$ & $f_K$ & ${\cal R}_K$ & 
  $m_K$ & $f_K$ & ${\cal R}_K$ \\\hline
Method (A) &
0.497 & 0.109 & 1.01 & 0.497 & 0.109 & 1.01 & 0.497 & 0.109 & 1.01 \\
$U_0$ only & 
0.469 & 0.117 & 0.96 & 0.482 & 0.117 & 0.95 & 0.475 & 0.113 & 1.02 \\
$U_0$ and $U_1$ & 
0.500 & 0.111 & 1.00 & 0.497 & 0.109 & 1.01 & 0.498 & 0.110 & 1.00 \\
$U_0$, $U_1$ and $U_2$ & 
0.497 & 0.109 & 1.01 & 0.497 & 0.109 & 1.01 & 0.496 & 0.109 & 1.01 
\\\hline\hline
$E$ only & & & & & & & & & \\\hline
Method (A) &
0.430 & 0.079 & 1.55 & 0.430 & 0.079 & 1.55 & 0.429 & 0.076 & 1.55 \\
$U_0$ only & 
0.380 & 0.077 & 1.54 & 0.401 & 0.076 & 1.51 & 0.415 & 0.073 & 1.55 \\
$U_0$ and $U_1$ & 
0.439 & 0.089 & 1.52 & 0.430 & 0.078 & 1.55 & 0.431 & 0.076 & 1.57
\\
$U_0$, $U_1$ and $U_2$ & 
0.430 & 0.078 & 1.55 & 0.430 & 0.078 & 1.55 & 0.427 & 0.076 & 1.55 
\\\hline\hline
$E$, $F$ & & & & & & & & & \\\hline
Method (A)
& 0.587 & 0.17  & 0.79  & 0.557 & 0.14  & 0.86  & 0.533 & 0.11 & 0.94\\
$U_0$ only & 
0.505 & 0.12  & 0.82 & 0.518 & 0.11  & 0.86 & 0.512 & 0.11  & 0.96 \\
$U_0$ and $U_1$ & 
 -    &  -     &  -   & 0.556 & 0.14  & 0.86 & 0.537 & 0.12  & 0.94  
\\
$U_0$, $U_1$ and $U_2$ & 
0.583 & 0.16  & 0.79 & 0.557 & 0.14  & 0.86 & 0.532 & 0.12  & 0.93 \\\hline\hline
$E$, $F$, $\rule{0mm}{4mm}\hat G$ & & & & & & & & & \\\hline
Method (A)
& 0.500 & 0.108 & 1.01  & 0.500 & 0.108 & 1.01  & 0.500 & 0.108 & 1.01\\
$U_0$ only & 
0.471 & 0.116 & 0.96 & 0.484 & 0.116 & 0.95 & 0.477 & 0.112 & 1.02 \\
$U_0$ and $U_1$ & 
0.504 & 0.110 & 1.00 & 0.500 & 0.108 & 1.01 & 0.502 & 0.109 & 1.00  \\
$U_0$, $U_1$ and $U_2$ & 
0.500 & 0.108 & 1.01 & 0.500 & 0.108 & 1.01 & 0.499 & 0.108 & 1.01 
\end{tabular}
\end{center}
\caption{Calculated properties of the $K$-meson for various values of the
momentum partitioning parameter, $\eta_P$; ``$-$'' means that no bound state
solution exists in this case.  The mass $(m_K^{\rm exp}=0.496)$ and decay
constant ($f_K^{\rm exp}=0.113$) are in GeV, ${\cal R}_K$ is dimensionless.
\label{resK}}
\end{table}

From Tables~\ref{respi} and \ref{resK}, and Eqs.~(\ref{gmorqbqM}),
(\ref{qbqM}), (\ref{params}) and (\ref{qbq1}), one calculates
\begin{equation}
\begin{array}{ccc}
 -\langle \bar q q \rangle^\pi_{\mu = 1\,{\rm GeV}} &  
-\langle \bar q q\rangle^K_{\mu = 1\,{\rm GeV}} & 
 -\langle\bar q q \rangle^{s\bar s}_{\mu = 1\,{\rm GeV}} \\
& &\\
(0.245 \, {\rm GeV})^3 &
(0.284 \, {\rm GeV})^3 &
(0.317 \, {\rm GeV})^3 
\end{array}
\end{equation}
showing that, for light pseudoscalars, the ``in-meson condensate'' I have
defined increases with increasing bound state mass; as does the leptonic
decay constant, $f_H$.\footnote[2]{$(-\langle \bar q q \rangle^H_{\mu })/f_H$ is
the residue of the bound state pole in the pseudoscalar vertex, just as $f_H$
is the residue of the bound state pole in the axial-vector vertex.  As
expected, $\langle\bar q q\rangle^\pi_{\mu=1\,{\rm GeV}}\approx
\left.\langle\bar q q\rangle^0_\mu\right|_{\mu=1\,{\rm GeV}}$.}  Both of
these trends are modified as one moves into the heavy-quark domain: $-\langle
\bar q q \rangle^H_{\mu } \to\,$const. and $f_H \to 0$ as ${\cal
M}_H\to\infty$.

\begin{figure}[t]
\hspace*{-4mm}\epsfig{figure=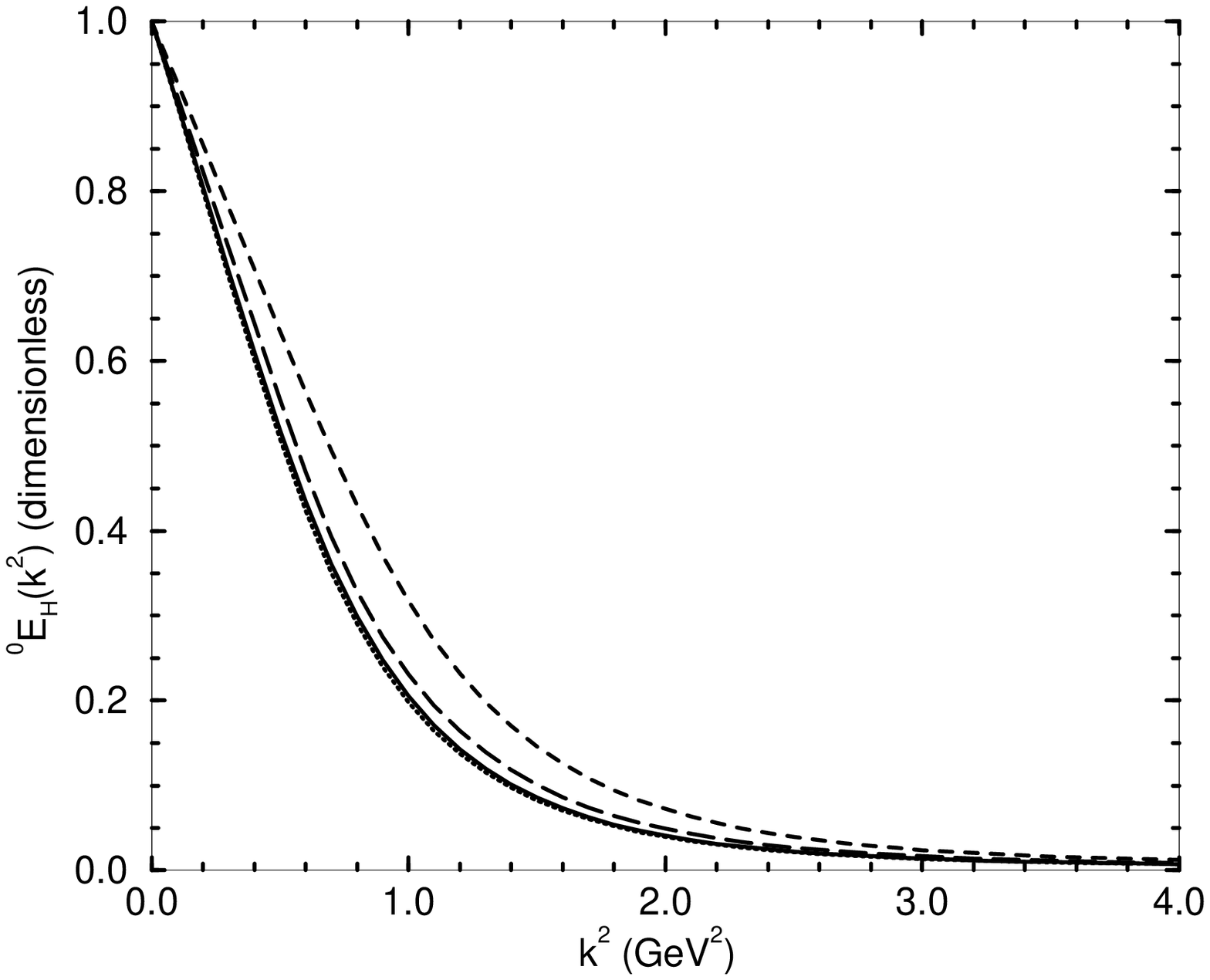,height=7.3cm} \vspace*{-73mm}

\hspace*{78mm}\epsfig{figure=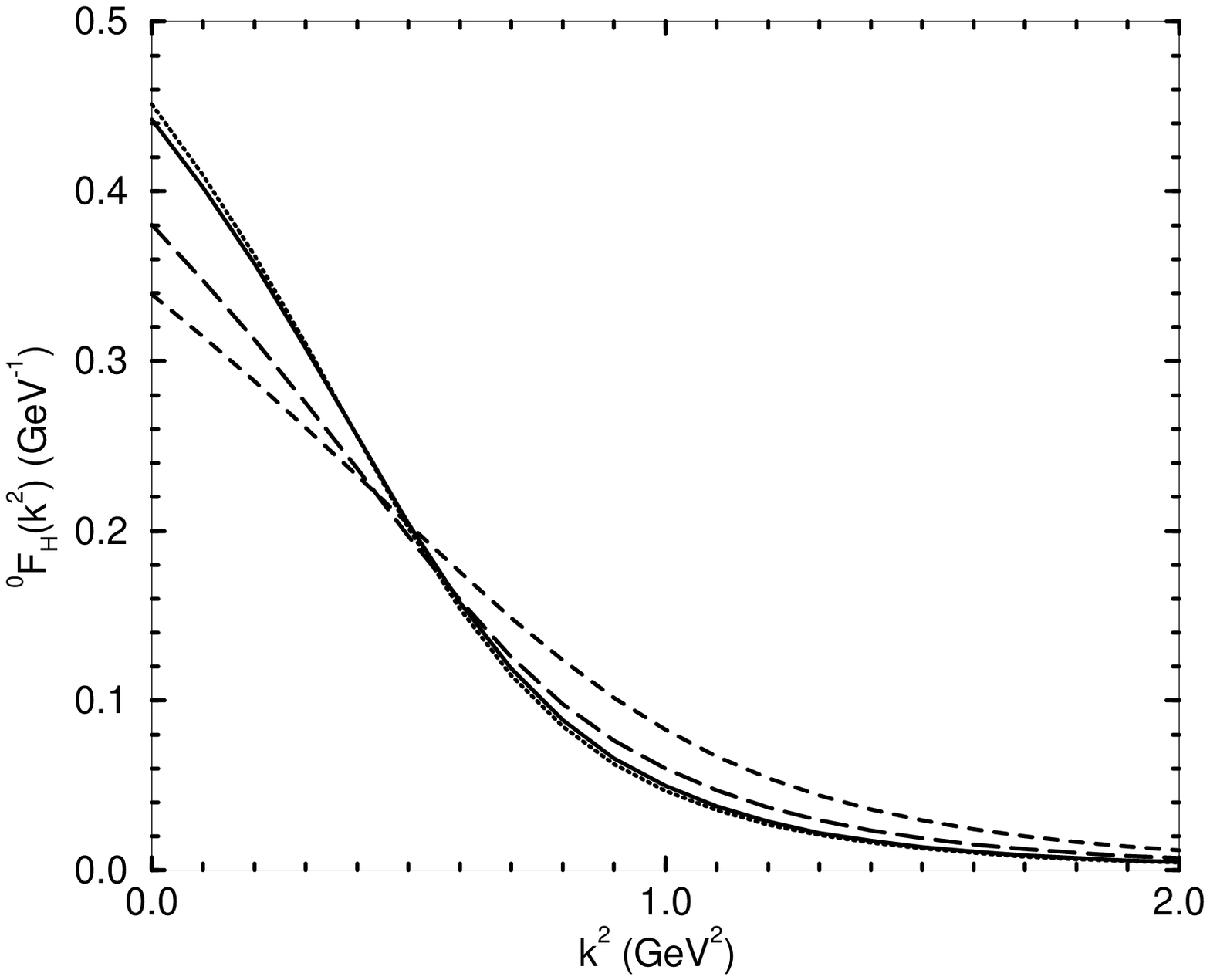,height=7.1cm} \vspace*{-8mm}

\hspace*{-4mm}\epsfig{figure=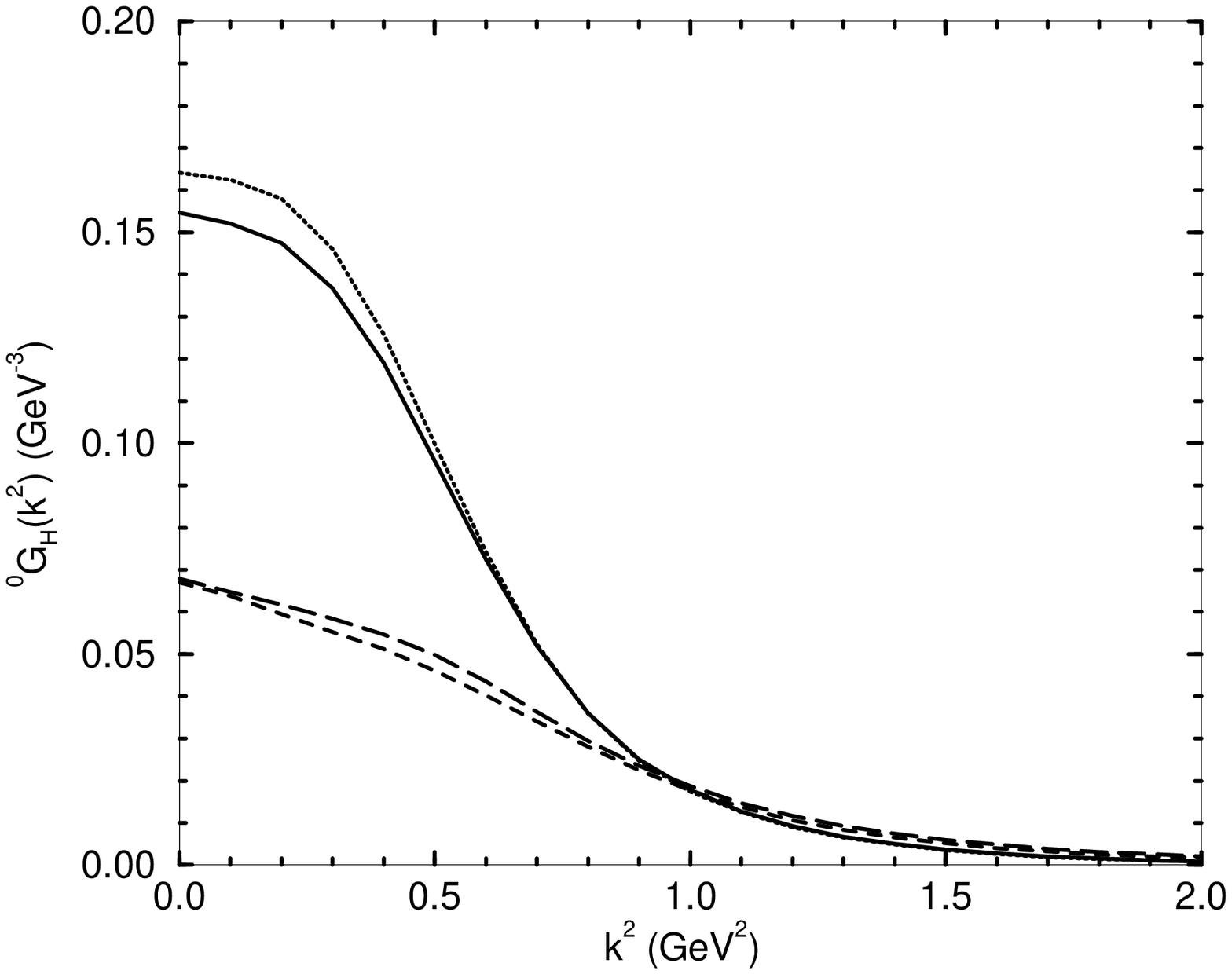,height=7.3cm} \vspace*{-73mm}

\hspace*{78mm}\epsfig{figure=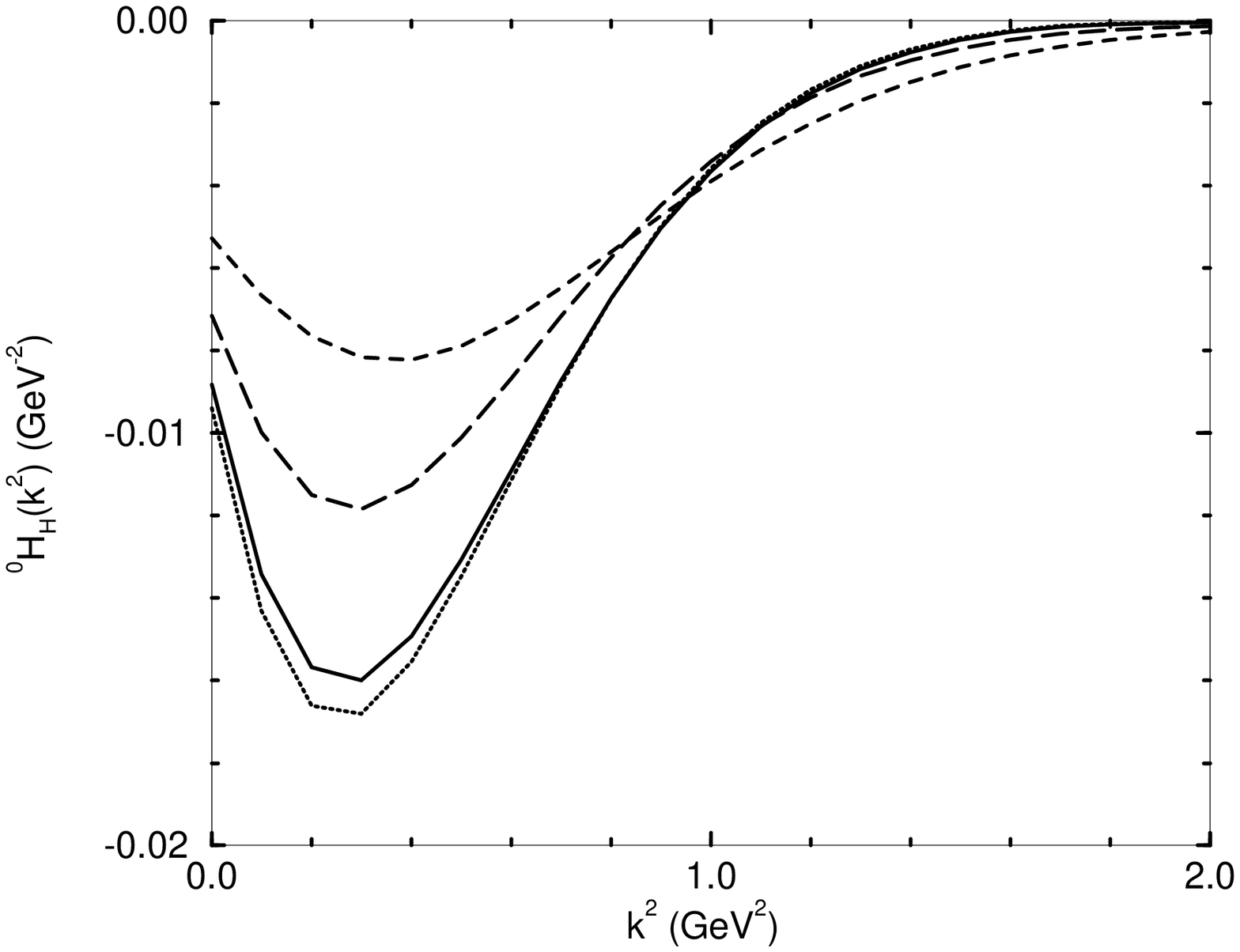,height=7.1cm} \vspace*{-8mm}
\caption{The zeroth Chebyshev moment of the scalar functions in the mesons'
Bethe-Salpeter amplitude: chiral limit (dotted line); $\pi$-meson (solid
line); $K$-meson (long-dashed line); and fictitious, $s\bar s$ bound state
(dashed line).  For ease of comparison the amplitudes are all rescaled so
that $^0\!E_H(k^2=0)=1$.
\label{bsapics}}
\end{figure}

Figure~\ref{bsapics} depicts the scalar functions in the Bethe-Salpeter
amplitude obtained as solutions of Eq.~(\ref{bsemod}), focusing on the zeroth
Chebyshev moment of each function, which is obtained via
\begin{equation}
^0\!E_H(k^2) := 
        \frac{2}{\pi}\int_0^\pi\,d\beta\,\sin^2\beta\,U_0(\cos\beta)\,
                        E_H(k^2,k\cdot P;P^2)\,,
\end{equation}
and similarly for $F$, $G$ [$\hat G$ for the $K$-meson] and $H$.  I note
that: the momentum-space width of $^0\!E_H(k^2)$ increases as the
current-quark mass of the bound state constituents increases;
$^0\!F_H(k^2=0)$ decreases with increasing current-quark mass but that
$^0\!F_H(k^2)$ is still larger at $k^2>0.5\,$GeV$^2$ for bound states of
higher mass; $^0\!G_H(k^2)$ [$^0\!\hat G_K(k^2)$] behaves similarly; and the
same is true for $H_H(k;P)$ but it is uniformly small in magnitude thereby
explaining its quantitative insignificance.

\begin{figure}[t]
\centering{\
\epsfig{figure=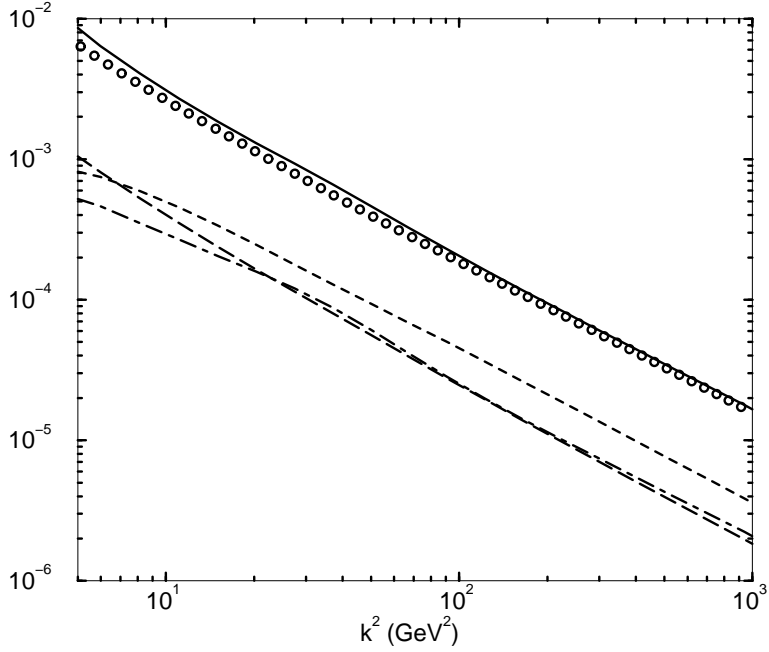,height=10.0cm} }\vspace*{-1.5\baselineskip}
\caption{Asymptotic behaviour of the 0th Chebyshev moments of the functions
in the $\pi$-meson Bethe-Salpeter amplitude: $f_\pi\,^0\!E_\pi(k^2)$ (GeV,
solid line); $f_\pi\,^0\!F_\pi(k^2)$ (dimensionless, long-dashed line);
$k^2\,f_\pi\,^0\!G_\pi(k^2)$ (dimensionless, dashed line); and
$k^2\,f_\pi\,^0\!H_\pi(k^2)$ (GeV, dot-dashed line).  The $k^2$-dependence is
identical to that of the chiral-limit quark mass function, $M(p^2)$,
Eq.~(\protect\ref{Mchiral}).  For other pseudoscalar mesons the momentum
dependence of these functions is qualitatively the same, although the
normalising magnitude differs.
\label{figUV}}
\end{figure}
Figure~\ref{figUV} depicts the large-$k^2$ behaviour of the scalar functions
in the pseudoscalar Bethe-Salpeter amplitude.  The momentum dependence of
$^0\!E_H(k^2)$ at large-$k^2$ is identical to that of the chiral-limit quark
mass function, $M(p^2)$ in Eq.~(\ref{Mchiral}),$^{\protect\ref{m90}}$ and
characterises the form of the quark-quark interaction in the ultraviolet.
Figure~\ref{figUV} elucidates that this is also true of $^0\!F_H(k^2)$,
$k^2\,^0\!G_H(k^2)$ [$k^2\,^0\!\hat G_K(k^2)$ for the $K$-meson] and
$k^2\,^0\!H_H(k^2)$.  Each of these functions reaches its ultraviolet limit
by $k^2 \simeq 10\,$GeV$^2$, which is very-much-less-than the renormalisation
point, $\mu^2=361\,$GeV$^2$.  As I will illustrate below, this result has
important implications for the behaviour of pseudoscalar meson form factors.

A direct verification of Eqs.~(\ref{bwti})-(\ref{gwti}) is possible in this
concrete model.  Consider the inhomogeneous axial-vector vertex equation,
Fig.~\ref{avdse}, in the ladder truncation:
\begin{equation}
\label{ihbseav}
\Gamma_{5\mu}^H(k;P) = 
Z_2 \gamma_5\gamma_\mu \frac{T^H}{2}-
\int^\Lambda_q
{\cal G}((k-q)^2)\, D_{\mu\nu}^{\rm free}(k-q)
 \frac{\lambda^a}{2}\gamma_\mu {\cal S}(q_+)\Gamma_{5\mu}^H(q;P){\cal S}(q_-)
\frac{\lambda^a}{2}\gamma_\nu\,.
\end{equation}
From the homogeneous BSE one already has the equations satisfied by
$E_H(k;0)$, $F_H(k;0)$, $G_H(k;0)$ and $H_H(k;0)$.  To proceed, one
substitutes Eq.~(\ref{truavv}) for $\Gamma_{5\mu}^H(k;P)$ in
Eq.~(\ref{ihbseav}).  Using the coupled equations for $E_H(k;0)$, etc., one
can identify and eliminate each of the pole terms associated with the
pseudoscalar bound state.  [That the homogeneous BSE is linear in the
Bethe-Salpeter amplitude allows this.]  That yields a system of coupled
equations for $F_R(k;0)$, $G_R(k;0)$ and $H_R(k;0)$, which can be solved
without complication.  [The factor of $Z_2$ automatically ensures that
$F_R(k^2=\mu^2;P=0)=1$.]  The realisation of the first two identities,
Eqs.~(\ref{bwti}) and (\ref{fwti}), is illustrated in Fig.~\ref{wtiplot}.
The remaining two identities, Eqs.~(\protect\ref{rgwti}) and
(\protect\ref{gwti}), are realised in a similar fashion.
\begin{figure}[t]
\centering{\
\epsfig{figure=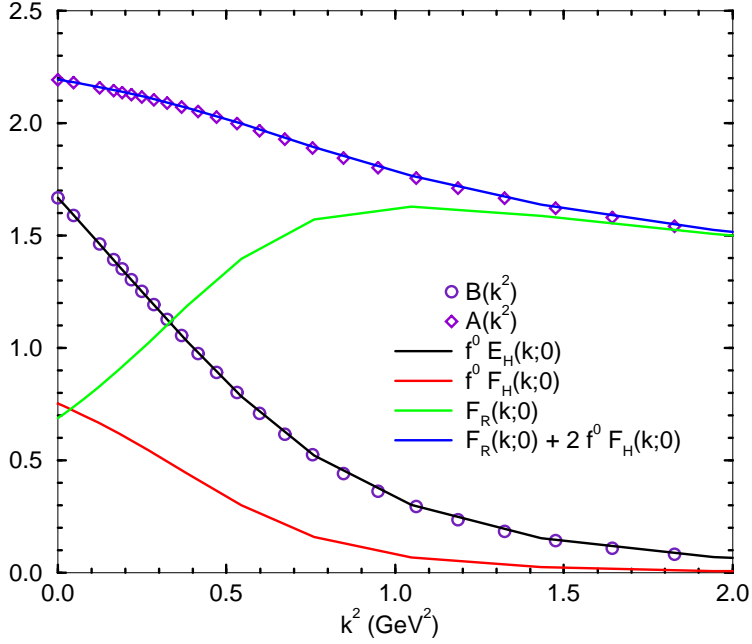,height=10.0cm} }\vspace*{-1.5\baselineskip}
\caption{An illustration of the realisation of the identities
Eqs.~(\protect\ref{bwti}) and (\protect\ref{fwti}), which are a necessary
consequence of preserving the axial vector Ward-Takahashi identity.
\label{wtiplot}}
\end{figure}

\vspace*{\baselineskip}

\addtocounter{section}{1}
\setcounter{subsection}{1}
{\large\bf \arabic{section}.~Additional Phenomenological
Applications.}\\[0.7\baselineskip]  
\addcontentsline{toc}{section}{\arabic{section}.~Additional Phenomenological
Applications} 
In the model illustration of Sec.~4) an algebraic {\it Ansatz} was developed
for the dressed-gluon propagator based on the qualitative behaviour of
solutions of the gluon DSE.  From this basic beginning, I illustrated how one
can proceed to calculate hadronic observables.  A number of qualitatively
significant features emerged, among them DCSB and confinement, all of which
are related to the strong momentum dependence of the quark mass function,
Eq.~(\ref{Mpp}), in the infrared.

That last observation suggests an alternative phenomenological approach:
develop an algebraic {\it Ansatz} for the dressed-quark propagator.  This is
not as fundamental as the approach in Sec.~4) because many, apparently
distinct features of the dressed-quark propagator are encoded in a few
parameters of the {\it Ansatz} for the dressed-gluon propagator; modelling
the dressed-quark propagator directly requires additional parameters to
describe correlated effects.  However, it has a significant merit: with an
algebraic as opposed to a numerical representation of the dressed-quark
propagator it is possible to study scattering observables more quickly and
easily.  The approach can yield quantitatively reliable results provided the
{\it Ansatz} exhibits those essential qualitative features manifest in a
direct solution of the quark DSE using a realistic {\it Ansatz} for the
dressed-gluon propagator.  Further, in allowing a rapid analysis of a broad
range of observable phenomena, it provides a means of exploring the
hypothesis that the bulk of hadronic phenomena are simply a manifestation of
the nonperturbative, momentum-dependent dressing of the elementary Schwinger
functions in QCD.

A simple and widely used model is\ucite{mrpion} 
\begin{eqnarray}
\label{SSM}
\bar\sigma_S(\xi)  & =  & 
        2 \bar m {\cal F}(2 (\xi + \bar m^2))
        + {\cal F}(b_1\, \xi) {\cal F}(b_3\, \xi) 
        \left( b_0 + b_2 {\cal F}(\varepsilon\, \xi)\right)\,,\\
\label{SVM}
\bar\sigma_V(\xi) & = & \frac{2 (\xi+\bar m^2) -1 
                + e^{-2 (\xi+\bar m^2)}}{2 (\xi+\bar m^2)^2}\,,
\end{eqnarray}
with $\bar\sigma_S(\xi):= \lambda\,\sigma_S(p^2)$, $\bar\sigma_V(\xi):=
\lambda^2\,\sigma_V(p^2)$, where $p^2 = \lambda^2\,\xi$, $\lambda$ is a
mass-scale, and ${\cal F}(y) := [1- \exp(-y)]/y$.  This five-parameter form,
where $\bar m$ is the current-quark mass, combines the effects of
confinement\footnote[2]{The representation of $S(p)$ as an entire function is
motivated by the algebraic solutions of Eq.~(\protect\ref{gendse}) in
Refs.~[\protect\ref{munczek},\ref{burden}].} and DCSB with free-particle
behaviour at large, spacelike $p^2$.\footnote[3]{At large-$p^2$: $\sigma_V(p^2)
\sim 1/p^2$ and $\sigma_S(p^2)\sim m/p^2$.  Therefore the parametrisation
does not incorporate the additional $\ln p^2$-suppression characteristic of
QCD: it corresponds to $\gamma_m \to 1$ in Eq.~(\protect\ref{Mchiral}).  It
is a useful but not necessary simplification, which introduces model
artefacts that are easily identified and accounted for.  $\varepsilon=0.01$
is introduced only to decouple the large- and intermediate-$p^2$ domains.}
Applying Eq.~(\ref{cbqbq}) in this case:
\begin{eqnarray}
-\langle \bar q q \rangle^0_\mu & := &
\lim_{M^2\to\infty}\,Z_4(\mu^2,M^2)\,
        \frac{3}{4\pi^2}\,
        \int_0^{M^2}\,ds\,s\,\sigma_S^0(s)\,,
\label{condensate}\\
& = & \lambda^3\,\ln\frac{\mu^2}{\Lambda_{\rm
QCD}^2}\,\frac{3}{4\pi^2}\, 
        \frac{b_0}{b_1\,b_3}\,,
\end{eqnarray}
and the pion mass is given by 
\begin{equation}
\label{gmor}
m_\pi^2\, f_\pi^2 = 2\,m \,\langle \bar q q \rangle^0_{1\, {\rm GeV}}\,.
\end{equation}
In Sec.~4.1) we saw that when all the components of $\Gamma_\pi$ are
retained, Eq.~(\ref{gmor}) yields an approximation to the pion mass found in
a solution of the Bethe-Salpeter equation that is accurate to 2\%.

The model has been used for both $u/d$- and $s$-quark propagators with the
difference between flavours manifest in $b_0^s\neq b_0^{u/d}$, $b_2^s\neq
b_2^{u/d}$ and $m_s \neq m_{u/d}$: the first allows for a difference between
the $K$ and $\pi$ in-meson condensates and the second for $M^E_{s} \neq
M^E_{u/d}$, and all three are phenomenological constraints observed in the
previous section.  As emphasised above, in a solution of the quark DSE using
an {\it Ansatz} for the dressed-gluon propagator, the parameters in
Eq.~(\ref{SSM}) are correlated and one can anticipate this crudely when
fitting them.
\vspace*{0.5\baselineskip}

{\it \arabic{section}.\arabic{subsection})~Pion Electromagnetic Form
Factor.}\\[0.3\baselineskip] 
\addcontentsline{toc}{subsection}{\arabic{section}.\arabic{subsection})~Pion
Electromagnetic Form Factor} 
\addtocounter{subsection}{1}
The renormalised impulse approximation to the electromagnetic pion form
factor is\ucite{mrpion} 
\begin{eqnarray}
\label{pipiA}
\lefteqn{(p_1 + p_2)_\mu\,F_\pi(q^2):= \Lambda_\mu(p_1,p_2)  }\\
& & \nonumber
= \frac{2 N_c}{N_\pi^2}\,\int\,\frac{d^4k}{(2\pi)^4}
        \,{\rm tr}_D\left[ \bar{\cal G}_\pi(k;-p_2)
S(k_{++})\, i\Gamma^\gamma_\mu(k_{++},k_{+-})\,S(k_{+-})\,
{\cal G}_\pi(k-q/2;p_1)\,S(k_{--})\right],
\end{eqnarray}
$k_{\alpha\beta}:= k + \alpha p_1/2 + \beta q/2$ and $p_2:= p_1 + q$,
illustrated in Fig.~\ref{impulse}.
\begin{figure}[t]
  \centering{\
     \epsfig{figure=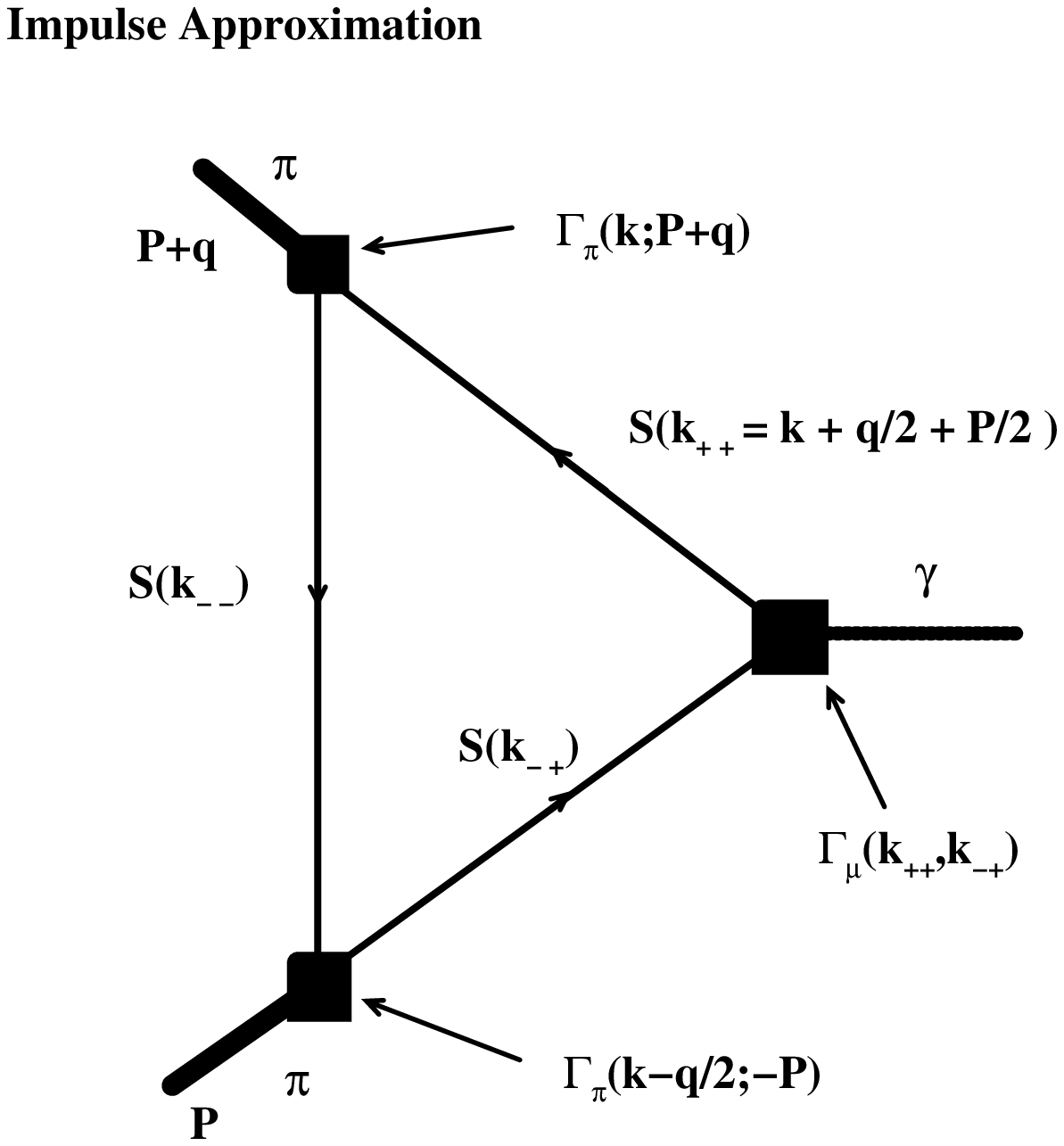,height=10cm}  }\vspace*{-20mm}
\caption{Impulse approximation to $F_\pi(q^2)$: $S$ labels the
dressed-quark propagator; $\Gamma_\pi$ the pion Bethe-Salpeter amplitudes;
and $\Gamma_\mu$ the dressed-quark-photon vertex.
\label{impulse}}
\end{figure}
Here ${\cal G}_\pi(k;P)$ is the pion Bethe-Salpeter amplitude normalised such
that $E(0;0) = B(0)$ in the chiral limit, in which case the consistent
canonical normalisation of the Bethe-Salpeter amplitude is
\begin{eqnarray}
\label{pinorm}
\lefteqn{ 2 \delta^{ij} N_\pi^2\,P_\mu = 
\int^\Lambda_q \left\{\rule{0mm}{5mm}
{\rm tr} \left[ 
\bar{\cal G}_\pi^i(q;-P) \frac{\partial S(q_+)}{\!\!\!\!\!\!\partial P_\mu} 
{\cal G}_\pi^j(q;P) S(q_-) \right]
\right. 
} \\
& & \nonumber \left.  
\;\;\;\;\;\;\;\;\;\;\;\;\;\;\;\;\;\;\;
 + {\rm tr} \left[ 
\bar{\cal G}_\pi^i(q;-P) S(q_+) {\cal G}_\pi^j(q;P) 
        \frac{\partial S(q_-)}{\!\!\!\!\!\!\partial P_\mu}\right]
\rule{0mm}{5mm}\right\}  \,,
\end{eqnarray}
where $\bar{\cal G}_\pi(q;-P)^t := C^{-1}\,{\cal G}_\pi(-q;-P)\,C$ with
$C=\gamma_2\gamma_4$, the charge conjugation matrix, and $X^t$ denotes the
matrix transpose of $X$.  

Given $S$ and Eqs.~(\ref{bwti})-(\ref{gwti}), what form does the
Bethe-Salpeter amplitude take?  

In Sec.~4.2) we saw that the zeroth Chebyshev moments of the pion
Bethe-Salpeter amplitude provided results for $m_\pi$ and $f_\pi$ that were
indistinguishable from those obtained with the full solution.  Also $H_\pi
\simeq 0$ and hence it was quantitatively unimportant in the calculation of
$m_\pi$ and $f_\pi$.  These results are not specific to that particular
model; in the latter case because the right-hand-side of Eq.~(\ref{gwti}) is
zero and hence in general there is no ``seed'' term for $H_\pi$.  We also saw
that at large-$k^2$, independent of assumptions about the form of $K$, one
has
\begin{equation}
\label{uvbeh}
E_\pi^0(k^2;P^2) \propto \,-\langle \bar q q \rangle^0_{k^2}
                \frac{\alpha(k^2)}{k^2}\,, 
\end{equation}
and that the same is true of $F_\pi^0(k^2;P^2)$, $k^2\,G_\pi^0(k^2;P^2)$ and
$k^2\,H_\pi^0(k^2;P^2)$.  This makes manifest the ``hard-gluon'' contribution
to $F_\pi(q^2)$ in Eq.~(\ref{pipiA}).  In addition, in an asymptotically free
theory, one has\ucite{mr97}
\begin{equation}
k^2\,G_\pi^0(k^2;P^2) = 2 F_\pi^0(k^2;P^2)\,,\;
        k^2
\mathrel{\rlap{\lower4pt\hbox{\hskip0pt$\sim$}}
\raise2pt\hbox{$>$}} M_{\rm UV}^2\,,
\end{equation}
with $M_{\rm UV} := 10\, \Lambda_{\rm QCD}$.

These observations, combined with Eqs.~(\ref{bwti})--(\ref{gwti}), motivate a
model for ${\cal G}_\pi$:
\begin{equation}
E_\pi(k;P) = B_0(k^2)
\end{equation}
with 
$F_\pi(k;P)= E_\pi(k;P)/(110\,f_\pi) $,
$G_\pi(k;P)= 2 F_\pi(k;p)/[k^2 + M_{\rm UV}^2]$
and $H_\pi(k;P)\equiv 0$.
The relative magnitude of these functions at large $k^2$ is chosen to
reproduce the numerical results in Fig.~\ref{figUV}.

The final element in Eq.~(\ref{pipiA}) is $\Gamma^\gamma_\mu(p_1,p_2)$, the
renormalised, dressed-quark-photon vertex, and it is because this vertex must
satisfy the vector Ward-Takahashi identity:
\begin{equation}
\label{vwti}
(p_1 - p_2)_\mu \, i\Gamma^\gamma_\mu(p_1,p_2) = 
S^{-1}(p_1) - S^{-1}(p_2)\,,
\end{equation}
that $(p_1 - p_2)_\mu\,\Lambda_\mu(p_1,p_2)=0$ and no renormalisation
constants appear explicitly in Eq.~(\ref{pipiA}).
$\Gamma^\gamma_\mu(p_1,p_2)$ has been much studied\ucite{ayse97} and,
although its exact form remains unknown, its robust qualitative features have
been elucidated so that a phenomenologically efficacious {\it Ansatz} has
emerged\ucite{bc80}
\begin{eqnarray}
\label{bcvtx}
i\Gamma^\gamma_\mu(p,q) &:= &
i\Sigma_A(p^2,q^2)\,\gamma_\mu
+ (p+q)_\mu\,\left[\case{1}{2}i\gamma\cdot (p+q) \, \Delta_A(p^2,q^2)
        + \Delta_B(p^2,q^2)\right];\\
\Sigma_f(p^2,q^2) &:= & \case{1}{2}\,[f(p^2)+f(q^2)]\,,\;
\Delta_f(p^2,q^2)  :=  \frac{f(p^2)-f(q^2)}{p^2-q^2}\,,
\end{eqnarray}
where $f= A, B$.  A feature of Eq.~(\ref{bcvtx}) is that the vertex is
completely determined by the renormalised dressed-quark propagator.  In
Landau gauge the quantitative effect of modifications, such as that canvassed
in Ref.~[\ref{cp92}], is small and can be compensated for by small changes in
the parameters that characterise a given model study.\ucite{hawes}

The model parameters were determined\ucite{mrpion} by optimising a
least-squares fit to $f_\pi$, $r_\pi$ and $\langle \bar q q
\rangle^0_{1\,{\rm GeV}}$, and a selection of pion form factor data on the
domain $q^2\in [0,4]\,$GeV$^2$.  The procedure does not yield a unique
parameter set with, for example, the two sets:
\begin{equation}
\label{Sparams}
\begin{array}{lcccccc}
        & \lambda ({\rm GeV}) & \bar m & b_0 & b_1 & b_2 & b_3 \\
{\sf A} & 0.473               & 0.0127 & 0.329 & 1.51 & 0.429 & 0.430\,,\\
{\sf B} & 0.484               & 0.0125 & 0.314 & 1.63 & 0.445 & 0.405\,,
\end{array}
\end{equation}
providing equally good fits, as illustrated in Table~\ref{tabres}.  
\begin{table}[t]
\begin{center}
\begin{tabular}{cll}
        &       Calculated      &       Experiment \\\hline
$f_\pi$ &       0.092$\,$GeV    &       0.092  \\
$(-\langle \bar q q\rangle^0_{1\,{\rm GeV}})^{1/3}$
        &       0.236           &       0.236 $\pm$ 0.008\ucite{derek} \\
$m_{u/d}$ &     0.006           &       0.008 $\pm$ 0.004\ucite{pdg96} \\
$m_\pi$ &       0.1387          &       0.1385 \\
$r_\pi$ &       0.55$\,$fm      &       0.663 $\pm$ 0.006\ucite{amend}\\
$r_\pi f_\pi$ & 0.25$\,$(dimensionless)&       0.310 $\pm$ 0.003 \\\hline  
\end{tabular}
\end{center}
\caption{Comparison between the calculated values of low-energy pion
observables and experiment or, in the case of $(-\langle \bar q
q\rangle^0_{1\,{\rm GeV}})^{1/3}$ and $m_{u/d}$, the values estimated using
other theoretical tools.  Each set in Eq.~(\protect\ref{Sparams}) yields the
same calculated values. $\Lambda_{\rm QCD}=0.234\,$GeV.
\label{tabres}}
\end{table}
There is a domain of parameter sets that satisfy the fitting criterion and
they are distinguished only by the calculated magnitude of the pion form
factor at large-$q^2$.  The two sets in Eq.~(\ref{params}) delimit reasonable
boundaries and illustrate the model dependence in the result.  In the chiral
limit: $f_\pi^0= 0.090\,$GeV.

The quark propagator obtained with these parameter values is pointwise little
different to that obtained in Ref.~[\protect\ref{cdrpion}].  One gauge of
this is the value of the Euclidean constituent quark mass.  Here $M^E_{u/d}=
0.32\,$GeV whereas $M^E_{u/d}= 0.30\,$GeV in Ref.~[\protect\ref{cdrpion}].
It is also qualitatively similar to the numerical solution described in
Sec.~4.1)\ucite{mr97} where $M^E_{u/d}= 0.56\,$GeV.  Indeed, the results are
not sensitive to details of the fitting function: fitting with different
confining, algebraic forms yields $S(p)$ that is pointwise little changed,
and the same results for observables.  The earlier
parametrisation\ucite{cdrpion} has been applied more widely, as reviewed in
Ref.~[\ref{pctrev}], and Table~\ref{oldresults} summarises the results.
\begin{table}[t]
\begin{center}
\begin{tabular}{cll}
   &   Calculated  &   ``Experiment''  \\  \hline  
  $f_{\pi} \; $    &  ~0.0924 GeV &   ~0.0924 $\pm$ 0.001     \\  
  $f_{K} \; $    &  ~0.113  &   ~0.113 $\pm$ 0.001     \\  
  $m_{\pi} \; $    & ~0.139  & ~0.138  \\   
  $m_{K} \; $    & ~0.494  & ~0.494  \\   
  $m^{\rm ave}_{1\,{\rm GeV}^2}$ & ~0.0045 &   ~0.008 \\  
   $m^s_{1\,{\rm GeV}^2}$ & ~0.113 &   ~0.1 $\sim$ 0.3 \\    
   $-\langle \bar q q \rangle^{\frac{1}{3}}_{1\,{\rm GeV}^2}$ & ~0.247 &
          ~$0.236 \pm 0.008$   \\  
  $r_{\pi^\pm} \;$ & ~0.55 fm & ~0.663 $\pm$ 0.006  \\   
 $r_{K^\pm} \;$
         & ~0.49\ucite{kaonFF}  & ~0.583 $\pm$ 0.041  \\   
 $r_{K^0}^2 \;$
        & -0.020 fm$^2$ & -0.054 $\pm$ 0.026  \\   
 $g_{\pi^0\gamma\gamma}\;$ 
        & ~0.50\ucite{cdrpion} (dimensionless) & ~0.504 $\pm$ 0.019\\ 
 $F^{3\pi}(4m_\pi^2)\;$
        & ~$1.04$\ucite{gppp} & ~$1$ (Anomaly)\\ 
 $a_0^0 \;  $
        & ~0.19\ucite{pipi}  & ~0.26 $\pm$ 0.05 \\  
 $a_0^2 \;  $ & -0.054 & -0.028 $\pm$ 0.012 \\  
 $2 a_0^0-5 a_0^2\;$ & ~0.65 & ~0.66 $\pm$ 0.12 \\  
 $a_1^1 \;  $ & ~0.038 & ~0.038 $\pm$ 0.002\\  
 $a_2^0 \; $  & ~0.0017 & ~0.0017 $\pm$ 0.0003\\  
 $a_2^2 \;  $ & -0.00029 & \\  
  $f_{K}/f_\pi \; $    &  ~1.22 &   ~1.22 $\pm$ 0.01     \\  
 $r_{K^\pm} /r_{\pi^\pm}$ & ~0.87  & ~0.88 $\pm$ 0.06 \\\hline    
\end{tabular}
\end{center}
\caption{Summary of results obtained using the parametrisation of $S(p)$
introduced in Ref.~[\protect\ref{cdrpion}], which differs little from that
specified by Eqs.~(\protect\ref{SSM}) and (\protect\ref{SVM}).  $a^I_J$ are
$\pi$-$\pi$ scattering lengths, whose experimental values are discussed in
Ref.~[\protect\ref{pocanic}], and $F^{3\pi}(4m_\pi^2)$ is the value of the
$\gamma\pi \to \pi\pi$ transition form factor at the softest point
kinematically accessible.  The citations for the calculated results specify
the article in which the annotated study is described.  The ``experimental''
values of the current-quark masses and the quark condensate are estimates
made using other theoretical tools: see Table~\protect\ref{tabres}.
\label{oldresults}}
\end{table}

In the calculations $f_\pi r_\pi$ is 20\% too small.  This discrepancy cannot
be reduced in impulse approximation because the nonanalytic contributions to
the dressed-quark-photon vertex associated with $\pi$-$\pi$ rescattering and
the tail of the $\rho$-meson resonance are ignored.\ucite{ewff} It can only
be eliminated if these contributions are included.  This identifies a
constraint on realistic, impulse approximation calculations: they should not
reproduce the experimental value of $f_\pi r_\pi$ to better than $\sim20$\%,
otherwise the model employed has unphysical degrees-of-freedom.

The pion form factor calculated\ucite{mrpion} using Eqs.~(\ref{SSM}) and
(\ref{SVM}) with (\ref{Sparams}) is compared with available data in
Figs.~\ref{smallF} and \ref{midF}.  It is also compared with the result
obtained in Ref.~[\ref{cdrpion}], wherein the calculation is identical {\em
except} that the pseudovector components of the pion were neglected.
\begin{figure}[t]
\centering{\
\epsfig{figure=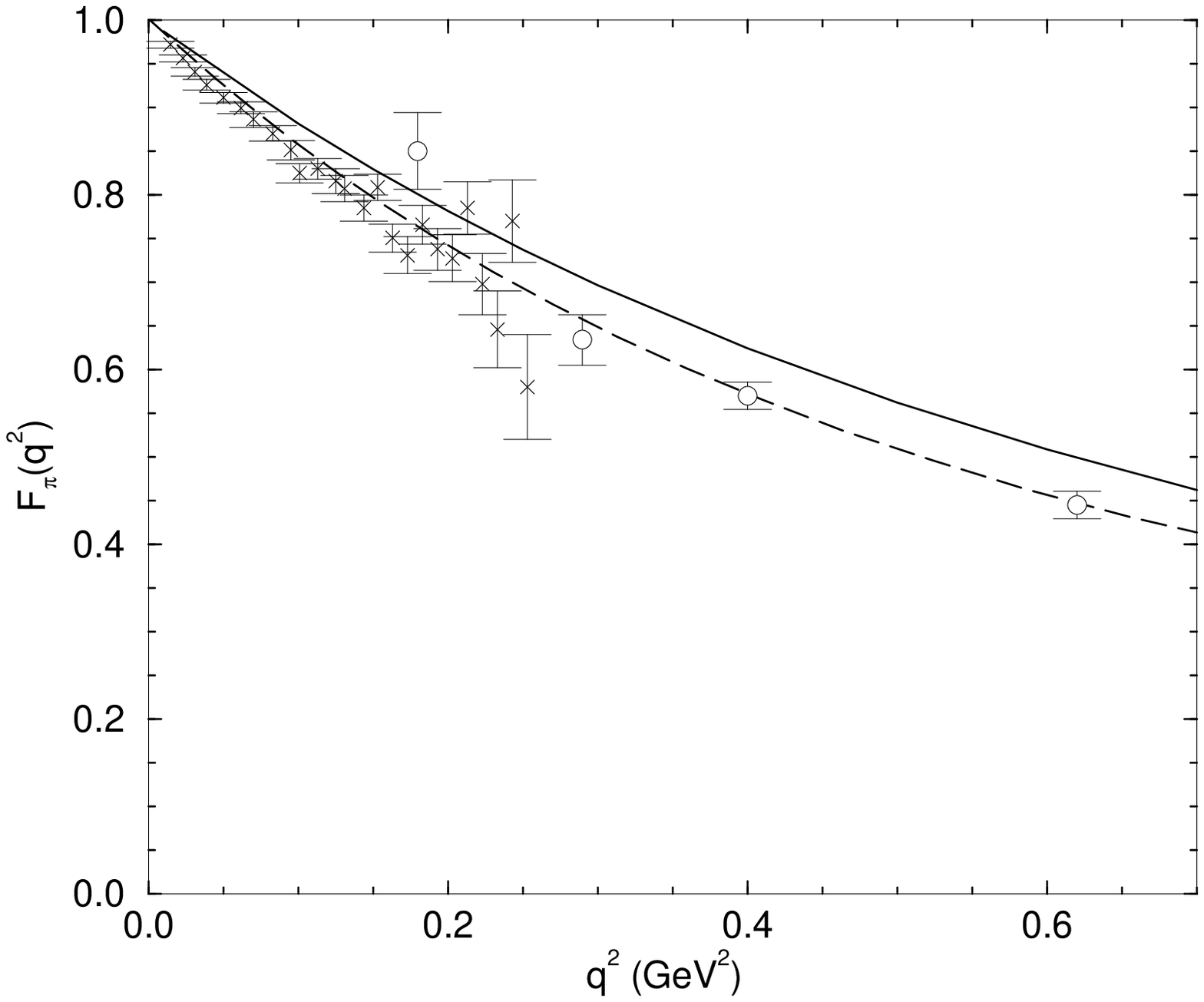,height=10.0cm}}
\caption{Calculated pion form factor compared with data at small $q^2$.  The
data are from Refs.~[\protect\ref{amend}] (crosses) and [\protect\ref{bebek}]
(circles).  The solid line is the result obtained when the pseudovector
components of the pion Bethe-Salpeter amplitude are included, the dashed-line
when they are neglected.$\,^{\protect\ref{cdrpion}}$ On the scale of this
figure, both parameter sets in Eq.~(\protect\ref{Sparams}) yield the same
calculated result.
\label{smallF}}
\end{figure}
Figure~\ref{smallF} shows a small, systematic discrepancy between the
calculations and the data at low $q^2$, which is due to the underestimate of
$r_\pi$ in impulse approximation.\footnote[2]{Just as in the present
calculation, $f_\pi r_\pi = 0.25$ in Ref.~\protect\ref{cdrpion}.  However,
the mass-scale is fixed so that $f_\pi=0.084$, which is why this result
appears to agree better with the data at small-$q^2$: $r_\pi$ is larger.}
The results obtained with or without the pseudovector components of the pion
Bethe-Salpeter amplitude are quantitatively the same, which indicates that
the pseudoscalar component, $E_\pi$, is dominant in this domain.

\begin{figure}[t]
\centering{\
\epsfig{figure=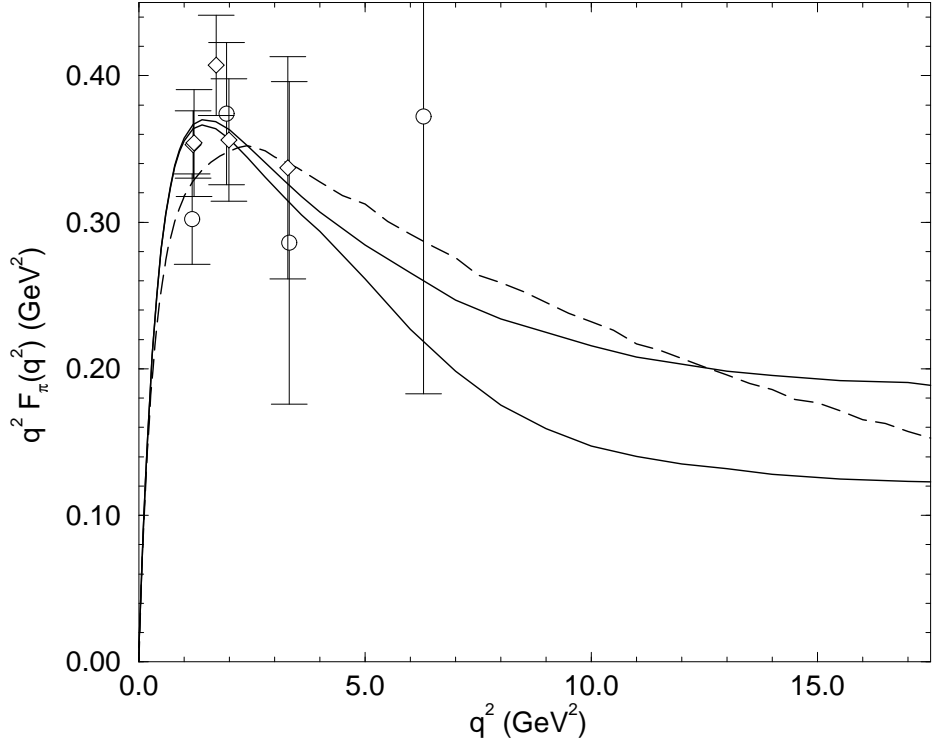,height=10.0cm}}
\caption{Calculated pion form factor compared with the largest $q^2$ data
available: diamonds - Ref.~[\protect\ref{bebek}]; and circles -
Ref.~[\protect\ref{bebekB}].  The solid lines are the results obtained when
the pseudovector components of the pion Bethe-Salpeter amplitude are included
(lower line - set {\sf A} in Eq.~(\protect\ref{Sparams}); upper line - set
{\sf B}), the dashed-line when they are
neglected.$\,^{\protect\ref{cdrpion}}$
\label{midF}}
\end{figure}
The increasing uncertainty in the experimental data at intermediate $q^2$ is
apparent in Fig.~\ref{midF}, as is the difference between the results
calculated with or without the pseudovector components of the pion
Bethe-Salpeter amplitude.  These components provide the dominant contribution
to $F_\pi(q^2)$ at large pion energy because of the multiplicative factors:
$\gamma\cdot P$ and $\gamma\cdot k\,k\cdot P$, which contribute an additional
power of $q^2$ in the numerator of those terms involving $F^2$, $FG$ and
$G^2$ relative to those proportional to $E$.  Using the method of
Ref.~[\ref{cdrpion}] and the model-independent asymptotic behaviour indicated
by Eq.~(\ref{uvbeh}) one finds
\begin{equation}
\label{FUV}
F_\pi(q^2) \propto \frac{\alpha(q^2)}{q^2}\,
        \frac{(-\langle \bar q q\rangle^0_{q^2})^2}{f_\pi^4}\,;
\end{equation}
i.e., $q^2 F_\pi(q^2) \approx {\rm const.}$, up to calculable $\ln
q^2$-corrections.  If the pseudovector components of $\Gamma_\pi$ are
neglected, the additional numerator factor of $q^2$ is missing and one
obtains\ucite{cdrpion} $q^4 F_\pi(q^2)\approx {\rm const.}$

With this model the behaviour identified in Eq.~(\ref{FUV}) becomes apparent at
$q^2 \mathrel{\rlap{\lower4pt\hbox{\hskip0pt$\sim$}}
\raise2pt\hbox{$>$}} 2\,M_{UV}^2
$.
This is the domain on which the asymptotic behaviour indicated by
Eq.~(\ref{uvbeh}) is manifest.  The calculated results, obtained with the two
sets of parameters in Eq.~(\ref{params}), illustrate the model dependent
uncertainty:
\begin{equation}
\left.q^2 F_\pi(q^2)\right|_{q^2 \sim 10-15\,{\rm GeV}^2} \sim 0.12 -
0.19\,{\rm GeV}^2\,.
\end{equation}
It arises primarily because the model allows for a change in one parameter to
be compensated by a change in another.  In this example: $b_2^{\sf B} >
b_2^{\sf A}$ but $b_0^{\sf B}+b_2^{\sf B}=b_0^{\sf A}+b_2^{\sf A}$; and
$b_1^{\sf A}\,b_3^{\sf A}= b_1^{\sf B}\,b_3^{\sf B}$, which allows an equally
good fit to low-energy properties but alters the intermediate-$q^2$ behaviour
of $F_\pi(q^2)$.  As emphasised, in a solution of Eq.~(\ref{gendse}) these
coefficients of the $1/p^4$ and $1/p^6$ terms are correlated and such
compensations cannot occur.

As a comparison, evaluating the leading-order perturbative-QCD result with
the asymptotic quark distribution amplitude:
$\phi_{\rm as}(x) := \surd 3\,f_\pi\, x (1-x)$, 
yields 
\begin{equation}
q^2 F_\pi(q^2) = 16\,\pi f_\pi^2 \,\alpha(q^2) \approx 0.15\,{\rm GeV}^2\,,
\end{equation} 
assuming a value of $\alpha(q^2\sim 10\,{\rm GeV}^2)\approx 0.3$.  However,
the perturbative analysis neglects the anomalous dimension accompanying
condensate formation; e.g., Eqs.~(\ref{bwti})-(\ref{gwti}) are not satisfied
in Ref.~[\protect\ref{FJ}].
\vspace*{0.5\baselineskip}

{\it \arabic{section}.\arabic{subsection})~Electroproduction of Vector
Mesons.}\\[0.3\baselineskip] 
\addcontentsline{toc}{subsection}{\arabic{section}.\arabic{subsection})~Electroproduction
of Vector Mesons} 
\addtocounter{subsection}{1}
There is an extensive body of literature describing Pomeron phenomenology,
all derived from the observation that the total cross section in high-energy
scattering: $p$-$p$, $\bar p$-$p$, $\pi^\pm$-$p$, $\gamma$-$p$, etc., is
forward-peaked and rises slowly with $\surd s$.  This is
illustrated\ucite{pdg96} in Fig.~\ref{pptotal} and can be
described$\,^{\ref{pichowsky}}$
\begin{figure}[t]
\centering{\ \epsfig{figure=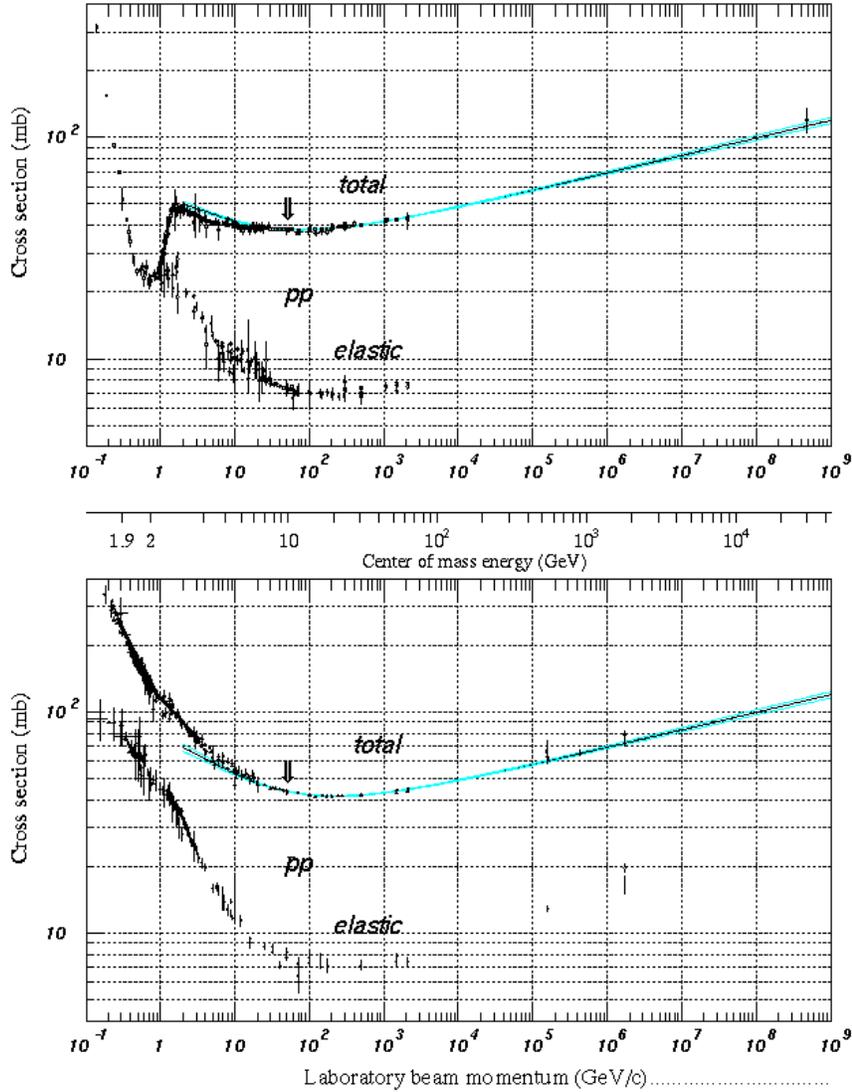,height=150mm}}
\caption{Total and elastic cross sections for $p$-$p$ and $\bar p$-$p$
scattering.  The slow increase of the total cross section with $\surd s$ at
high energy is obvious.
\label{pptotal}}
\end{figure}
by a Pomeron-exchange model of the quark-nucleon interaction with the
following features:
\begin{enumerate}
\item The quark-Pomeron coupling is $\bar q^f(p_2) \Gamma_\mu^f q^f(p_1)$,
where $\Gamma_\mu:= \beta_f\, \gamma_\mu$ with $\beta_f$ a flavour-dependent
coupling constant.  It is the only flavour-dependence in the interaction.
\item The Pomeron ``propagator'' is characterised by a Regge trajectory:
\begin{equation}
G(s,t):= (\alpha_1 s)^{\alpha_0 + \alpha_1 t}
\end{equation}
with $\alpha_0>0$, which ensures the increase with $s$, and the
Pomeron-nucleon coupling is $3 \beta_u F_1(t)$, where $F_1(t)$ is the Dirac
form factor of the proton.
\item The interaction is used in impulse approximation so that, for example,
the $\pi N \to \pi N$ interaction is completely described by
\begin{equation}
\langle P; p_2 m_s^\prime | T_{\pi N \to \pi N} | q; p_1 m_s\rangle
:= 
\Lambda_\mu(q,P)\,3\beta_{u/d}\,F_1(t)\,G(s,t)\,
        \bar u_{m_s^\prime}(p_2) \gamma_\mu u_{m_s}(p_1)\,,
\end{equation}
where $u_{m_s}(p_1)$ is a nucleon spinor and 
$\Lambda_\mu(q,P):= 2 \Lambda_\mu^u(q,P) + 2 \Lambda_\mu^{\bar d}(q,P)$ with
\begin{equation}
\Lambda_\mu^f(q,P):= N_c\,{\rm tr}_D\,
        \int\case{d^4k}{(2\pi)^4} S_u(k_{-+}) \Gamma_\pi(k_{0-})
                        S_d(k_{--})\bar\Gamma_\pi(k)S_u(k_{++})
                \beta_f i\gamma_\mu
\end{equation}
describing the interaction of the Pomeron with the $f$-quark in the pion.
\end{enumerate}
The parameters: $\beta_f$, $\alpha_0$, $\alpha_1$, in this model were
fixed$\,^{\ref{pichowsky}}$ by requiring a good description of $\pi N$ and $K
N$ scattering, and this is achieved with
\begin{equation}
\begin{array}{cccc}
\beta_{u/d}  = 2.35 \,{\rm GeV}^{-2},\; 
        & \beta_s = 1.50 \,{\rm GeV}^{-2},\; & 
\alpha_0= 0.10,\; &  \alpha_1 = 0.33\,{\rm GeV}^{-2}\,.
\end{array}
\end{equation}

In the diffractive regime the process $e^{-}\, p \to e^{-\prime}\, p^\prime\,
V$, where $V= \rho$, $\phi$, $\psi$, is also expected to proceed via
soft-Pomeron exchange and the model introduced above can be applied directly.
The matrix element is
\begin{eqnarray}
\langle p_2 m_2;k \lambda_{\rho} | J_{\mu} | p_1 m_1 \rangle 
& = & 2 \beta_f t_{\mu \nu \lambda}(q,k) \,
\varepsilon_{\lambda}^{\lambda_{\rho}}(k)  \, 
 G_{\cal P}(\bar{w}^2,t)\, 3\beta_u F_1(t)
\, \bar{u}(p_2) \gamma_\nu u(p_1) \,,
\end{eqnarray}
depicted in Fig.~\ref{electro},
\begin{figure}[t] 
\centering{\ 
\epsfig{figure=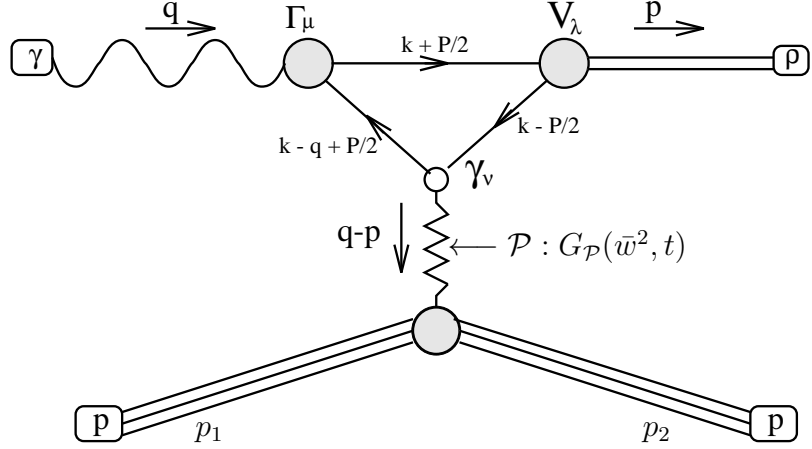,height=6.0cm} }
        \vspace*{-30mm}

        \hspace*{40mm} $\longleftarrow  {\cal P}: 
                         G_{\cal P}(\bar w^2,t)$ 
        \vspace*{19mm}

\hspace*{3mm} $p_1$ \hspace*{53mm} $p_2$

\caption{$\rho$-meson electroproduction matrix element.  $\bar \omega^2:= -
(q - P/2 + p_1)^2$, $W^2:=-(P+p_2)^2$.\label{electro}}
\end{figure}
where the $\gamma {\cal P} \to \rho$ transition form factor is\\
\parbox{160mm}{
\begin{eqnarray}
\label{gPV}
\lefteqn{t_{\mu \nu \lambda}(q,P) = 3 e_0 \,\int \frac{d^{4}k}{(2\pi)^4} \; 
{\rm tr} \left\{ S(k+\case{1}{2}P) \; \times \right. }\\
&  &  \nonumber 
\left. \Gamma_{\mu}^\gamma(k+\case{1}{2}P,k-q+\case{1}{2}P)
        \,S(k-q+\case{1}{2}P) \;
 \gamma_{\nu} \; S(k-\case{1}{2}P) \; V_{\lambda}(k;-P)\right\}.
\end{eqnarray}}

\begin{figure}[t]
\centering{\ 
\epsfig{figure=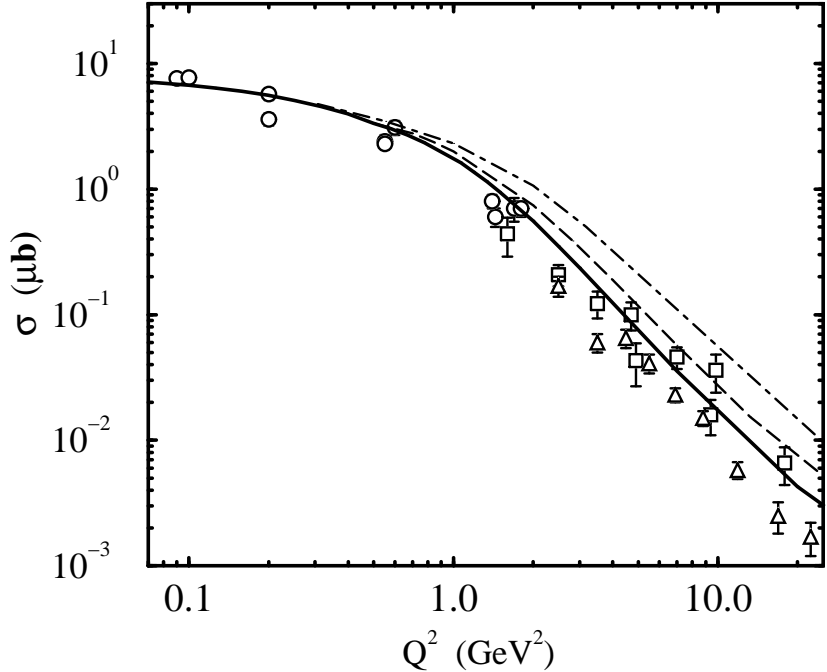,height=10.0cm}}
\caption{$\rho$-meson electroproduction cross section at $W=15\,$GeV: solid
line -- calculated result; long-dashed line -- result if $\bar m_{u/d}\to
10\, \bar m_{u/d}$; dash-dot line -- result if $\bar m_{u/d}\to 25\, \bar
m_{u/d} \simeq \bar m_s$.  The data are: circles --
Ref.~[\protect\ref{sham}]; squares -- Ref.~[\protect\ref{aubert}]; triangles
-- Ref.~[\protect\ref{arneodo}].
\label{rhoEP}}
\end{figure}
\begin{figure}[t]
\centering{\ 
\epsfig{figure=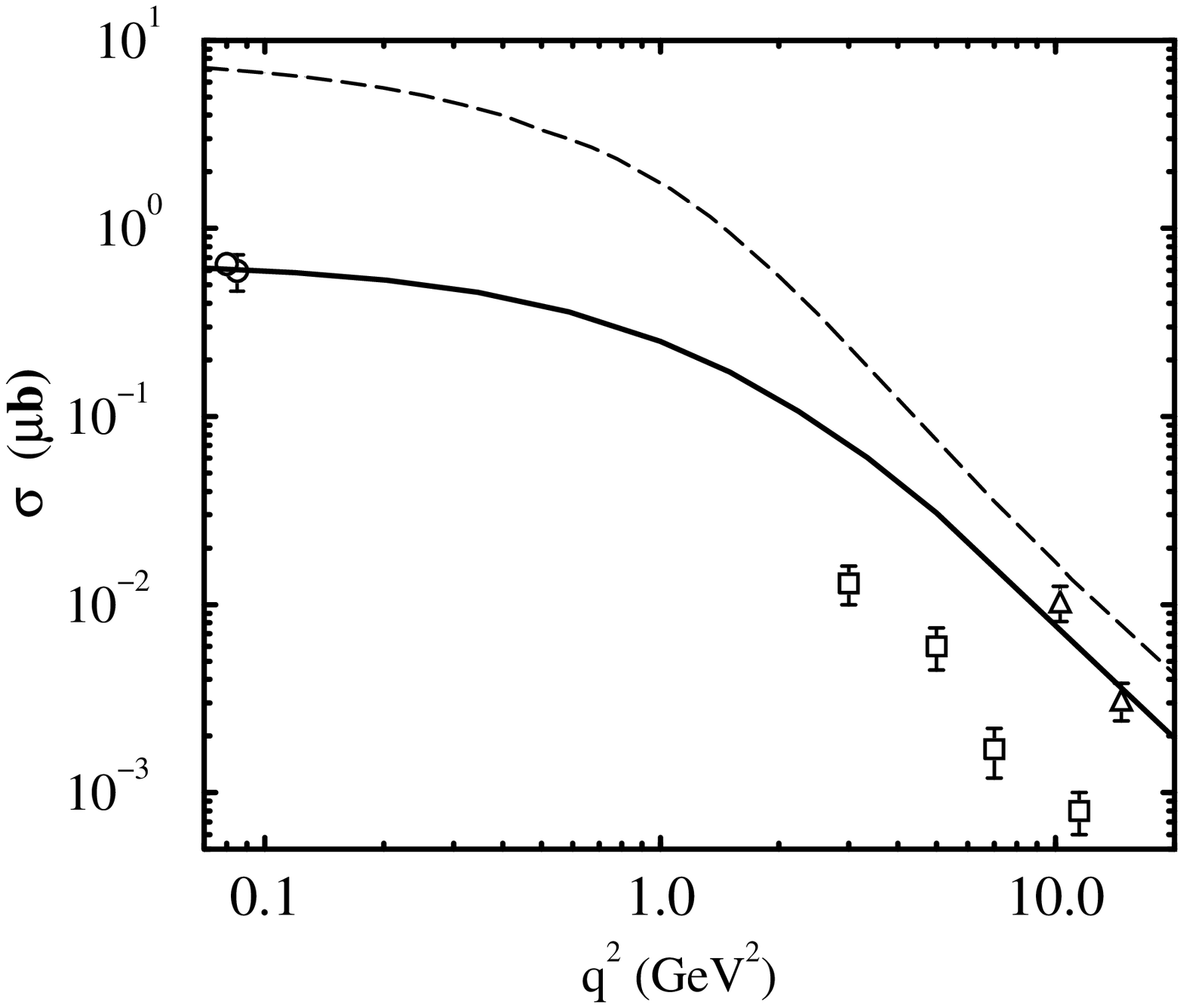,height=10.0cm}}
\caption{$\phi$-meson electroproduction cross section at $W=70\,$GeV: solid
line; the dashed line is the $\rho$-meson result for comparison.  The data
are: triangles -- Ref.~[\protect\ref{arneodo}]; circles --
Ref.~[\protect\ref{derrick}]; squares -- Ref.~[\protect\ref{derrickb}].
\label{phiEP}}
\end{figure}

The unknown quantity in Eq.~(\ref{gPV}) is the vector meson Bethe-Salpeter
amplitude, $V_{\nu}(k;-P)$.  In the absence of a solution of the associated
Bethe-Salpeter equation, an oft used and phenomenologically efficacious
procedure\ucite{pctrev} is to parametrise the amplitude in a manner similar
to that employed for the pion in Sec.~5.1):
\begin{equation}
V_\nu(k;P) 
        = \left(\gamma_\nu + \frac{P_\nu \gamma\cdot P}{m_V^2}\right)
        \case{1}{N_V}\,
        \left\{ {\rm e}^{-k^2/a^2_V} + \frac{c_V}{1+k^2/b_V^2}\right\}\,,
\end{equation}
where $N_V$ is fixed via the canonical normalisation condition: clearly,
$P\cdot V(k;P)=0$.  The parameters are 
\begin{equation}
\begin{array}{llll}
        &       a\,({\rm GeV}) &       b\,({\rm GeV}) &       c       \\
\rho    & 0.40         &         0.008 &        125.0 \\
\phi    & 0.45          &       0.6     &       0.3     \\
\psi    & 1.10          &       0.0     &       0.0     
\end{array}
\end{equation}
which were fixed$\,^{\ref{pichowsky}}$ by requiring the fit to the
dimensionless coupling constants in Eq.~(\ref{vmobs}).
\begin{equation}
\label{vmobs}
\begin{array}{lccccc}
&g_{\rho\to e^+ e^-} &
g_{\rho\to \pi^+ \pi^-} & 
g_{\phi\to e^+ e^-} & 
g_{\phi\to K \bar K}&
g_{\psi\to e^+ e^-}       \\
{\rm Theory} &
4.6  & 6.8 & 12.7 & 3.9 & 11.5\\
{\rm Experiment} &
5.0 & 6.1 &  12.9 & 4.6 & 11.5
\end{array}
\end{equation}
The experimental values of the decay constants were calculated from the
widths in Ref.~[\ref{pdg96}], and the fit is acceptable given the simplicity
of the {\it Ansatz} for $V_\nu$, which includes only one of the eight scalar
functions necessary to completely specify a vector meson bound state.  At
this point there are no free parameters in the calculation of the
electroproduction cross sections.

Figure~\ref{rhoEP} depicts the $Q^2$-dependence of the $\rho$-meson
electroproduction cross section and the magnitude is a prediction.  There is
complete agreement on the entire range of accessible $Q^2$, with the large
$Q^2$ behaviour:\ucite{pichowsky} $1/Q^4$, which becomes evident at $Q^2
\simeq 1$-$2\,$GeV$^2$.  Below that point the nonperturbative character of
the dressed-quark propagator dominates the evolution of the cross section.
It is important to observe the prediction that, the larger the current-quark
mass of the constituents, the larger the value of $Q^2$ at which the
asymptotic behaviour is manifest.

The calculated $\phi$-meson electroproduction cross section is depicted in
Fig.~\ref{phiEP}.  It is in excellent agreement with Refs.~[\ref{derrick}]
and [\ref{derrickb}], which used a nucleon target, as opposed to
Ref.~[\ref{arneodo}], which averaged over variety of nuclear targets.  As
could be anticipated from Fig.~\ref{rhoEP}, the onset of the asymptotic
$1/Q^4$ behaviour is pushed to larger-$Q^2$ for the $\phi$-meson because the
current-quark mass of the constituents, the $s$-quark, is larger.

\begin{figure}[t]
\centering{\ 
\epsfig{figure=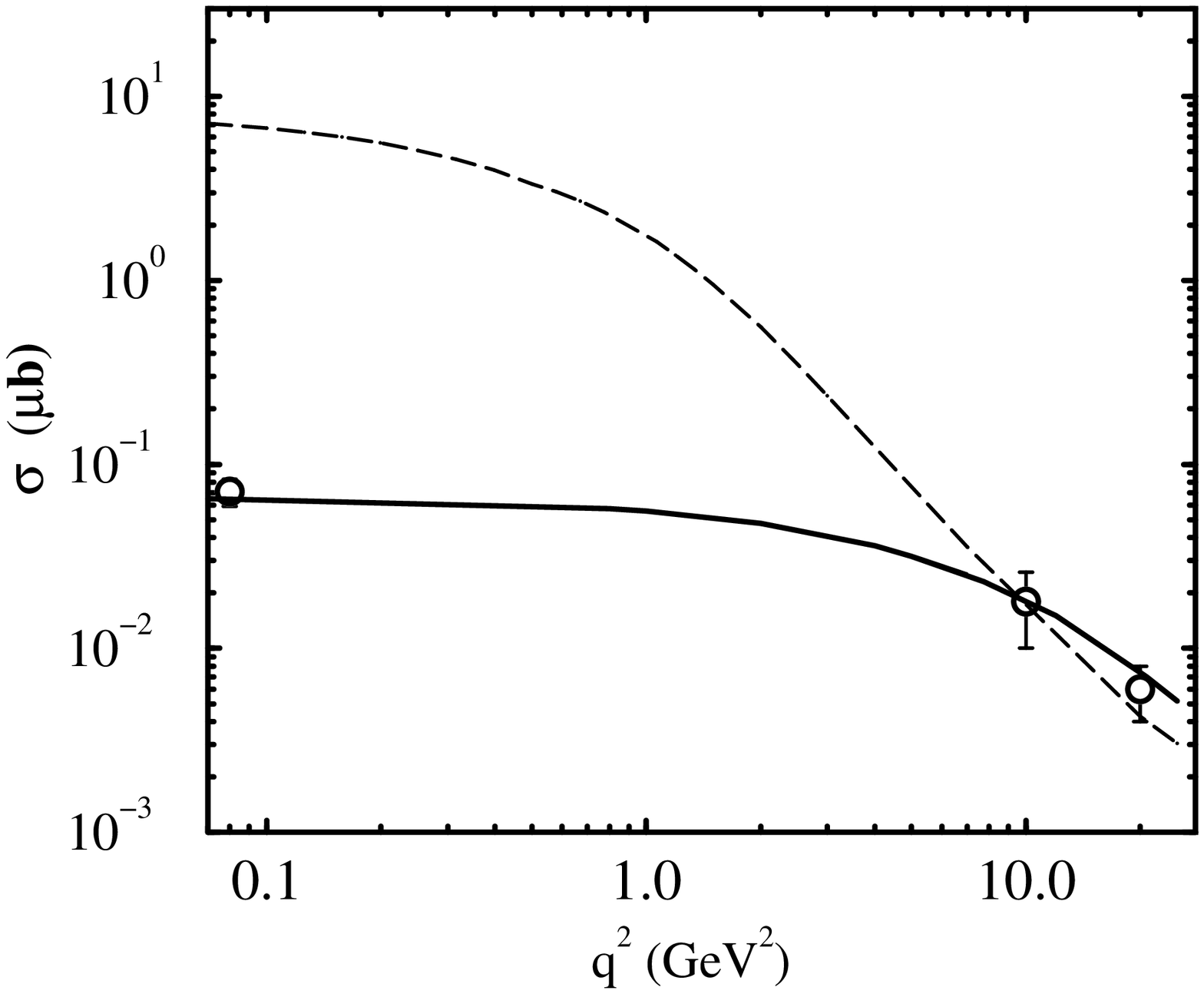,height=10.0cm}}
\caption{$\psi$-meson electroproduction cross section at $W=100\,$GeV: solid
line; the dashed line is the $\rho$-meson result at $W=15\,$GeV for
comparison.  The data are from
Refs.~[\protect\ref{derrickc},\protect\ref{aid}].
\label{psiEP}}
\end{figure}
In calculating the $\psi$-meson electroproduction cross section a very simple
form was used for the $c$-quark propagator:
\begin{equation}
S_c(k):= \,\frac{1}{m_c^2}\,(-i\gamma\cdot k + m_c)\, {\cal F}(1+k^2/m_c^2)
\end{equation}
with $m_c = 1.2\,$GeV ($\sim m_c^{1\,{\rm GeV}}$ in Eq.~(\ref{monegev})).
This and the simple form of the $\psi$-meson Bethe-Salpeter amplitude
anticipate the successful application of DSEs to heavy-meson
observables.\ucite{misha} The calculated cross section is depicted in
Fig.~\ref{psiEP}.  The striking prediction, confirmed by recent data, is that
although two-orders of magnitude smaller than the $\rho$-meson cross section
at the photoproduction point, the $\psi$-meson cross section is equal to the
$W=100\,$GeV, $\rho$-meson cross section at $Q^2=15\,$GeV$^2$.  This is
because the large $c$-quark mass shifts the onset of the asymptotic
$1/Q^4$-behaviour to larger-$Q^2$.

\vspace*{\baselineskip}

\addtocounter{section}{1}
\setcounter{subsection}{1}
{\large\bf \arabic{section}.~Finite Temperature and Chemical
Potential.}\\[0.7\baselineskip] 
\addcontentsline{toc}{section}{\arabic{section}.~Finite Temperature and
Chemical Potential} 
As we have seen, at zero temperature and chemical potential the low-energy
and small-$q^2$ behaviour of QCD is characterised by confinement and DCSB.
The internal scale that marks the boundary between small and large energy in
QCD is $M_\chi\sim \Lambda_{\rm QCD}$.  As the energy and/or momentum
transfer increases, QCD is characterised by asymptotic freedom; i.e., the
coupling evolves
\begin{equation}
\alpha_{\rm S}(Q^2,T=0=\mu) \stackrel{Q^2\to \infty}{\longrightarrow} 0
\end{equation}
and quarks and gluons behave as weakly interacting, massless particles in
high-energy and/or large-$Q^2$ processes.  

The study of QCD at finite temperature and baryon number density proceeds via
the introduction of the intensive variables: temperature, $T$; and quark
chemical potential, $\mu$.  These are additional mass-scales, with which the
coupling can {\it run} and hence, for $T\gg \Lambda_{\rm QCD}$ and/or $\mu\gg
\Lambda_{\rm QCD}$, $\alpha_{\rm S}(Q^2=0,T,\mu)\sim 0$.  It follows that at
finite temperature and/or baryon number density there is a phase of QCD in
which quarks and gluons are weakly interacting, {\em irrespective} of the
momentum transfer\ucite{collinsperry}; i.e., a quark-gluon plasma.  Such a
phase of matter existed approximately one microsecond after the big-bang.  In
this phase confinement and DCSB are absent and the nature of the strong
interaction spectrum is qualitatively different.  The contemporary
expectation for the position of the phase boundary in the $(\mu,T)$-plane is
illustrated in Fig.~\ref{wisdom}.
\begin{figure}[t]
\centering{\ \epsfig{figure=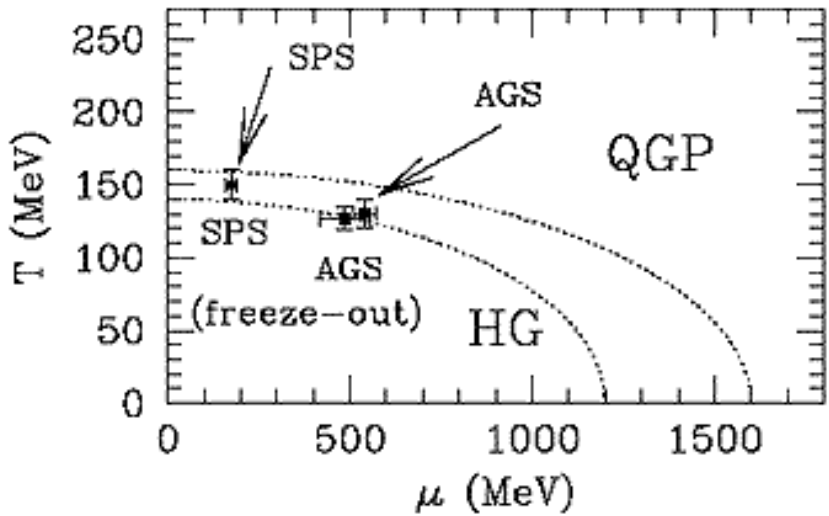,height=70mm}}
\caption{The anticipated quark-gluon phase boundary in the $(\mu_N,T)$-plane.
``HG'' - hadron gas, ``QGP'' - quark-gluon plasma.  The nucleon chemical
potential: $\mu_N := 3 \mu$; i.e., three-times the quark chemical potential.
``SPS'' and ``AGS'' mark the points in the plane that it is estimated these
facilities explore.
\label{wisdom}}
\end{figure} 

The path followed in the transition to the plasma is important because it
determines some observational consequences of the plasma's existence.  For
example,\ucite{krishna} the time-scale for the expansion of the early
universe: $\sim 10^{-5}\, {\rm s}$, is large compared with the natural
time-scale in QCD: $1/\Lambda_{\rm QCD} \sim 1\,{\rm fm}/c \sim
10^{-23}\,{\rm s}$, hence thermal equilibrium is maintained throughout the
QCD transition.  Therefore, if the transition is second-order, the ratio $B
:=\,$baryon-number/entropy, remains unchanged from that value attained at an
earlier stage in the universe's evolution.  However, a first-order transition
would be accompanied by a large increase in entropy density and therefore a
reduction in $B$ after the transition.  Hence the order of the QCD transition
constrains the mechanism for baryon number generation in models describing
the formation of the universe, since with a second-order transition this
mechanism is only required to produce the presently observed value of $B$ and
need not allow for dilution.  In the absence of quarks, QCD has a first-order
deconfinement transition, and with three or four massless quarks a
first-order chiral symmetry restoration transition is
expected.\ucite{krishna} A current, primary question is what happens in the
realistic case of two light quark flavours?

Nonperturbative methods are necessary to study the phase transition, which is
characterised by qualitative changes in order parameters such as the quark
condensate.  One widely used approach is the numerical simulation of finite
temperature lattice-QCD, with the first simulations in the early eighties and
extensive efforts since then.\ucite{karsch97} Here I focus on the application
of DSEs.  This is a new usage and much remains to be learnt: these are
exploratory studies.  One goal is to develop DSE models of QCD at finite-$T$
and $\mu$ (QCD$^T_\mu$) that can be used to check the results of numerical
simulations, and be constrained by them.  These models can then be employed
to extrapolate into that domain presently inaccessible to lattice studies,
such as finite chemical potential and the effects of $T$ and $\mu$ on bound
state properties, the latter of which are expected to provide the signatures
of quark-gluon plasma formation in relativistic heavy ion collisions.

Before discussing details it is interesting to provide a human scale for the
temperatures and densities involved.  The natural scale in QCD is
$\Lambda_{\rm QCD}\sim 200\,$MeV and temperatures of this order of magnitude
will be necessary to create the plasma.  $\Lambda_{\rm QCD} \sim 10^{10}
\times\,$room-temperature!  It represents a temperature on the astrophysical
and cosmological scale.  Nuclear matter density $\rho_0 \approx 3\times
10^{14}$ g/cm$^3$ = 0.16 N/fm$^3$ and this is more-than $10^{13}\times\,$ the
density of solid lead!  The density at the core of a neutron star is expected
to be approximately $4\,\rho_0$\ucite{wiringa} and it is densities on this
order that are anticipated to be required for plasma formation.

The expectation of the existence of a new phase of matter, the quark-gluon
plasma, has led to the construction of a Relativistic Heavy Ion Collider
(RHIC) at Brookhaven National Laboratory.  Construction is due to be
completed in 1999.  It will use counter-circulating, colliding 100$\,$A$\,$GeV
$^{197}\!$Au beams to generate a total centre-of-mass energy of $\sim
40\,$TeV, in an effort to produce an equilibrated quark-gluon plasma.  It is
anticipated to approach the quark-gluon plasma via a low baryon density
route.  Contemporary, fixed target experiments at the Brookhaven-AGS and
CERN-SpS explore a high baryon density environment at much lower
centre-of-mass energies.  These experiments are crucial in developing the
expertise necessary for operating detectors under RHIC conditions but they
are not expected to ``discover'' the plasma.  The CERN-SpS experiments have
nevertheless produced some interesting results.

\begin{figure}[t]
\centering{\ 
\epsfig{figure=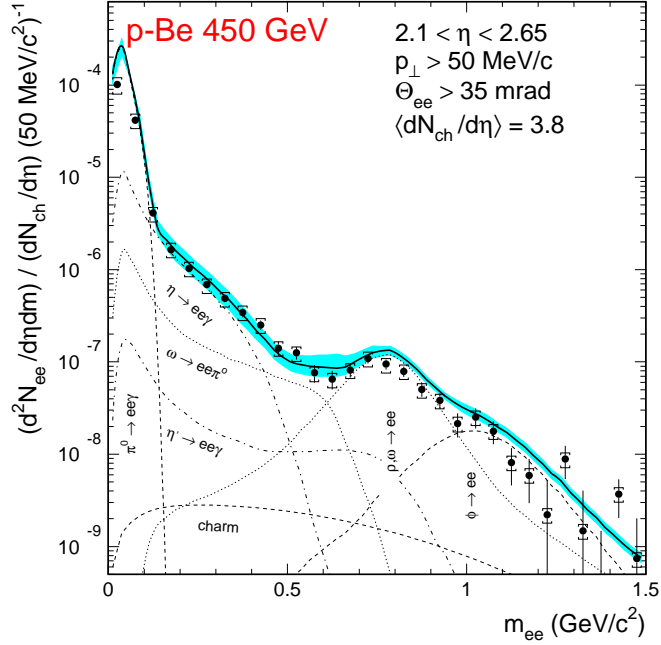,height=120mm}}\vspace*{-30mm}
\caption{Mass spectra for inclusive $e^+$-$\,e^-$ pairs in 450$\,$GeV p-Be
collisions showing the data and various contributions from hadron decays.
The shaded region indicates the systematic error on the summed
contributions.$\,^{\protect\ref{na45home}}$
\label{cerespA}}
\end{figure}
One example is the ``NA45-CERES'' experiment,\ucite{drees97} which studies
$e^+$-$\,e^-$ pair production in heavy ion collisions.  $e^\pm$ pairs leave
the interaction region without interacting strongly and hence they are a
probe of the early stages of the interaction.  In Fig.~\ref{cerespA} I
illustrate the dilepton spectrum obtained in high-energy p-Be collisions.  It
is well described by known hadron decays.  The same is true of p-Au
collisions.
\begin{figure}[t]
\centering{\ 
\epsfig{figure=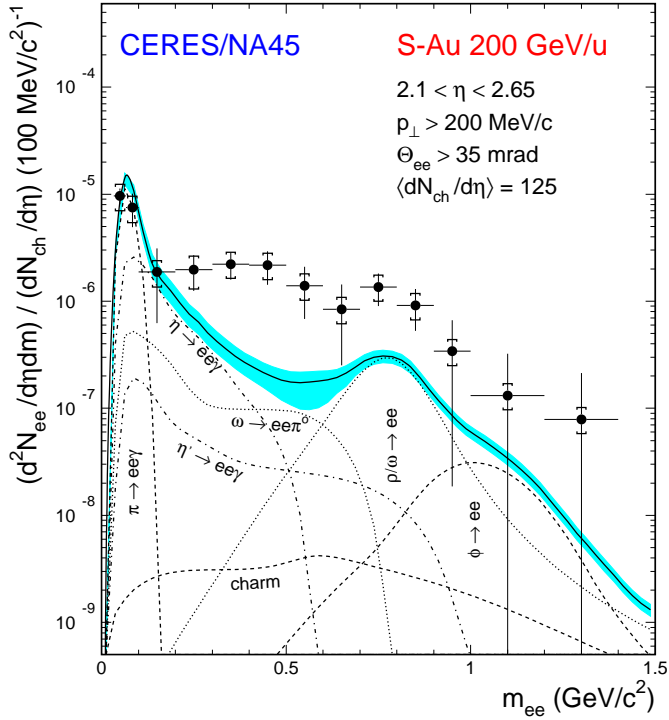,height=120mm}}\vspace*{-25mm}
\caption{Mass spectra for inclusive $e^+$-$\,e^-$ pairs in 200$\,$GeV S-Au
collisions showing the data and various contributions from hadron decays.
The shaded region indicates the systematic error on the summed
contributions.$\,^{\protect\ref{na45home}}$
\label{ceresSAu}}
\end{figure}
However, this is not the case in S-Au collisions, as illustrated in
Fig.~\ref{ceresSAu}.  There the known hadron decays describe the data only
for $m_{ee}<300\,$MeV, which is the region dominated by $\pi^0$ Dalitz
decays.  At higher energies the shape of the spectrum is different and shows
a strong enhancement in the dilepton yield.  Integrating over the region
$0.2<m_{ee}<1.5\,$GeV the enhancement factor is
\begin{equation}
5.0 \pm 0.7\,{\rm stat.} \pm 2.0\,{\rm syst.}
\end{equation}
The enhancement persists in Pb-Au collisions.\ucite{drees97}  What
explanation can be offered?

One model calculation\ucite{brown95} shows that this enhancement can be
explained by a medium-induced reduction of the $\rho$-meson's mass,
another\ucite{rapp97} that it follows from an increase in the $\rho$-meson's
width.  A decrease in the $\rho$-meson's mass is consistent with the QCD sum
rules analysis of Ref.~[\ref{derek95}] but inconsistent with that of
Ref.~[\ref{klingl97}], which employs a more complex phenomenological model
for the in-medium spectral density used in matching the two sides of the sum
rule.  In Ref.~[\ref{klingl97}] there is no shift in the $\rho$-meson mass
but a significant increase in its width.  The consistency between
Refs.~[\ref{rapp97}] and [\ref{klingl97}] is not surprising since, in
contrast to Ref.~[\ref{derek95}], they both rely heavily on effective
Lagrangians with elementary hadron degrees-of-freedom.  These are
possibilities that can be explored using DSEs, which focusing on
dressed-quark and -gluon degrees of freedom is an approach most akin to
Ref.~[\ref{derek95}].  A first attempt,\ucite{schmidt98} summarised in
Sec.~7.6), predicts a 15\% suppression of $m_\rho$ at nuclear matter density
but employs a model that is inadequate to address $\Gamma_\rho$.
\vspace*{0.5\baselineskip}

{\it \arabic{section}.\arabic{subsection})~Notes on Field
Theory.}\\[0.3\baselineskip]
\addcontentsline{toc}{subsection}{\arabic{section}.\arabic{subsection})~Notes
on Field Theory} \addtocounter{subsection}{1}
Equilibrium statistical field theory can be understood by analogy with
equilibrium statistical mechanics.  For a particle moving in a potential $V$
the density matrix is given by the path integral
\begin{eqnarray}
\rho(x,x^\prime;{ T:=1/\beta}) & := &
\int_{x(0)=x}^{x(\beta)=x^\prime}\,{\cal
D}x(\tau)\,\exp\left\{-\int_0^{\beta}\,d\tau 
        \left[\,
        \underline{ \case{1}{2} m \dot x(\tau)^2 - V(x(\tau))}
        \right]\right\}\,,
\end{eqnarray}
where the underlined term is just the Lagrangian.  All of the thermodynamic
information about this system can be obtained from the partition function
\begin{equation}
Z({ T})  :=  \int_V\,dx\,\rho(x,x,{ T})\,;
\end{equation}
for example, the pressure $P= T\ln\,Z(T)/V$ and the baryon density $\rho^B=
(1/3) \partial P/\partial \mu$.

The equilibrium thermodynamics of a quantum field theory is also completely
specified by a partition function, or generating functional.  In the
particular case of a self-interacting scalar field this is given by the
functional integral: 
\begin{eqnarray}
\label{phiZ}
{\cal Z}[T] &:=& 
\int\,
\Pi_{\tilde{x},\tau\in[0,\beta]}\,D\phi(\tilde{x},\tau)\,
\exp\left(- \int_{0}^{\beta}\,d\tau\,\int\,d^3x\,{\cal L}^E(x;\phi)\right)\,,
\end{eqnarray}
where ${\cal L}^E(x;\phi)$ is the Euclidean Lagrangian density describing the
interaction of $\phi(\tilde x,\tau)$, whose boundary conditions are periodic:
\begin{equation}
\phi(\tilde x,\tau=0)=\phi(\tilde x,\tau={ \beta})\,.
\end{equation}
The boundary conditions for fermions are antiperiodic:
\begin{equation}
\psi(\tilde x,\tau=0)= -\psi(\tilde x,\tau={ \beta})\,.
\end{equation}
This difference in boundary conditions is the reason for the difference
between the Matsubara frequencies of fermions and bosons and hence why
fermions acquire a screening mass at finite temperature.

It is immediately obvious that the $O(4)$ invariance of the Euclidean theory
is lost: at finite temperature (and/or chemical potential) the theory
exhibits only an $O(3)$ symmetry corresponding to spatial rotations and
translations.  This is why the formalism, necessarily used in lattice
simulations, is only applicable to equilibrium systems - there is no
generator of translations in time.  One also notes from Eq.~(\ref{phiZ}) that
as $T\to\infty$ one dimension disappears completely and hence the
corresponding $(d-1)$-dimensional theory is a candidate to describe the
infinite-temperature limit of a $d$-dimensional theory.

The finite temperature, free fermion Lagrangian density is
\begin{equation}
{\cal L}^E_{\rm Free}(\bar \psi, \psi) = \bar\psi(\vec{x},\tau)\,
\left(\vec{\gamma}\cdot \vec{\partial} + \gamma_4 \,\partial_\tau
+m\right)\,\psi(\vec{x},\tau)\,.
\end{equation}
Introducing a Fourier decomposition: 
\begin{equation}
\psi(\vec{x},\tau) = 
T\sum_{n=-\infty}^{\infty}\,\int\,\frac{d^3p}{(2\pi)^3}\,\psi(\vec{p},\omega_n)
{\rm e}^{i\vec{p}\cdot\vec{x} + i \omega_n \tau}\,,
\end{equation}
antiperiodicity entails that the fermion Matsubara frequencies are
\begin{equation}
\omega_n = (2 n + 1) \,\pi\,T\,,\;n\in {\rm Z}\!\!\!{\rm Z}
\end{equation}
and the free fermion action is
\begin{equation}
\label{Sfree}
S_\beta^E[\bar\psi,\psi]_{\rm Free}= 
T \sum_{n=-\infty}^{\infty}\,\int\,\frac{d^3p}{(2\pi)^3}\,
\bar\psi(\vec{p},\omega_n)\,
\left(i\vec{\gamma}\cdot \vec{p}+i\gamma_4\omega_n + m\right)\,
        \psi(\vec{p},\omega_n)\,.
\end{equation}
From this one identifies the finite temperature, free fermion propagator
\begin{equation}
S(p) = \frac{1}{i\vec{\gamma}\cdot \vec{p}+i\gamma_4\,\omega_n + m}\,.
\end{equation}

Analogous arguments, using the periodic boundary conditions, lead to an
identification of the free boson propagator
\begin{equation}
D(p,\Omega_n) = \frac{1}{|\vec{p}|^2+ \Omega_n^2 + m^2}\,,
\end{equation}
where $\Omega_n= 2\,\pi\,n\,T$.  Having obtained the free particle
propagators one can proceed to define a perturbation theory.  As an example,
in massless $\phi^4$ theory the one-loop correction to the $\phi$ propagator
is
\begin{equation}
\propto T\,\sum_{n=-\infty}^{\infty}\,
        \int\frac{d^3p}{(2\pi)^3}\,\frac{1}{\Omega_n^2 + |\vec{p}|^2}\,.
\end{equation}
The sum can be evaluated:
\begin{equation}
T\sum_{n=-\infty}^{\infty}\,\frac{1}{\Omega_n^2 + |\vec{p}|^2}
= \frac{1}{|\vec{p}|}\,\frac{1}{\exp(|\vec{p}|/T) -1 }
+ T{\rm -independent~piece}\,,
\end{equation}
to yield the Bose-Einstein factor.  This is a source of problems: for large
temperatures
\begin{equation}
\frac{1}{\exp(|\vec{p}|/T) -1 } = \frac{T}{|\vec{p}|}
\end{equation}
and one can thereby encounter additional infrared divergences.

These are particularly serious in QCD.  To illustrate this\ucite{kapusta}
consider an $(\ell + 1)$-gluon-loop diagram and focus on the $n=0$ mode,
which obviously yields the dominant infrared behaviour.  The estimate is made
easier if one neglects the tensor structure and notes that: from the vertices
there is a factor of $g^{2\ell}\,p^{2\ell}$; the loop-sum-integral gives
$(T\int_{p^2\in[0,T]} d^3p)^{\ell+1}$; and the propagators,
$(p^2+m^2)^{-3\ell}$ where $m$ is a possible, dynamically generated
mass-scale.  A little thought and calculation shows that the net order of a
given diagram in perturbation theory is
\begin{equation}
\begin{array}{ccc}
\ell = 1,2      &       \ell = 3        & \ell \geq 4\\
g^{2\ell}\, T^4   &       g^6\, T^4 \,\ln(T/m)  & g^6\, T^4\, [g^2T/m]^{\ell-3}
\end{array}
\end{equation}
Clearly, if $m=0$ the diagrams are infrared divergent for $\ell>2$.  The
divergences may cancel when all diagrams of a given order are summed but that
is difficult to verify.  Suppose instead that the mass-scale $m\sim g\,T$, as
does the Debye mass in QED, then no problem arises: at each order above
$\ell=3$ the diagrams are suppressed by powers of the coupling and a
self-consistent calculation of the mass-scale is straightforward.  This
underlies the successful application of the method of ``hard thermal
loops''.\ucite{htl} However, suppose that $m\sim g^2 T$, which is the next
possibility.  In this case every diagram above $\ell=2$ contributes with the
same strength: $g^6$, which presents a serious impediment to the application
of perturbation theory!

The introduction of a quark chemical potential modifies Eq.~(\ref{Sfree}): 
\begin{equation}
S_\beta^E[\bar\psi,\psi]_{\rm Free}  := 
T\sum_{n=-\infty}^{\infty}\,\int\,\frac{d^3p}{(2\pi)^3}\,
\bar\psi(\vec{p},\omega_n)\,
\left(i\vec{\gamma}\cdot \vec{p}+i\gamma_4\,\omega_n - \gamma_4  \mu 
        + m\right)\,        \psi(\vec{p},\omega_n)
\end{equation}
so that even the free Dirac operator is not hermitian and hence the QCD
action is necessarily complex.  As such it does not specify a probability
measure, which precludes the straightforward application of Monte-Carlo
methods in the evaluation of the partition function.  However, the
application of DSEs remains straightforward.  The propagators and vertices
are complex, so twice as many functions are required to represent them but
otherwise there are no complications.  Thus they provide a nonperturbative
means of exploring this domain, which is presently inaccessible in lattice
simulations.
\vspace*{0.5\baselineskip}

{\it \arabic{section}.\arabic{subsection})~Some Lattice
Results.}\\[0.3\baselineskip]
\addcontentsline{toc}{subsection}{\arabic{section}.\arabic{subsection})~Some
Lattice 
Results} \addtocounter{subsection}{1}
Since the early eighties, as one branch of the extensive application of
lattice methods in many areas of QCD, Monte-Carlo simulations have been used
to estimate the finite temperature QCD partition function.\ucite{detar} These
studies have contributed considerably to the current understanding of the
nature of the quark-gluon plasma.  Due to the persistent limitation of
computational power many analyses have focused on the pure gauge sector,
which exhibits a first-order deconfinement transition at a critical
temperature of $T_c^{N_f=0}\approx 270\,$MeV.\ucite{ukawa90} In studying the
chiral transition this commonly used quenched approximation is inadequate
because the details depend sensitively on the number of active (light)
flavours.  It is therefore necessary to include the fermion determinant.

That is even more important when $\mu\neq 0$ because the Dirac operator is
not hermitian and thus the fermion determinant acquires an explicit imaginary
part, in addition to those terms associated with axial anomalies.  The QCD
action being complex entails that the study of finite density is
significantly more difficult than that of finite temperature.  Simulations
that ignore the fermion determinant at $\mu\neq 0$ encounter a forbidden
region, which begins at $\mu = m_\pi/2$,\ucite{dks96} and since $m_\pi\to 0$
in the chiral limit this is a serious limitation, preventing a reliable study
of chiral symmetry restoration.  The phase of the fermion determinant is
essential in eliminating this artefact.\ucite{adam}

\begin{figure}[t]
\centering{\ \epsfig{figure=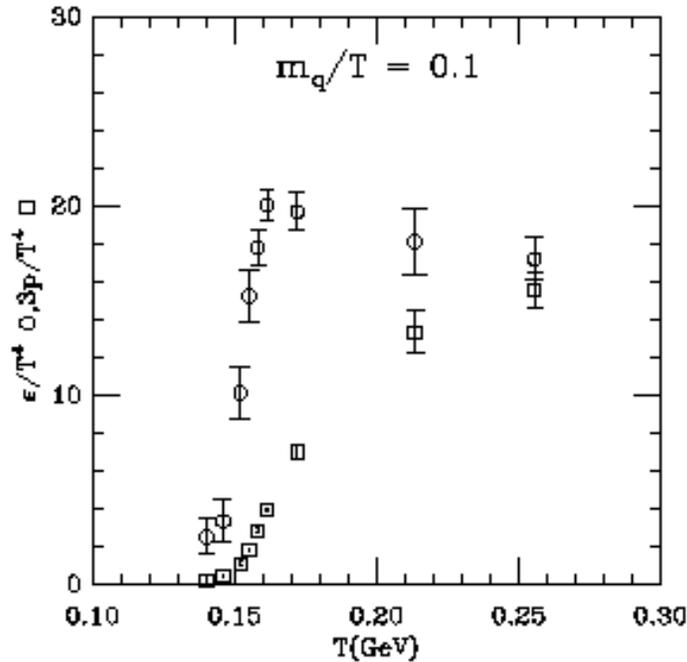,height=90mm}}
\caption{Energy density and pressure for 2-light-flavour QCD on lattices with
four temporal lattice sites, from Ref.~[\protect\ref{blum}].
\label{lattEp}}
\end{figure}
QCD with dynamical quarks is a contemporary focus and for two flavours of
light quarks the theory appears\ucite{karsch95} to exhibit a second-order
transition at a critical temperature $T_c^{N_f=2}\approx 150\,$MeV.  This is
illustrated in Fig.~\ref{lattEp}, which shows a rapid change in the energy
density in a small region around $150\,$MeV.  For three or more light quark
flavours the continuum theory is expected to have a first order chiral
symmetry restoration transition.

\begin{figure}[t]
\centering{\ \epsfig{figure=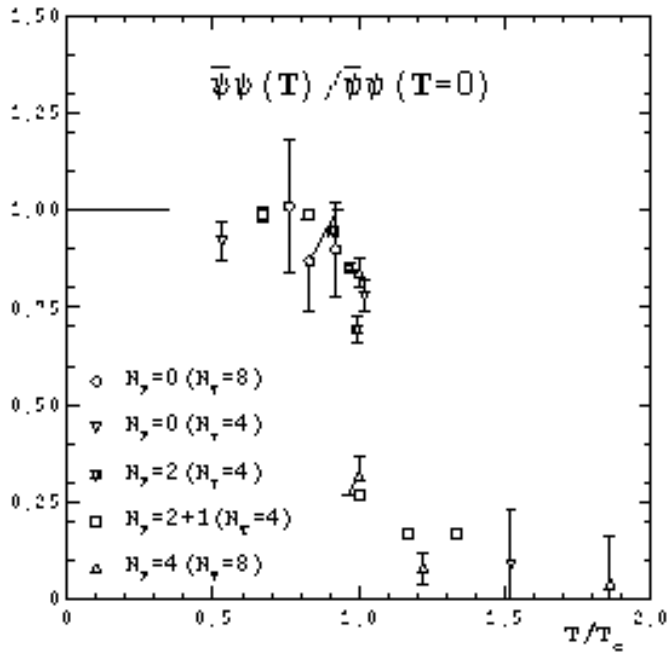,height=90mm}}
\caption{Chiral condensate calculated$\,^{\protect\ref{karsch95}}$ using
staggered fermions and normalised to its zero temperature value.  Only the
$N_f=0$ results are extrapolated to zero quark mass.
\label{lattqbq}}
\end{figure}
The quark condensate is an order parameter for chiral symmetry breaking, with
its nonzero value at $T=0$ responsible for the pion mass being proportional
to the square-root of the light current-quark masses.  Its behaviour near the
critical temperature is depicted in Fig.~\ref{lattqbq}, and the rapid,
qualitative change with increasing $T$ is easily apparent.  Very important is
that, independent of the number of light-quark flavours, the condensate is
{\it unchanged} for $T<0.9\,T_c$.  It suggests that hadron properties are
insensitive to $T$ until very near the phase boundary.

The simulations with dynamical fermions are still preliminary, and many
uncertainties remain.  For example, a review\ucite{laermann} of recent
results obtained with larger lattices and lighter quarks reports a
significant lattice-volume-dependence for the critical exponents of the two
light-flavour chiral symmetry restoration transition: the transition may even
be first order!  This might be an artefact of introducing lighter dynamical
quarks, which drive the simulations to stronger coupling.  The order of the
transition with three and four flavours also remains unclear.  With these
uncertainties it is apparent that the lattice study of the phase transition
will require further, even more computer-intensive simulations.

\vspace*{\baselineskip}

\addtocounter{section}{1}
\setcounter{subsection}{1}
{\large\bf \arabic{section}.~DSEs at Finite $T$ and
$\mu$.}\\[0.7\baselineskip]  
\addcontentsline{toc}{section}{\arabic{section}.~DSEs at Finite $T$ and
$\mu$} 
The contemporary application of DSEs at finite temperature and chemical
potential is a straightforward extension of the $T=0=\mu$ studies.  The
direct approach is to develop a finite-$T$ extension of the {\it Ansatz} for
the dressed-gluon propagator.  The quark DSE can then be solved and, having
the dressed-quark and -gluon propagators, the response of bound states to
increases in $T$ and $\mu$ can be studied.  As a nonperturbative approach
that allows the simultaneous study of DCSB and confinement, the DSEs have a
significant overlap with lattice simulations: each quantity that can be
estimated using lattice simulations can also be calculated using the DSEs.
This means they can be used to check the lattice simulations, and
importantly, that lattice simulations can be used to constrain their
model-dependent aspects.  Once agreement is obtained on the common domain,
the DSEs can be used to explore phenomena presently inaccessible to lattice
simulations.
\vspace*{0.5\baselineskip}

{\it \arabic{section}.\arabic{subsection})~Quark DSE.}\\[0.3\baselineskip] 
\addcontentsline{toc}{subsection}{\arabic{section}.\arabic{subsection})~Quark
DSE} 
\addtocounter{subsection}{1}
The renormalised dressed-quark propagator at finite-$(T,\mu)$ has the general
form
\begin{eqnarray}
\label{genformS}
S(\vec{p},\tilde\omega_k)  & = & \frac{1}
{i\vec{\gamma}\cdot \vec{p}\,A(\vec{p},\tilde\omega_k)
+i\gamma_4\,\tilde\omega_k C(\vec{p},\tilde\omega_k)+
B(\vec{p},\tilde\omega_k)}\\
& \equiv&  -i\vec{\gamma}\cdot \vec{p}\,\sigma_A(\vec{p},\tilde\omega_k)
-i\gamma_4\,\tilde\omega_k \sigma_C(\vec{p},\tilde\omega_k)+
\sigma_B(\vec{p},\tilde\omega_k)\, 
\end{eqnarray}
where $\tilde\omega_k := \omega_k + i \mu $.  The complex scalar functions:
$A(\vec{p},\tilde\omega_k)$, $B(\vec{p},\tilde\omega_k)$ and
$C(\vec{p},\tilde\omega_k)$ satisfy:
\begin{equation}
{\cal F}(\vec{p},\tilde\omega_k)^\ast = {\cal
        F}(\vec{p},\tilde\omega_{-k-1})\,,
\end{equation}
${\cal F}=A,B,C$, and although not explicitly indicated they are functions
only of $|\vec{p}|^2$ and $\tilde\omega_k^2$.  

The DSE for the renormalised dressed-quark propagator is
\begin{equation}
\label{qDSE}
S^{-1}(\vec{p},\tilde\omega_k) = Z_2^A \,i\vec{\gamma}\cdot \vec{p} + Z_2 \,
(i\gamma_4\,\tilde\omega_k + m_{\rm bm})\, 
        + \Sigma^\prime(\vec{p},\tilde\omega_k)\,,
\end{equation}
where $Z_2^A$ and $Z_2$ are renormalisation constants, $m_{\rm bm}$ is the
bare mass, and the regularised self energy is
\begin{eqnarray}
\Sigma^\prime(\vec{p},\tilde\omega_k) & = & i\vec{\gamma}\cdot \vec{p}
\,\Sigma_A^\prime(\vec{p},\tilde\omega_k) + i\gamma_4\,\tilde\omega_k
\,\Sigma_C^\prime(\vec{p},\tilde\omega_k) +
\Sigma_B^\prime(\vec{p},\tilde\omega_k)\; ,
\end{eqnarray}
with
\begin{eqnarray}
\Sigma_{\cal F}^\prime(\vec{p},\tilde\omega_k) & = &
\int_{l,q}^{\bar\Lambda}\, \case{4}{3}\,g^2\,
D_{\mu\nu}(\vec{p}-\vec{q},\tilde\omega_k-\tilde\omega_l)\case{1}{4} {\rm
tr}\left[{\cal P}_{\cal F} \gamma_\mu
S(\vec{q},\tilde\omega_l)
\Gamma_\nu(\vec{q},\tilde\omega_l;\vec{p},\tilde\omega_k)\right]\,,
\label{regself}
\end{eqnarray}
where ${\cal P}_A:= -(Z_1^A/p^2)i\gamma\cdot p$, ${\cal P}_B:= Z_1 $, ${\cal
P}_C:= -(Z_1/\tilde\omega_k)i\gamma_4$, $Z_1^A$ and $Z_1$ are vertex
renormalisation constants, and $\int_{l,q}^{\bar\Lambda}:=\, T
\,\sum_{l=-\infty}^\infty\,\int^{\bar\Lambda}\frac{d^3q}{(2\pi)^3}$.  This
last is a mnemonic to represent a translationally invariant regularisation of
the integral with $\bar\Lambda$ the regularisation mass scale.

In Eq.~(\ref{regself}) the Landau-gauge, finite-$(T,\mu)$ dressed-gluon
propagator has the form
\begin{equation}
g^2 D_{\mu\nu}(\vec{p},\Omega) = 
P_{\mu\nu}^L(\vec{p},\Omega) \,\Delta_F(\vec{p},\Omega) + 
P_{\mu\nu}^T(\vec{p})\, \Delta_G(p,\Omega) 
\end{equation}
where
\begin{eqnarray}
P_{\mu\nu}^T(\vec{p}) & := &\left\{
\begin{array}{ll}
0; \; & \mu\;{\rm and/or} \;\nu = 4,\\
\displaystyle
\delta_{ij} - \frac{p_i p_j}{p^2}; \; & \mu,\nu=1,2,3
\end{array}\right.\,,\\
P_{\mu\nu}^L(\vec{p},\Omega) & := & \delta_{\mu\nu} 
        - \frac{p_\mu p_\nu}{\sum_{\alpha=1}^4 \,p_\alpha p_\alpha}
        - P_{\mu\nu}^T(p);\;\mu,\nu= 1,\ldots, 4\, .
\end{eqnarray}
A ``Debye-mass'' for the gluon appears as a $T$-dependent contribution to
$\Delta_F$.

In renormalising we require that
\begin{equation}
\left. S^{-1}(\vec{p},\tilde\omega_0)
\right|_{\vec{p}^2+\tilde\omega_0^2=\zeta^2}^{\mu =0} =  
        i\vec{\gamma}\cdot \vec{p} + i\gamma_4\,\omega_0 + m_R^\zeta\,,
\end{equation}
where $\zeta$ is the renormalisation point and $m_R^\zeta$ is the
renormalised current-quark mass.  This entails that the renormalisation
constants are:
\begin{eqnarray}
Z_2^A(\zeta^2,\bar\Lambda^2) & = & 1 -
\left.\Sigma^\prime_A(\vec{p},\tilde\omega_0) 
\right|^{\mu=0}_{|\vec{p}|^2+\tilde\omega_0^2=\zeta^2} \,,\\
Z_2(\zeta^2,\bar\Lambda^2) & = &1 - \left.\Sigma^\prime_C(\vec{p},\tilde\omega_0)
\right|^{\mu=0}_{|\vec{p}|^2+\tilde\omega_0^2=\zeta^2} \,,\\ 
m_R^\zeta & = & Z_2 m_{\rm bm} + \left.\Sigma^\prime_B(\vec{p},\tilde\omega_0)
\right|^{\mu=0}_{|\vec{p}|^2+\tilde\omega_0^2=\zeta^2} \,,
\end{eqnarray}
and yields the renormalised self energies:
\begin{equation}
{\cal F}(\vec{p},\tilde\omega_k) = 
\xi_{\cal F} + \Sigma^\prime_{\cal F}(\vec{p},\tilde\omega_k) 
        - \left.\Sigma^\prime_{\cal F}(\vec{p},\tilde\omega_0)
                \right|^{\mu=0}_{|\vec{p}|^2+\tilde\omega_0^2=\zeta^2}\,,
\end{equation}
where ${\cal F}=A$, $B$, $C$; $\xi_A = 1 = \xi_C$ and $\xi_B = m_R^\zeta$.

In studying confinement one cannot assume that the analytic structure of a
dressed propagator is the same as that of the free particle propagator: it
must be determined dynamically.  Indeed, one knows that the $\tilde
p_k:=(\vec{p},\tilde\omega_k)$-dependence of $A$ and $C$ is qualitatively
important since it can conspire with that of $B$ to eliminate free-particle
poles in the dressed-quark propagator.\ucite{burden} In this case the
propagator does not have a Lehmann representation so that, in general, the
Matsubara sum cannot be evaluated analytically.  More importantly, it either
complicates or precludes a real-time formulation of the finite temperature
theory, which makes the study of nonequilibrium thermodynamics a very
challenging problem.  In addition, as we will see, this $\tilde
p_k$-dependence of $A$ and $C$ can be a crucial factor in determining the
behaviour of bulk thermodynamic quantities such as the pressure and entropy;
being responsible for these quantities reaching their respective
Stefan-Boltzmann limits only for very large values of $T$ and $\mu$.  It is
therefore important in any DSE study to retain $A(\tilde p_k)$ and $C(\tilde
p_k)$, and their dependence on $\tilde p_k$.
\vspace*{0.5\baselineskip}

{\it \arabic{section}.\arabic{subsection})~Phase Transitions and Order
Parameters.}\\[0.3\baselineskip]  
\addcontentsline{toc}{subsection}{\arabic{section}.\arabic{subsection})~Phase
Transitions and Order Parameters} 
\addtocounter{subsection}{1}
Phase transitions are characterised by the behaviour of an order parameter,
$\langle X \rangle$, the expectation value of an operator.  In the ordered
phase of a system: $\langle X \rangle \neq 0$, whereas in the disordered
phase $\langle X \rangle = 0$.  A phase transition is first-order if $\langle
X \rangle \to 0$ discontinuously, whereas it is second-order if $\langle X
\rangle \to 0$ continuously.  For a second-order transition, the length-scale
associated with correlations in the system diverges as $\langle X \rangle \to
0$ and one can define a set of critical exponents that characterise the
behaviour of certain macroscopic properties at the transition point.  For
example, in a system that is ferromagnetic for temperatures less than some
critical value, $T_c$, the magnetisation, $M$, is an order parameter and in
the absence of an external magnetic field $M \propto (T_c-T)^\beta$ for
$T\sim T_c^-$, where $\beta$ is the critical exponent.  At $T=T_c$ the
behaviour of the magnetisation in the presence of an external field, $h\to
0^+$, defines another critical exponent, $\delta$: $M \propto
h^{(1/\delta)}$.  In a system that can be described by mean field theory
these critical exponents are
\begin{eqnarray}
\beta^{\rm MF}= 0.5\,,\; & & \delta^{\rm MF} = 3.0\,.
\end{eqnarray}
The problem is that it can be difficult to identify the order parameter
relevant to the discussion of a phase transition.

One order parameter for the chiral symmetry restoration transition is well
known - it is the quark condensate, defined via the renormalised
dressed-quark propagator:\ucite{mr97}
\begin{equation}
\label{qbarq}
-\langle \bar q q\rangle_\zeta:= N_c\,
         \lim_{\bar\Lambda\to \infty}
        Z_4(\zeta,\bar\Lambda)\,
        \int_{l,q}^{\bar\Lambda}
        \frac{B_0(\tilde p_k )}
        {|\vec{p}|^2 A_0(\tilde p_k )^2
        + \tilde\omega_l^2 C_0(\tilde p_k )^2 
        + B_0(\tilde p_k )^2}\,,
\end{equation}
for each massless quark flavour, where the subscript ``$0$'' denotes that the
scalar functions: $A_0$, $B_0$, $C_0$, are obtained as solutions of
Eq.~(\ref{qDSE}) in the chiral limit.  The functions have an implicit
$\zeta$-dependence.  An equivalent order parameter is
\begin{equation}
\label{chiorder}
{\cal X} := {\sf Re}\,B_0(\vec{p}=0,\tilde \omega_0)\,,
\end{equation}
which was used in Refs.~[\ref{prl}-\ref{greg}].  Thus the zeroth Matsubara mode
determines the character of the chiral phase transition, a conjecture
explored in Ref.~[\ref{jackson96}].

What is an order parameter for deconfinement?  

In Sec.~2.3) I observed that the analytic properties of Schwinger functions
play an important role in confinement.  For illustrative simplicity, set
$\mu=0$, the generalisation to $\mu\neq 0$ is not difficult, and consider
\begin{eqnarray}
\label{schwinger}
\Delta_{B_0}(x,\tau=0) & := & T \sum_{n=-\infty}^\infty\,\frac{1}{4\pi x}\,
\frac{2}{\pi}\int_0^\infty dp\,p\,\sin(px)\,\sigma_{B_0}(p,\omega_n) \\
& := & \frac{T}{2 \pi x}\, \sum_{n=0}^\infty\,\Delta_{B_0}^n(x)\,.
\end{eqnarray}
For a free fermion of mass $M$, $\sigma_{B_0}(p,\omega_n)=
M/(\omega_n^2+p^2+M^2)$ and
\begin{equation}
\Delta_B^n(x)= M \,{\rm e}^{-x\,\sqrt{\omega_n^2+M^2}}\,:
\end{equation}
the $n=0$ term dominates the sum.  In this case the ``mass-function''
\begin{eqnarray}
M(x;T) & := & \frac{d}{dx} \left(-\ln\left|\Delta_{B_0}^{{
0}}(x)\right|\right) =  \sqrt{\pi^2 T^2 + M^2}\,.
\end{eqnarray}

The most important observation is that for a free particle $M(x,T)$ has a
fixed, real value, which identifies the mass-pole in the propagator.  It also
exhibits the fermion ``screening mass'' $= \pi T$, which becomes important
for $T \sim M/\pi$.  In the context of dynamical mass generation: $M\sim
M^E$.  Since $M^E_{u/d}\approx 450\,$MeV one anticipates that finite-$T$
effects will become important at $T\sim 150\,$MeV (or finite $\mu$ effects at
$\mu \sim 450\,$MeV).  For a boson of mass $M_b$, $M(x;T)=M_b$: there is no
screening mass.

How does $\Delta^0(x)$ behave if the dressed-propagator does not have a
Lehmann representation?  An example\ucite{stingl} is
\begin{equation}
{\cal D}(p,\Omega)  = \frac{p^2+\Omega_n^2+M^2}
        {(p^2+\Omega_n^2+M^2)^2 + 4\, b^4}\,,
\end{equation}
which has complex conjugate poles.  In this case 
\begin{equation}
\Delta^{{ 0}}_{{\cal D}}(x) = {\rm e}^{- M x}\, \cos[ b x]\,;
\end{equation}
i.e., the Schwinger function oscillates and the mass-function has
singularities, which is an unambiguous signal for the absence of a Lehmann
representation and hence confinement!

An order parameter for confinement is now obvious.\ucite{hrw94} Denote the
position of the first zero in $\Delta_{B_0}^0(x)$ by $r^{z_1}_0$, which is
inversely proportional to the distance of the poles from the real axis.
Define $\kappa_0 := 1/r^{z_1}_0$, then $\kappa_0 \propto b$ and deconfinement
is observed if, for some $T=T_c$, $\kappa_0(T_c)=0$: at this point thermal
fluctuations have overwhelmed the confinement scale-parameter and the poles
have migrated to the the real-axis.  This criterion generalises easily to the
case of $\mu\neq 0$ and to situations in which the dressed-propagator has an
essential singularity rather than complex conjugate poles.  It is also valid
for both light and heavy quarks.

An analogue of this criterion, with 
\begin{equation}
\Delta(t) := \frac{1}{2\pi} 
        \int_{-\infty}^\infty\,dp_4\,{\rm e}^{ip_4 t}\,\sigma_S(\vec{p}=0,p_4)\,,
\end{equation}
has been used to very good effect in an analysis\ucite{m95} of QED$_3$ at
$T=0$.  QED$_3$ is confining in quenched approximation but not when massless
fermions are allowed to influence the propagation of the photon.  In that
case complete charge screening is possible.  Confinement is recovered in the
theory if the fermion in the photon vacuum polarisation loop is massive.
\begin{figure}[t]
\centering{\ 
\epsfig{figure=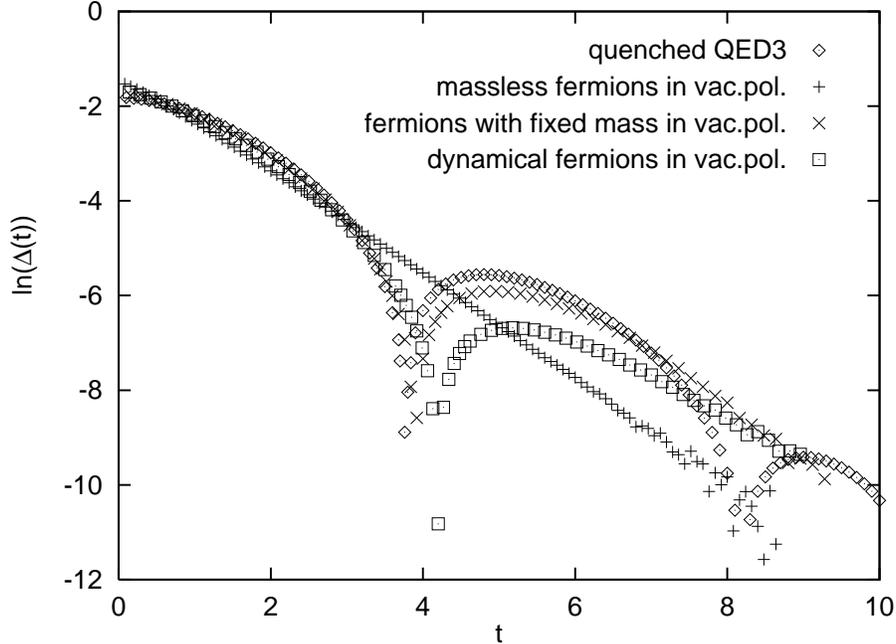,height=90mm}  }
\caption{$-E(t):= \ln\Delta(t)$ in QED$_3$.  Here the analogue of the
mass-function is $E^\prime(t)$ and the difference between the unconfined
theory: ``$+$'', and the confining theories is unmistakable.
\label{pm95}}
\end{figure}
This application is summarised in Fig.~\ref{pm95}.
\vspace*{0.5\baselineskip}

{\it \arabic{section}.\arabic{subsection})~Illustration at $(T\neq
0,\mu=0)$.}\\[0.3\baselineskip]
\addcontentsline{toc}{subsection}{\arabic{section}.\arabic{subsection})
~Illustration at $(T\neq 0,\mu=0)$} \addtocounter{subsection}{1}
As a first example I summarise a study\ucite{prl} that uses a one-parameter,
model dressed-gluon propagator.  This parameter, $m_t$, is a mass-scale that
marks the boundary between the perturbative and nonperturbative domains, and
its value, $m_t=0.69\;$GeV, was fixed in $T=0$ studies.\ucite{fr96} The
extension of the model to finite-$T$ involves no additional parameters and is
defined with: $\Delta_F(p,\Omega) := {\cal D}(p,\Omega;m_D)$ and
$\Delta_G(p,\Omega) := {\cal D}(p,\Omega;0)$;
\begin{eqnarray}
{\cal D}(p,\Omega;m) & := & \case{16}{9} \pi^2   \left[ 
\frac{2\pi}{T} m_t^2 \delta_{0\,n} \delta^3(p) + 
\frac{1-\rm{e}^{
\left[-\right(p^2+\Omega^2+ m^2 \left)/(4m_t^2)\right]}}
        {p^2+\Omega^2+ m^2} \right]\; ,
\label{delta}
\end{eqnarray}
where $m_{\rm D}^2 = (8/3)\, \pi^2 T^2$ is the perturbatively evaluated
``Debye-mass''.\footnote[2]{The influence of the Debye-mass on finite-$T$
observables is qualitatively unimportant, even in the vicinity of the chiral
symmetry restoration transition.  The ratio of the coefficients in the two
terms in Eq.~(\protect\ref{delta}) is such that the long-range effects
associated with $\delta_{0\,k} \delta^3(p)$ are completely cancelled at
short-distances; i.e., for $|\vec{x}|^2\,m_t^2\ll 1$.}  The quark DSE was
solved using the rainbow approximation
\begin{equation}
\label{rainbow}
\Gamma_\mu(q,\omega_l;p,\omega_k) = \gamma_\mu\,.
\end{equation}
I have discussed this truncation in Secs.~2.4) and~4), and here only note
that in $T=0$ studies it has proven to be reliable in Landau gauge; i.e., at
this level an efficacious phenomenology with a more sophisticated vertex {\it
Ansatz} only requires a small quantitative modification of the parameters
that characterise the small-$k^2$ behaviour of the dressed-gluon
propagator.\ucite{hrw94} Using this truncation, mutually consistent
constraints are $Z_1^A=Z_2^A$ and $Z_1=Z_2$.

The quark DSE was solved numerically with $m_R^\zeta = 1.1\,{\rm MeV}$,
$\zeta = 9.47\,$GeV.  The $T=0$ fitting of $m_t$ and $m_R$ ensured a best
$\chi^2$-fit to a range of pion observables, yielding
\begin{equation}
\begin{array}{cccc}
f_\pi = 92.4 & m_\pi = 139.5 & r_\pi N_\pi= 0.24 & g_{\pi^0\gamma\gamma} =0.45\\
(92.4 \pm 0.3) & (138.3\pm 0.5) & (0.31 \pm 0.004) & (0.50 \pm 0.02)\\
a_0^0 =0.16 & a^2_0= -0.041 & a_1^1= 0.028 & a_2^0 = 0.0022 \\
(0.21\pm 0.02) & (-0.040 \pm 0.003) & (0.038 \pm 0.002) & (0.0017\pm 0.0003) 
\end{array}
\end{equation}
with the experimental values listed in parentheses.\footnote[3]{In Sec.~5.1)
I discussed why $r_\pi N_\pi \approx 0.25$ in impulse approximation.  The
$\pi$-$\pi$ scattering lengths fitted in Ref.~[\protect\ref{fr96}] were taken
from Ref.~[\protect\ref{sevior92}].}  The finite-$T$ study reproduces these
results to within 6\% at $T=5\,$MeV, using the finite-$T$ generalisations of
the formulae in Ref.~[\ref{fr96}]:
\begin{equation}
\label{pimass}
m_\pi^2\,N_\pi^2 = \langle m_R^\zeta \, (\bar q q)_\zeta \rangle_\pi\,;
\end{equation}
\begin{eqnarray}
\nonumber
\langle m_R^\zeta \, (\bar q q)_\zeta \rangle_\pi  & := & 
8 N_c\int_{k,p}^{\bar\Lambda}\,B_0\,\left(
        \sigma_{B_0} - 
        B_0\,\left[
        \omega_k^2 \sigma_C^2 +
        p^2 \sigma_A^2 + 
        \sigma_B^2 \right] \right)\;,
\label{mqbq}
\end{eqnarray}
which vanishes linearly with $m_R^\zeta$; the canonical normalisation
constant is
\begin{eqnarray}
\label{pinormT}
\lefteqn{N_\pi^2  =  2 N_c \int_{k,p}^{\bar\Lambda}\,
B_0^2\,
\left\{\sigma_A^2 - 2 \left[ 
        \omega_k^2\sigma_C\sigma_C^\prime + 
        p^2 \sigma_A\sigma_A^\prime \right. \right.
 + \left. \sigma_B\sigma_B^\prime\right]} \\
& & \nonumber
   - \case{4}{3}\,p^2\,\left(
        \left[ \omega_k^2\left(\sigma_C\sigma_C^{\prime\prime} -
        (\sigma_C^\prime)^2\right) \right. \right.
+ \left. \left. \left.
        p^2\left(\sigma_A\sigma_A^{\prime\prime} -
        (\sigma_A^\prime)^2\right) + 
        \sigma_B\sigma_B^{\prime\prime} -
        (\sigma_B^\prime)^2 \right] \right) \right\}\;,
\end{eqnarray}
with $\sigma^\prime_B \equiv \partial\sigma_B(p^2,\omega_k)/\partial p^2$,
etc.; and the pion decay constant is obtained from
\begin{eqnarray}
f_\pi\,N_\pi  &= &   4 N_c\int_p^\Lambda\,B_0\,
\left\{ \sigma_A \sigma_B + \case{2}{3} |\vec{p}|^2 
        \left(\sigma_A^\prime \sigma_B - 
         \sigma_A \sigma_B^\prime\right)\right\}\;.
\label{fpi}
\end{eqnarray}

Equations~(\ref{pimass})-(\ref{fpi}) were derived under the assumption that
$\Gamma_\pi= i \gamma_5 B_0$.  Some of the limitations of this assumption
were discussed in Secs.~4.2) and~5.1), and they are considered further in
Ref.~[\ref{mr97}].  It is quantitatively unreliable near the transition
temperature, however, the qualitative behaviour of $N_\pi$ and $f_\pi$ is the
same, see Table~5.  Only after these studies were completed was it understood
that $N_\pi$ provides the best approximation to the leptonic decay constant
when $\Gamma_\pi= i \gamma_5 B_0$ is assumed.

\begin{table}[t]
\begin{center}
\begin{tabular}{lcc} 
                &      $\alpha$                 &       $\beta$   \\\hline
${\cal X}$          &       1.1 GeV         &       0.33    \\
${\kappa_0} $   &       0.16 GeV        &       0.30    \\
$N_\pi^2$       &       (0.18 GeV)$^2$  &       1.1     \\
$f_\pi N_\pi$   &       (0.15 GeV)$^2$  &       0.93     \\
$\langle m_R (\bar q q)\rangle$
                &       (0.15\,{\rm GeV})$^4$
                                        &       0.92    \\
$m_\pi$         &       0.12 GeV        &       -0.11   \\
$f_\pi$         &       0.12 GeV        &       0.36    \\
\end{tabular}
\end{center}
\caption{Parameters characterising the behaviour of the listed quantities,
fitted to \mbox{$\alpha\,(1-T/T_c)^\beta$}, near $T_c= 150\;$MeV.
\label{critparam}}
\end{table}
\begin{figure}[t]
\centering{\ 
\epsfig{figure=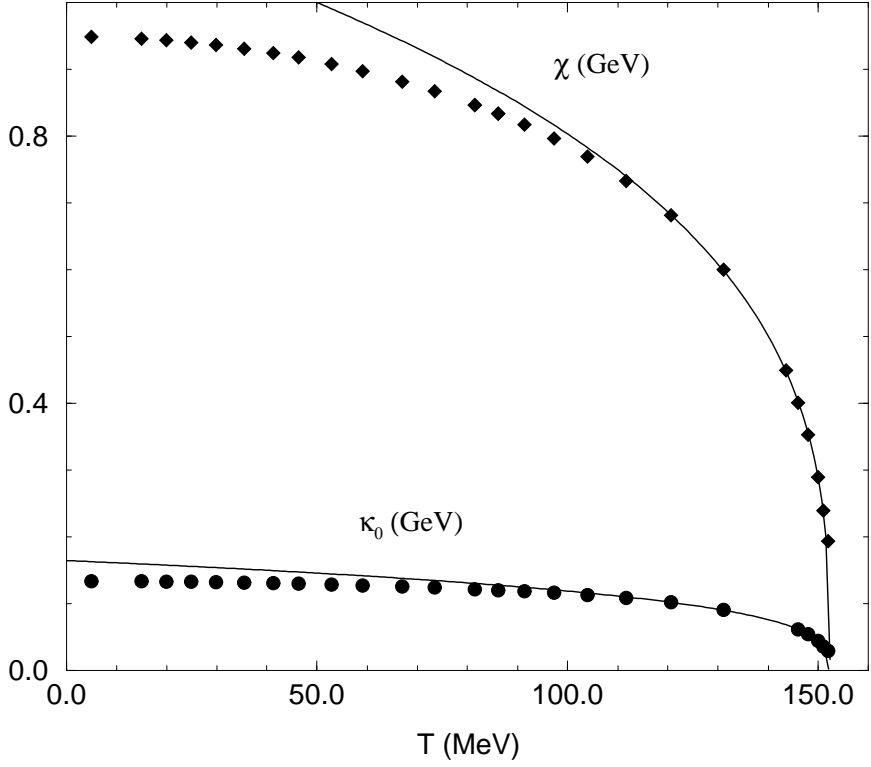,height=100mm}  }
\caption{The order parameters for chiral symmetry restoration (${\cal X}(T)$,
diamonds) and deconfinement ($\kappa_0(T)$, circles) both vanish at
$T_c=150\;$MeV.  The parameters for the fitted curves are presented in
Table.~\protect\ref{critparam}.
\label{coincidentT}}
\end{figure}
The calculated $T$-dependence of the chiral symmetry and deconfinement order
parameters is depicted in Fig.~\ref{coincidentT}.  The curves in the figure,
fitted on $T\in [120,150]\;$MeV, are of the form $\alpha\, (1-T/T_c)^\beta$
with $T_c \approx 150\;$ MeV and $\alpha$, $\beta$ given in
Table~\ref{critparam}.  The transitions are coincident and second-order with
$\beta_{\cal X} = \beta_{\kappa_0}$, within errors: $\sim 10$\%.  This
estimate of $\beta_{\cal X}$ is not a mean field value and it agrees with a
lattice estimate:\ucite{kl94} $\beta^{\rm lat}=0.30 \pm 0.08$.  It has been
argued\ucite{krishna} that two-light-flavour QCD is in the universality class
of the $N=4$ Heisenberg magnet, for which $\beta^{\rm H}=0.38 \pm 0.01$ and
both the DSE and lattice results are broadly consistent with this value.
However, neither of these estimates of $\beta$ survives more exhaustive
study,$\,^{\ref{laermann},\ref{arne}}$ and the most recent
analyses$\,^{\ref{arne},\ref{mPrivate}}$ suggest that in DSE models whose
long-range part is described by the regularised singularity in
Eq.~(\ref{delta}) the chiral symmetry restoration transition at finite-$T$ is
described by a mean-field value of $\beta$.

\begin{figure}[t]
\centering{\ 
\epsfig{figure=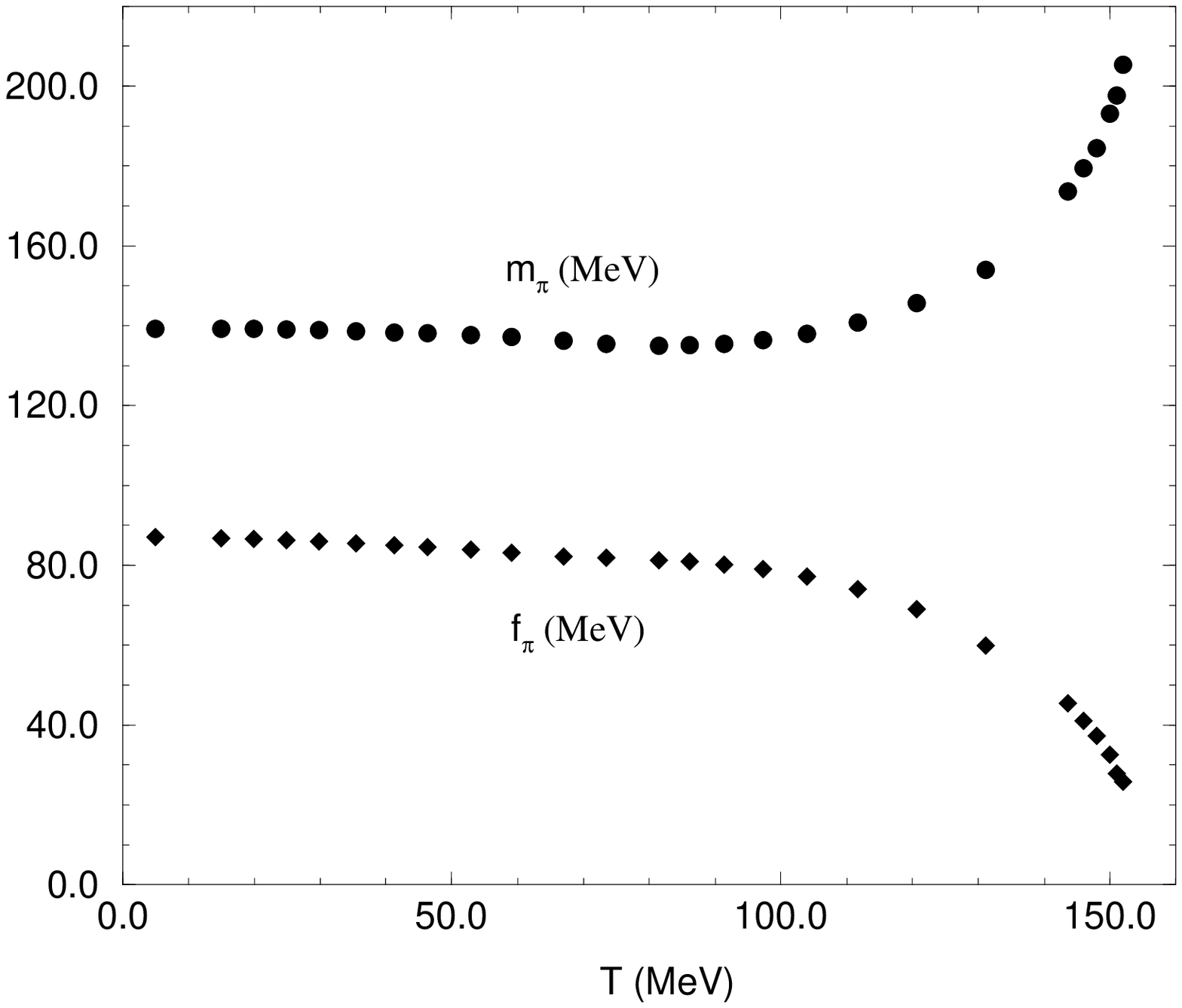,height=100mm}  }
\caption{Temperature dependence of the pion mass ($m_\pi(T)$, circles) and
pion weak-decay constant ($f_\pi(T)$, diamonds).
\label{pibehaviour}}
\end{figure}
\begin{figure}[t]
\centering{\ 
\epsfig{figure=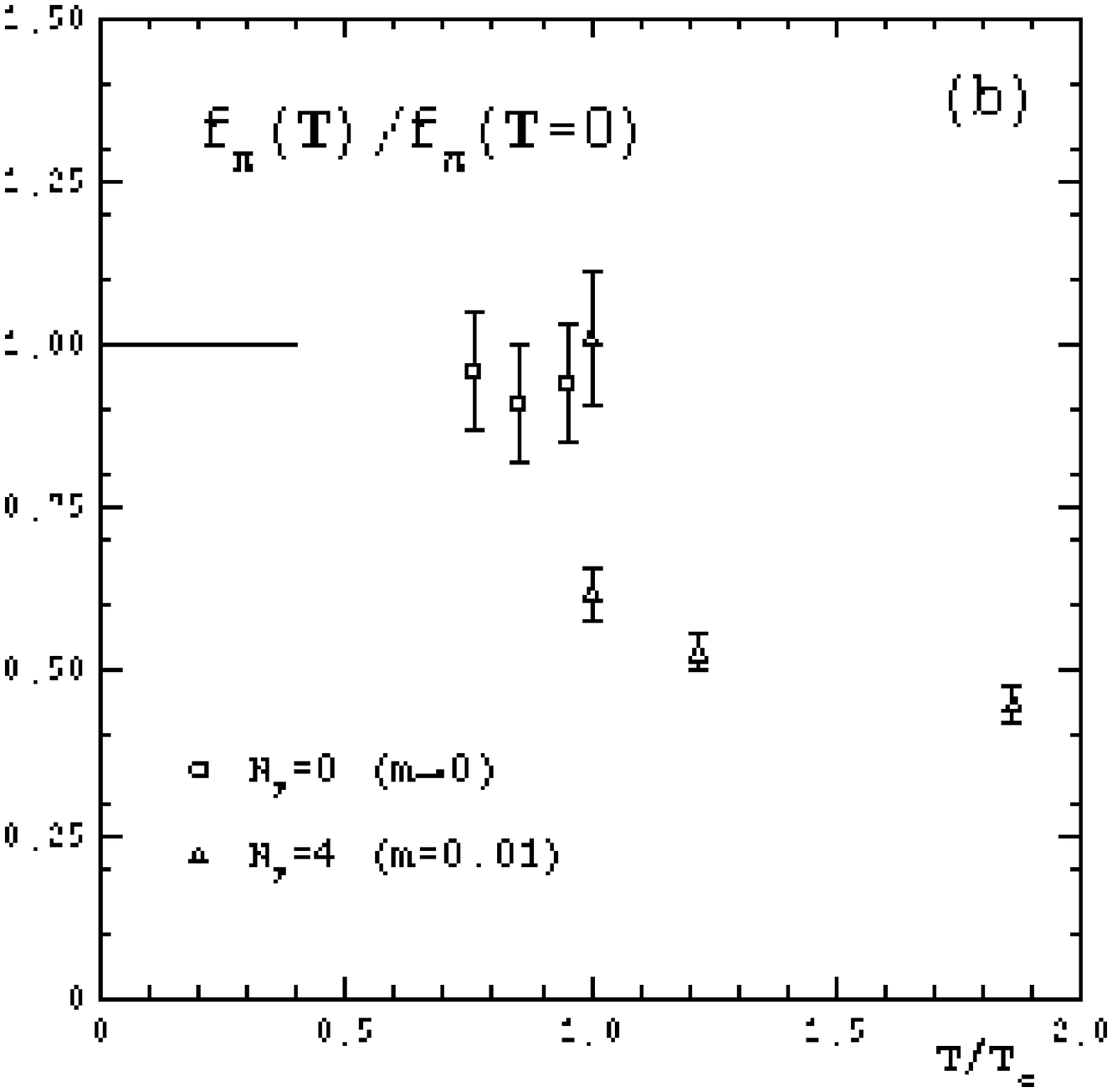,height=100mm}  }
\caption{Temperature dependence of the pion weak-decay constant on a
$32^3\times 8$ lattice.$\,^{\protect\ref{karsch95}}$
\label{lattfpi}}
\end{figure}
The behaviour of pion observables calculated from
Eqs.(\ref{pimass})-(\ref{fpi}) is depicted in Fig.\ref{pibehaviour}.  $f_\pi$
and $m_\pi$ are weakly sensitive to $T$ for $T< 0.7\,T_c^{\cal X}$, and this
is also seen in lattice simulations; e.g., the quark condensate in
Fig.~\ref{lattqbq} and $f_\pi$ in Fig.~\ref{lattfpi}.  However, as $T$
approaches $T_c^{\cal X}$, the mass eigenvalue in the pion Bethe-Salpeter
equation moves to increasingly larger values, as thermal fluctuations
overwhelm attraction in the channel, until at $T=T_c^{\cal X}$ there is no
solution and $f_\pi \to 0$.  This means that the pion-pole contribution to
the four-point, quark-antiquark correlation function disappears; i.e., there
is no quark-antiquark pseudoscalar bound state for $T>T_c^{\cal X}$.  That
may have important consequences for a wide range of physical
observables,\ucite{Betal95} if borne out by improved studies; e.g., such
$T$-dependence for $f_\pi$ and $m_\pi$ would lead to a 20\% reduction in the
$\pi \to \mu \nu_\mu$ decay widths at $T\approx 0.9\,T_c^{\cal X}$.
\vspace*{0.5\baselineskip}

{\it \arabic{section}.\arabic{subsection})~Complementary study at
$(T=0,\mu\neq 0)$.}\\[0.3\baselineskip]
\addcontentsline{toc}{subsection}{\arabic{section}.\arabic{subsection})~Complementary
study at $(T=0,\mu\neq 0)$} \addtocounter{subsection}{1}
The finite-$\mu$ behaviour of the same model\ucite{fr96} has also been
explored.\ucite{greg} The dressed-gluon propagator has the simple
form\ucite{fr96}
\begin{eqnarray}
g^2 D_{\mu\nu}(k) & = &
\left(\delta_{\mu\nu} - \frac{k_\mu k_\nu}{k^2}\right)
\frac{{\cal G}(k^2)}{k^2}\,;\\
\label{gksquare}
\frac{{\cal G}(k^2)}{k^2} & = &
\case{16}{9} \pi^2 \left[ 4 \pi^2 m_t^2 \delta^4(k)
+ \frac{1- {\rm e}^{-[k^2/(4 m_t^2)]}}{k^2}\right]\,,
\end{eqnarray}
and the rainbow approximation is used again.  Neither the dressed propagator
nor vertex have explicit $\mu$-dependence, which can arise through quark
vacuum polarisation insertions.  As such they may be inadequate at large
values of $\mu$, particularly near any critical chemical potential.  However,
in the absence of finite-$\mu$ studies of these quantities, the exploration
of such models is useful, and one can assess the results obtained in the
light of existing experiments and related theoretical studies.

The renormalised dressed-quark propagator is
\begin{equation}
S(p_{[\mu]}):= -i
\vec{\gamma}\cdot \vec{p}\, \sigma_A(p_{[\mu]} ) 
        - i \gamma_4\, \omega_{[\mu]} \,
\sigma_C(p_{[\mu]}) + \sigma_B(p_{[\mu]})\,, 
\end{equation}
where $p_{[\mu]}:= (\vec{p},\omega_{[\mu]})$, with $\omega_{[\mu]} := p_4 + i
\mu$.  The quark DSE and the renormalisation conditions are similar to those
discussed in the previous section, and the equation has two qualitatively
distinct solutions: a chirally symmetric Wigner-Weyl mode, characterised by
$B_0\equiv 0$; and a confining Nambu-Goldstone mode, characterised
$B_0\not\equiv 0$.

To explore the possibility of a phase transition one calculates the relative
stability of the different phases, which is measured by the difference in
pressure between them.  The pressure is obtained directly from the partition
function, ${\cal Z}$: it is the sum of all vacuum-to-vacuum transition
amplitudes.  In ``stationary phase'' approximation, the partition function is
given by the tree-level auxiliary-field effective action\ucite{haymaker} and
the pressure is:
\begin{equation}
\label{lnz}
P[S] := \frac{T}{V}\ln{\cal Z} = \frac{T}{V}\left\{ {\rm
TrLn}\left[\case{1}{T} S^{-1}\right] -\case{1}{2} {\rm Tr}\left[\Sigma
S\right]\right\}\,.
\end{equation}
It is a functional of $S(p_{[\mu]})$.  In the absence of interactions $\Sigma
\equiv 0$ and Eq.~(\ref{lnz}) yields the free fermion partition function.
[Additive gluon contributions cancel in the pressure difference and are
neglected.]  The contribution of hadrons and hadron-like correlations to the
partition function are neglected in Eq.~(\ref{lnz}).  At the level of
approximation consistent with Eq.~(\ref{lnz}) these terms are an additive
contribution that can be estimated using the {\it hadronisation} techniques
of Ref.~[\ref{cahill}].  After a proper normalisation of the partition
function; i.e., subtraction of the vacuum contribution, they are the only
contributions to the partition function in the confinement domain.  They are
easy to calculate and are considered no further here as they are not a
significant influence on the position of the phase boundary.

The pressure difference is
\begin{equation}
\label{bagpres}
\case{1}{2 N_f N_c} \,{\cal B}(\mu)  :=  
\int_p^\Lambda\,
\left\{ \ln\left[\frac{|\vec{p}|^2 A_0^2 + \omega_{[\mu]}^2 C_0^2 + B_0^2}
                {|\vec{p}|^2 \hat A_0^2 + \omega_{[\mu]}^2 \hat C_0^2}\right]
+  |\vec{p}|^2 \left(\sigma_{A_0} - \hat\sigma_{A_0}\right)
+  \omega_{[\mu]}^2 \left(\sigma_{C_0} - \hat\sigma_{C_0}\right)\right\}\,,
\end{equation}
which defines a $\mu$-dependent ``bag constant''.\ucite{reg85} In
Eq.~(\ref{bagpres}), $\hat A$ and $\hat C$ represent the solution of
Eq.(\ref{qDSE}) obtained when $B_0\equiv 0$; i.e., when DCSB is absent.  This
solution exists for all $\mu$.
\begin{figure}[t]
  \centering{\
  \epsfig{figure=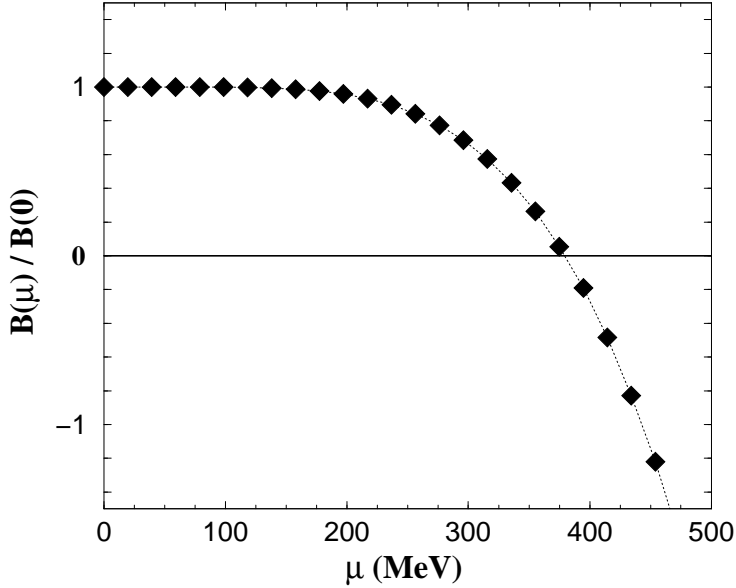,height=9.0cm}}
\caption{${\cal B}(\mu)$ from (\protect\ref{bagpres}); ${\cal B}(\mu)> 0$
marks the domain of confinement and dynamical chiral symmetry breaking.  The
zero of ${\cal B}(\mu)$ is $\mu_c=375\,$MeV.  ${\cal B}(0)= (0.104\,{\rm
GeV})^4$, which can be compared with the value $\sim (0.145\,{\rm GeV})^4$
commonly used in bag-like models of hadrons.$^{\protect\ref{cahill}}$
\label{bagpressure}}
\end{figure}
${\cal B}(\mu)$ is plotted in Fig.~\ref{bagpressure}.  It is positive when
the Nambu-Goldstone phase is dynamically favoured; i.e., has the highest
pressure, and becomes negative when the Wigner pressure becomes larger.  The
critical chemical potential is the zero of ${\cal B}(\mu)$; i.e.,
$\mu_c=375\,$MeV.  This abrupt switch from the Nambu-Goldstone to the
Wigner-Weyl phase signals a first order transition.

The order parameter for chiral symmetry restoration is that given in
Eq.~(\ref{chiorder}), while the confinement order parameter at $(T=0,\mu\neq
0)$ is derived from
\begin{equation}
\Delta_S(\tau) := \frac{1}{2 \pi}\int_{-\infty}^\infty\,dp_4\,
        {\rm e}^{i p_4 \tau}\,\sigma_{B_0}(\vec{p}=0,\omega_{[\mu]})\,,
\end{equation}
an analogue of Eq.~(\ref{schwinger}).  For a free, massive fermion
$\sigma_B(\vec{p}=0,\omega_{[\mu]})= M/(\omega_{[\mu]}^2+M^2)$.  This
function has poles at $p_4^2= -(M\pm \mu)^2$, which are associated with the
$\mu$-induced offset of the particle and antiparticle zero-point energies,
and
\begin{equation}
\Delta_S(\tau) = \case{1}{2}\, {\rm e}^{-(M-\mu)\, \tau}\, \theta(M-\mu)\,,
\end{equation}
which is positive-definite and monotonically decreasing.  In contrast, as
observed above, for a Schwinger function with complex-conjugate $p^2$-poles,
$\Delta_S(\tau)$ has zeros at $\tau>0$.

\begin{figure}[t]
 \centering{\
 \epsfig{figure=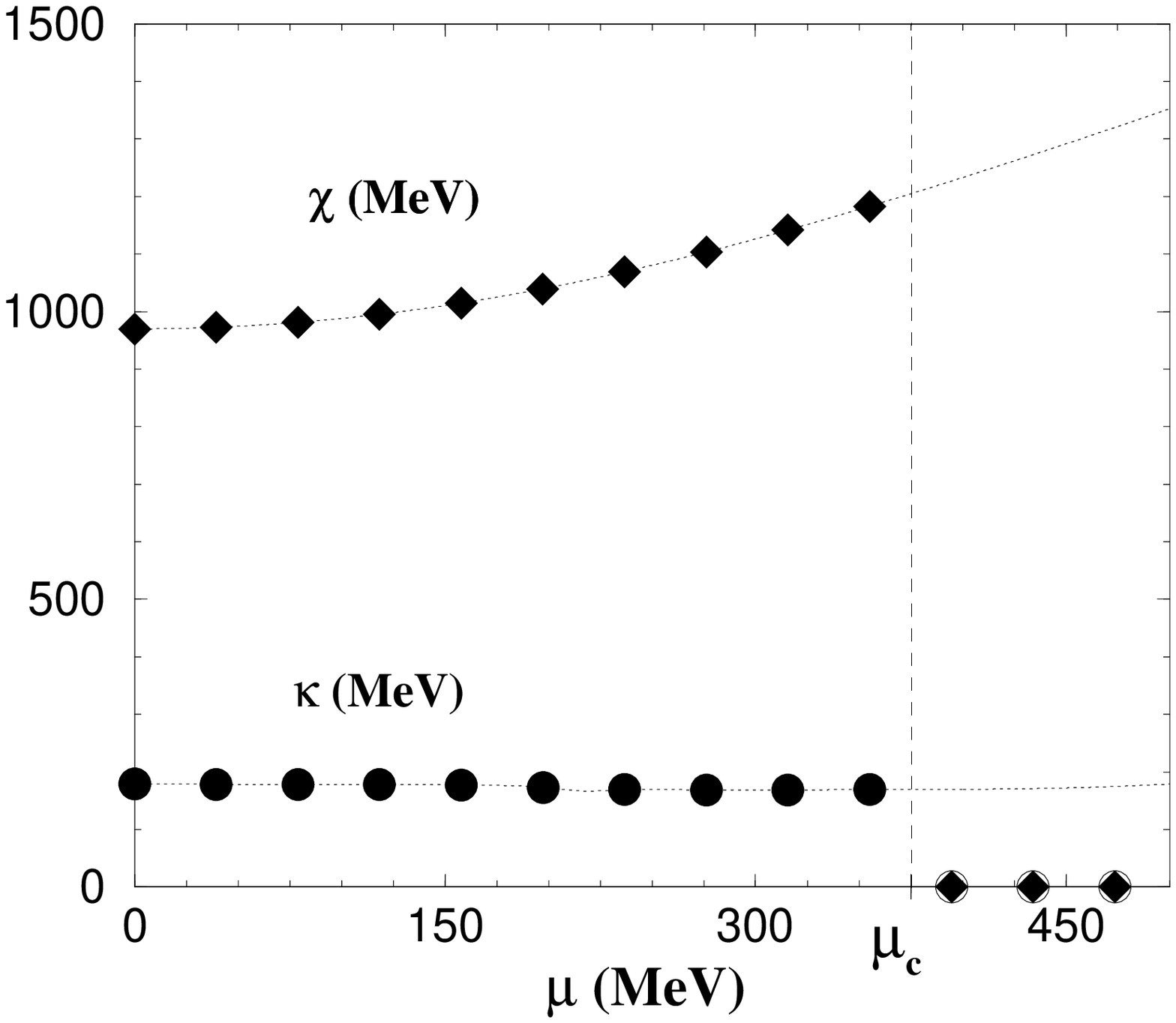,height=9.0cm}}
\caption{The order parameters for chiral symmetry restoration [${\cal X}$,
diamonds] and deconfinement [$\kappa$, circles].  $\mu_c=375\,$MeV.
\label{figcrit}}
\end{figure}
The $\mu$-dependence of the order parameters for chiral symmetry restoration
and deconfinement is depicted in Fig.~\ref{figcrit}.  The chiral order
parameter {\it increases} with increasing chemical potential up to $\mu_c $,
with ${\cal X}(\mu_c)/{\cal X}(0)\approx 1.2$, whereas $\kappa(\mu)$ is
insensitive to increasing $\mu$.  At $\mu_c$ they both drop immediately and
discontinuously to zero, as expected of a first-order phase transition.  The
increase of the chiral order parameter with $\mu$ is a necessary consequence
of the momentum dependence of the scalar piece of the quark self energy,
$B(p_{[\mu]})$, as is easily seen in Ref.~[\ref{thermo}] and in Secs.~7.5)
and~7.6).  The vacuum quark condensate behaves in qualitatively the same
manner as ${\cal X}$.

\begin{figure}[t]
\centering{\
\epsfig{figure=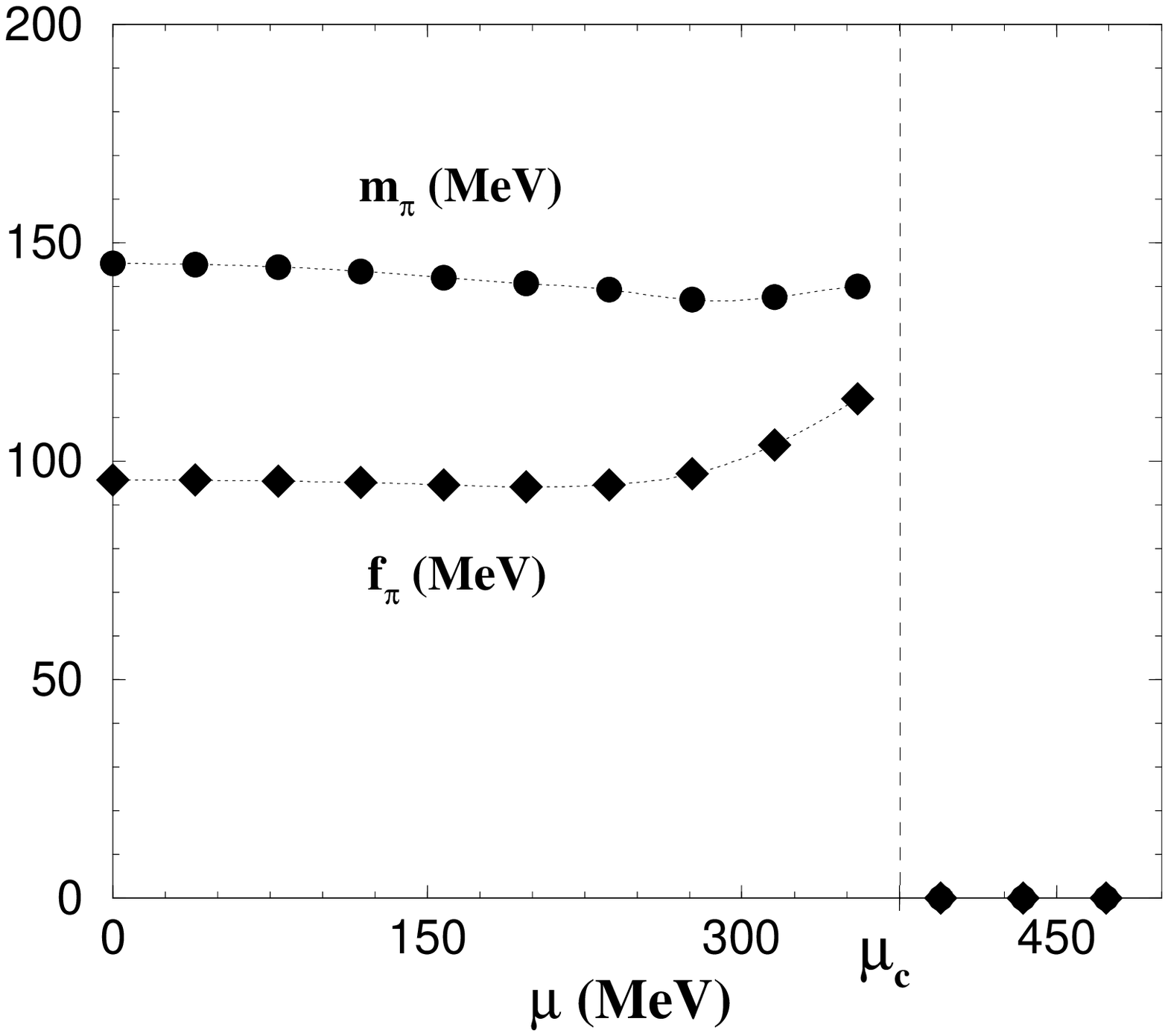,height=9.0cm}}
\vspace*{-\baselineskip}
 
\caption{Chemical potential dependence of the pion mass [$m_\pi$, circles]
and pion leptonic decay constant [$f_\pi$, diamonds].
\label{mpifpi}}
\end{figure}
The behaviour of $m_\pi$ and $f_\pi$ is illustrated in Fig.~\ref{mpifpi}.
One observes that although the chiral order parameter {\it increases} with
$\mu$, $m_\pi$ {\it decreases} slowly as $\mu$ increases.  This slow fall
continues until $\mu \approx 0.7\,\mu_c$, when $m_\pi(\mu)/m_\pi(0) \approx
0.94$.  At this point $m_\pi$ begins to increase although, for $\mu<\mu_c$,
$m_\pi(\mu)$ does not exceed $m_\pi(0)$.  This precludes pion condensation,
in qualitative agreement with Ref.~[\ref{kubodera}].  The behaviour of
$m_\pi$ results from mutually compensating increases in $\langle m_R^\zeta
(\bar q q)_\zeta\rangle_\pi$ and $N_\pi^2$.  This is a manifestation of the
manner in which dynamical chiral symmetry breaking protects pseudoscalar
meson masses against rapid changes with $\mu$.  The pion leptonic decay
constant is insensitive to the chemical potential until $\mu\approx
0.7\,\mu_c$, when it increases sharply so that $f_\pi(\mu_c^-)/f_\pi(\mu=0)
\approx 1.25$.  The relative insensitivity of $m_\pi$ and $f_\pi$ to changes
in $\mu$, until very near $\mu_c$, mirrors the behaviour of these observables
at finite-$T$.\ucite{prl} For example, it leads only to a $14$\% increase in
the $\pi\to \mu\nu$ decay width at $\mu\approx 0.9\,\mu_c$.  The universal
scaling conjecture of Ref.~[\ref{brown}] is inconsistent with the
anticorrelation we observe between the $\mu$-dependence of $f_\pi$ and
$m_\pi$.

Comparing the $\mu$-dependence of $f_\pi$ and $m_\pi$ with their
$T$-dependence, one observes an anticorrelation; e.g., at $\mu=0$, $f_\pi$
falls continuously to zero as $T$ is increased towards $T_c \approx
150\,$MeV.\ucite{prl} This too is a necessary consequence of the
momentum-dependence of the quark self-energy.  In calculating these
observables the natural dimension is mass-squared, and their behaviour at
finite $T$ and $\mu$ is determined by ${\sf Re}(\omega_{[\mu]}^2) \sim [\pi^2
T^2 - \mu^2]$, where the $T$-dependence arises from the introduction of the
fermion Matsubara frequency: \mbox{$p_4 \to (2n +1)\pi T$}.  Hence when such
a quantity decreases with $T$ it will increase with $\mu$, and vice-versa.
This is elucidated in Secs.~7.5)~and~7.6), and in Ref.~[\ref{schmidt98}].

The confined-quark vacuum consists of quark-antiquark pairs correlated in a
scalar condensate.  Increasing $\mu$ increases the scalar density: $(-\langle
\bar q q \rangle)$.  This result is an expected consequence of confinement,
which entails that each additional quark must be locally paired with an
antiquark thereby increasing the density of condensate pairs as $\mu$ is
increased.  For this reason, as long as $\mu<\mu_c$, there is no excess of
particles over antiparticles in the vacuum and hence the baryon number
density remains zero;\ucite{thermo} i.e., $ \rho_B^{u+d}=0\,,\; \forall \mu <
\mu_c $.  This is just the statement that quark-antiquark pairs confined in
the condensate do not contribute to the baryon number density.

The quark pressure, $P^{u+d}[\mu]$, can be calculated,\ucite{thermo} see
Sec.~7.5), and one finds that after deconfinement it increases rapidly, as
the condensate ``breaks-up'', and an excess of quarks over antiquarks
develops.  The baryon-number density, \mbox{$\rho_B^{u+d} = (1/3)\partial
P^{u+d}/\partial \mu$}, also increases rapidly, with
\begin{equation}
\rho_B^{u+d}(\mu\approx 2 \mu_c) \simeq 3 \,\rho_0\,,
\end{equation}
where $\rho_0=0.16\,{\rm fm}^{-3}$ is the equilibrium density of nuclear
matter.  For comparison, the central core density expected in a
$1.4\,M_\odot$ neutron star is $3.6$-$4.1\,\rho_0$.\ucite{wiringa} Finally,
at $\mu\sim 5 \mu_c$, the quark pressure saturates the ultrarelativistic
limit: $P^{u+d}= \mu^4/(2\pi^2)$, and there is a simple relation between
baryon-density and chemical-potential:
\begin{equation}
\label{fqg}
\rho_B^{u_F+d_F}(\mu) = \frac{1}{3} \, \frac{2 \mu^3}{\pi^2}\,,
\; \forall \mu \gsim 5 \mu_c \,,
\end{equation}
so that $\rho_B^{u_F+d_F}(5\mu_c)\sim 350\,\rho_0$.  Thus the quark pressure
in the deconfined domain overwhelms any finite, additive contribution of
hadrons to the equation of state, which anticipating this was neglected in
Ref.~[\ref{greg}].  This discussion suggests that a quark-gluon plasma may be
present in the core of dense neutron stars.
\vspace*{0.5\baselineskip}

{\it \arabic{section}.\arabic{subsection})~Simultaneous study of $(T\neq
0,\mu\neq 0)$.}\\[0.3\baselineskip] 
\addcontentsline{toc}{subsection}{\arabic{section}.\arabic{subsection})~Simultaneous
study of $(T\neq 0,\mu\neq 0)$} 
\addtocounter{subsection}{1}
This is the most difficult problem and the most complete study\ucite{thermo}
to date employs a simple {\it Ansatz} for the dressed-gluon propagator:
\begin{equation}
\label{mnprop}
g^2 D_{\mu\nu}(\vec{p},\Omega_k) = 
\left(\delta_{\mu\nu} 
- \frac{p_\mu p_\nu}{|\vec{p}|^2+ \Omega_k^2} \right)
2 \pi^3 \,\frac{\eta^2}{T}\, \delta_{k0}\, \delta^3(\vec{p})\,,
\end{equation}
which exhibits the infrared enhancement suggested by Ref.~[\ref{bp89}].  As
an infrared-dominant model that does not represent well the behaviour of
$D_{\mu\nu}(\vec{p},\Omega_k)$ away from $|\vec{p}|^2+ \Omega_k^2 \approx 0$,
some model-dependent artefacts arise.  However, there is significant merit in
its simplicity and, since the artefacts are easily identified, the model
remains useful as a means of elucidating many of the qualitative features of
more sophisticated {\it Ans\"atze}.

With this model, using the rainbow approximation, the \qcdtm\ gap equation,
or DSE for the dressed-quark propagator, is\ucite{bender96}
\begin{equation}
\label{mndse}
S^{-1}(\vec{p},\omega_k) = S_0^{-1}(\vec{p},\tilde\omega_k)
        + \case{1}{4}\eta^2\gamma_\nu S(\vec{p},\tilde\omega_k) \gamma_\nu\,.
\end{equation}
A simplicity inherent in Eq.~(\ref{mnprop}) is now apparent: it allows the
reduction of an integral equation to an algebraic equation, in whose solution
many of the qualitative features of more sophisticated models are manifest,
as will become clear.  In terms of the scalar functions introduced in
Eq.~(\ref{genformS}), Eq.~(\ref{mndse}) reads
\begin{eqnarray}
\label{beqnfour}
\eta^2 m^2 & = & B^4 + m B^3 + \left(4 \tilde p_k^2 - \eta^2 -
        m^2\right) B^2 -m\,\left( 2\,{{\eta }^2} + {m^2} +
        4\,\tilde p_k^2 \right)B   \,,     \\ 
A(\tilde p_k) & = & C(\tilde p_k) = \frac{2 B(\tilde p_k)}{m +B(\tilde p_k)}\,.
\end{eqnarray}

Of particular interest is the chiral limit, $m=0$.  In this case
Eq.~(\ref{beqnfour}) reduces to a quadratic equation for $B(\tilde p_k)$,
which has two qualitatively distinct solutions.  The ``Nambu-Goldstone''
solution, for which
\begin{eqnarray}
\label{ngsoln}
B(\tilde p_k) & = &\left\{
\begin{array}{lcl}
\sqrt{\eta^2 - 4 \tilde p_k^2}\,, & &{\sf Re}(\tilde p_k^2)<\case{\eta^2}{4}\\
0\,, & & {\rm otherwise}
\end{array}\right.\\
C(\tilde p_k) & = &\left\{
\begin{array}{lcl}
2\,, & & {\sf Re}(\tilde p_k^2)<\case{\eta^2}{4}\\
\case{1}{2}\left( 1 + \sqrt{1 + \case{2 \eta^2}{\tilde p_k^2}}\right)
\,,& & {\rm otherwise}\,,
\end{array}\right.
\end{eqnarray}
describes a phase of this model in which: 1) chiral symmetry is dynamically
broken, because one has a nonzero quark mass-function, $B(\tilde p_k)$, in
the absence of a current-quark mass; and 2) the dressed-quarks are confined,
because the propagator described by these functions does not have a Lehmann
representation.  The alternative ``Wigner'' solution, for which
\begin{eqnarray}
\label{wsoln}
\hat B(\tilde p_k)  \equiv  0 &,\;& 
\hat C(\tilde p_k)  = 
\case{1}{2}\left( 1 + \sqrt{1 + \case{2 \eta^2}{\tilde p_k^2}}\right)\,,
\end{eqnarray}
describes a phase of the model in which chiral symmetry is not broken and the
dressed-quarks are not confined. 
 
With these two ``phases'', characterised by qualitatively different,
momentum-dependent modifications of the quark propagator, this model can be
used to explore chiral symmetry restoration and deconfinement, and elucidate
aspects of the method in such studies.

In this model the relative stability of the different phases is measured by a
$(T,\mu)$-dependent ``bag constant'',\ucite{reg85}
\begin{eqnarray}
\label{bagconsgen}
{\cal B}(T,\mu) & := & P[S_{\rm NG}] - P[S_{\rm W}]\,, 
\end{eqnarray}
where $S_{\rm NG}$ means Eq.~(\ref{genformS}) obtained from
Eq.~(\ref{ngsoln}) and $S_{\rm W}$, Eq.~(\ref{genformS}) obtained from
Eq.~(\ref{wsoln}).  As above, ${\cal B}(T,\mu) >0$ indicates the stability of
the confined (Nambu-Goldstone) phase and hence the phase boundary is
specified by that curve in the $(T,\mu)$-plane for which
\begin{eqnarray}
\label{bagconsgenA}
{\cal B}(T,\mu) & \equiv & 0\,.
\end{eqnarray}

\begin{figure}[t]
\centering{\
\epsfig{figure=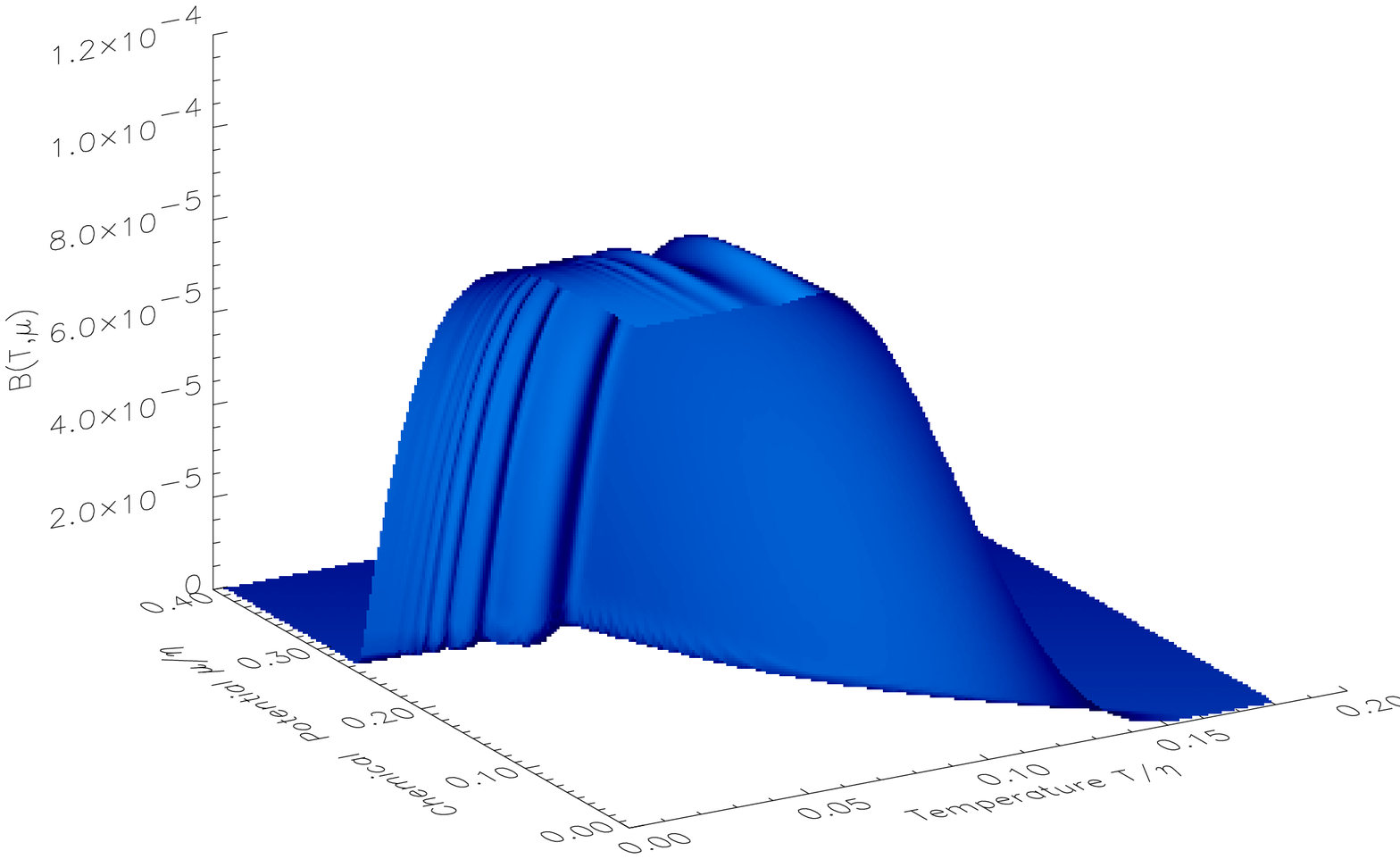,height=10.0cm}}
\caption{${\cal B}(T,\mu)$ from Eq.~(\protect\ref{bagcons}); \mbox{${\cal
B}(T,\mu)>0$} marks the confinement domain.  The scale is set by \mbox{${\cal
B}(0,0)= (0.102\,\eta)^4 = (0.109\,{\rm GeV})^4$};
$\eta=1.06\,$GeV.$^{\protect\ref{mn83}}$
\label{bagpresF}}
\end{figure}
In the chiral limit
\begin{equation}
\label{bagcons}
{\cal B}(T,\mu)  = 
 \eta^4\,2 N_c N_f \frac{\bar T}{\pi^2}\sum_{l=0}^{l_{\rm max}}
        \int_0^{\bar\Lambda_l}\,dy\,y^2\,
        \left\{{\sf Re}\left(2 \bar p_l^2 \right) 
                - {\sf Re}\left(\frac{1}{C(\bar p_l)}\right)
- \ln\left| \bar p_l^2 C(\bar p_l)^2\right|        \right\},
\end{equation}
with: $\bar T=T/\eta$, $\bar \mu=\mu/\eta$; $l_{max}$ is the largest value of
$l$ for which $\bar\omega^2_{l_{\rm max}}\leq \case{1}{4}+\bar\mu^2$ and this
also specifies $\omega_{l_{max}}$, $\bar\Lambda^2 = \bar\omega^2_{l_{\rm
max}}-\bar\omega_l^2$, $\bar p_l = (\vec{y},\bar\omega_l+i\bar\mu)$.  ${\cal
B}(T,\mu)$ is depicted in Fig.~\ref{bagpresF} and the critical line in
Fig.~\ref{critline}.  The deconfinement and chiral symmetry restoration
transitions are coincident.

\begin{figure}[t]
\centering{\
\epsfig{figure=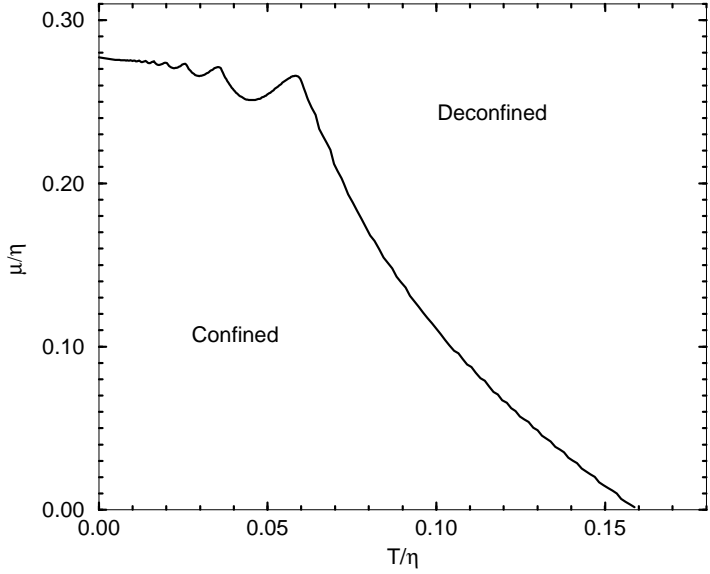,height=9.0cm}}
\caption{The phase boundary in the $(\bar T,\bar \mu)$-plane obtained from
(\protect\ref{bagconsgenA}) and (\protect\ref{bagcons}).  The ``structure''
in this curve, apparent for small-$T$, is an artefact of the inadequate
representation of the quark-quark interaction in the ultraviolet by
Eq.~(\protect\ref{mnprop}).
\label{critline}}
\end{figure}
For $\mu=0$ the transition is second order and the critical temperature is
$T_c^0 = 0.159\,\eta$, which using the value of $\eta=1.06\,$GeV obtained by
fitting the $\pi$ and $\rho$ masses\ucite{mn83} corresponds to $T_c^0 =
0.170\,$GeV.  This is only 12\% larger than the value obtained in
Sec.~7.3),\ucite{prl} and the order of the transition is the same.  However,
in the present case the critical exponent is $\beta=0.5$.  For any $\mu \neq
0$ the transition is first-order, as revealed by close scrutiny of
Fig.~\ref{bagpresF}.  For $T=0$ the critical chemical potential is
$\mu_c^0=0.3\,$GeV, which is \mbox{$\approx 30$\%} smaller than the result in
Sec.~7.4).\ucite{greg}  One notes from Fig.~\ref{critline} that $\mu_c(T)$
is insensitive to $T$ until $T\approx 0.3\,T_c^0$.  The discontinuity
in the order parameters vanishes as $\mu\to 0$.

In the deconfinement domain, illustrated clearly in Fig.~\ref{critline}, the
quarks contribute an amount
\begin{equation}
\label{pwigner}
P[S_{\rm W}]= \eta^4\,2 N_c N_f \frac{\bar T}{\pi^2}\sum_{l=0}^{\infty}
        \int_0^{\infty}\,dy\,y^2\,
        \left\{\ln\left| \beta^2 \tilde p_l^2 \hat C(\bar p_l)^2\right|
        - 1 +  {\sf Re}\left(\frac{1}{\hat C(\bar p_l)}\right)  \right\}
\end{equation}
to the pressure, which must be renormalised to zero on the phase boundary.
Just as for free fermions, this expression is formally divergent and one must
isolate and define the active, temperature-dependent contribution.  This is
difficult because, in general, $\hat C(\bar p_l)$ is only known numerically
and hence it is not possible to evaluate $P[S_{\rm W}]$ analytically.  A
method for the numerical evaluation of Eq.~(\ref{pwigner}) was developed in
Ref.~[\ref{thermo}].

Consider the derivative of the integrand in Eq.~(\ref{pwigner}):
\begin{eqnarray}
\label{presb}
\lefteqn{
\sum_{l=0}^\infty\,\frac{d}{d\bar T}\,\left\{\ln\left| \beta^2 \tilde p_l^2 \hat
        C(\bar p_l)^2\right| 
        - 1 +  {\sf Re}\left(\frac{1}{\hat C(\bar p_l)}\right)  \right\} = }\\
&& \nonumber
\sum_{l=0}^\infty\,
\left\{ -\frac{1}{\bar T} \left[
        \frac{(y-\bar\mu)^2}{(y-\bar\mu)^2+\bar\omega_l^2}
        + \frac{(y+\bar\mu)^2}{(y+\bar\mu)^2+\bar\omega_l^2}
        \right]
        +  {\sf Re}\left(
        \frac{ 2 \hat C(\bar p_l) - 1}{\hat C(\bar p_l)^2}\,
        \frac{d \hat C(\bar p_l)}{d \bar T\;\;\;\;}\right)
\right\}\,.
\end{eqnarray}
In the absence of interactions $C(\bar p_l)\equiv 1$, the second term is zero
and
\begin{equation}
\label{presa}
-\frac{2}{\bar T} \sum_{l=0}^\infty\,
\left[\frac{(y-\bar\mu)^2}{(y-\bar\mu)^2+\bar\omega_l^2}
        + \frac{(y+\bar\mu)^2}{(y+\bar\mu)^2+\bar\omega_l^2}\right]  
= \frac{d}{d\bar T}\left\{
        \frac{e(y)}{\bar T} + {\cal I}(e(y))\right\}\,,
\end{equation}
where in this case $e(y)=y$ and 
\begin{equation}
{\cal I}(\zeta)  = 
        \ln\left[1 + \exp\left(-\frac{\zeta-\bar\mu}{\bar T}\right)\right]
        + \ln\left[1 + \exp\left(-\frac{\zeta+\bar\mu}{\bar T}\right)\right]\,.
\end{equation}
Appropriately inserting Eq.~(\ref{presa}) for the parenthesised term in
Eq.~(\ref{pwigner}), and neglecting $T$-independent terms one obtains,
\begin{eqnarray}
P[S_0] & = & \eta^4\,N_c N_f \frac{\bar
        T}{\pi^2}\int_0^{\infty}\,dy\,y^2\, {\cal I}(y)\\
 & = & \label{sbpres}
        \eta^4\, N_c N_f \frac{1}{12\pi^2}\left(
        \bar\mu^4 + 2 \pi^2 \bar\mu^2 \bar T^2 + \frac{7}{15}\pi^4
                        \bar T^4\right)\,, 
\end{eqnarray}
which is the massless free particle pressure.

To proceed in the general case the assumption is made\ucite{thermo} that the
nontrivial momentum dependence of $\hat C(\bar p_l)$, which is manifest in
all DSE-models of \qcdtm, acts primarily to modify the usual massless, free
particle dispersion law.  One evaluates the sum on the right-hand-side of
(\ref{presb}) numerically and uses the form on the right-hand-side of
Eq.~(\ref{presa}) to fit a modified, $T$-independent dispersion law,
$\underline{e}(y,\bar\mu)= y +\kappa(y,\bar\mu)$, to the numerical results.
The existence of a $\kappa(y,\bar\mu)$ that provides a good $\chi^2$-fit on
the deconfinement domain is understood as an {\it a posteriori} justification
of the assumption.  In Ref.~[\ref{thermo}] the relative error between the fit
and the numerical results is $< 10$\% on the entire $T$-domain.

\begin{figure}[t]
\centering{\
\epsfig{figure=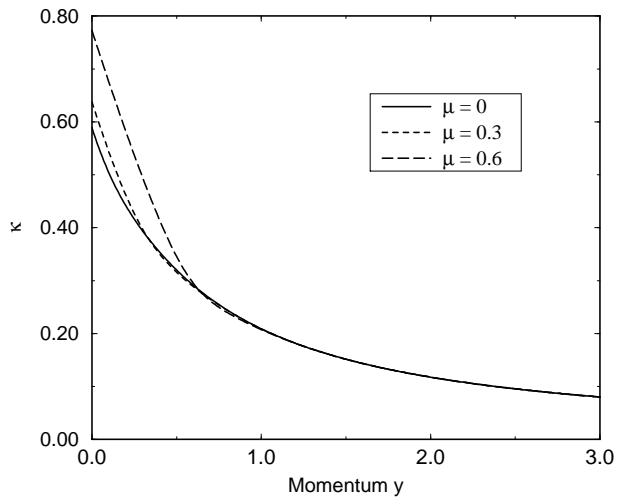,height=9.0cm,angle=-90}}
\caption{$\kappa(y,\bar\mu)$, which describes the nonperturbative
modification of the free particle dispersion law, for $\bar\mu=0, 0.3, 0.6$.
By assumption, it is independent of $T$.
\label{figrho}}
\end{figure}
The calculated form of $\kappa(y,\bar\mu)$ is depicted in Fig.~\ref{figrho};
it only depends weakly on $\bar\mu$.  The form indicates a persistence of
nonperturbative effects into the domain of deconfinement, evident in the
nontrivial momentum dependence of $\hat C(\bar p_l)$ and its slow evolution
to the asymptotic value $\hat C(\bar p_l)=1$.  The effect of this is to
generate a mass-scale in the massless dispersion law: $\kappa(0,0)\simeq 0.6
\sim 2 \bar\mu_c^0$.  This mass-scale is unrelated to the chiral-symmetry
order parameter, ${\cal X}$ in Eq.~(\ref{chiorder}), and is a qualitatively
new feature of the study.  For $\bar\mu> 5 \bar\mu_c^0$ the explicit
mass-scale introduced by the chemical potential overwhelms the dynamically
generated scale.

Using this result, Eq.~(\ref{pwigner}) becomes
\begin{eqnarray}
P[S_{\rm W}] & = & 
\eta^4\,N_c N_f \frac{\bar
        T}{\pi^2}\int_0^{\infty}\,dy\,y^2\, 
        {\cal I}(\underline{e}(y,\bar\mu))\,,
\end{eqnarray}
and the quark pressure in this DSE-model of \qcdtm\ is 
\begin{equation}
\label{qpres}
P_q(T,\mu)  =  \theta({\cal D})\left\{
        P[S_{\rm W}] - \left.P[S_{\rm W}]\right|_{\partial {\cal D}}\right\}
\end{equation}
where ${\cal D}$ is the domain marked ``Deconfined'' in Fig.~\ref{critline},
$\theta( {\cal D})$ is a step function, equal to one for $(T,\mu)\in {\cal
D}$, and $\left.P[S_{\rm W}]\right|_{\partial {\cal D}}$ indicates the
evaluation of this expression on the boundary of ${\cal D}$, as defined by
the intersection of a straight-line from the origin in the $(T,\mu)$-plane to
the argument-value.
\begin{figure}[t]
\centering{\
\epsfig{figure=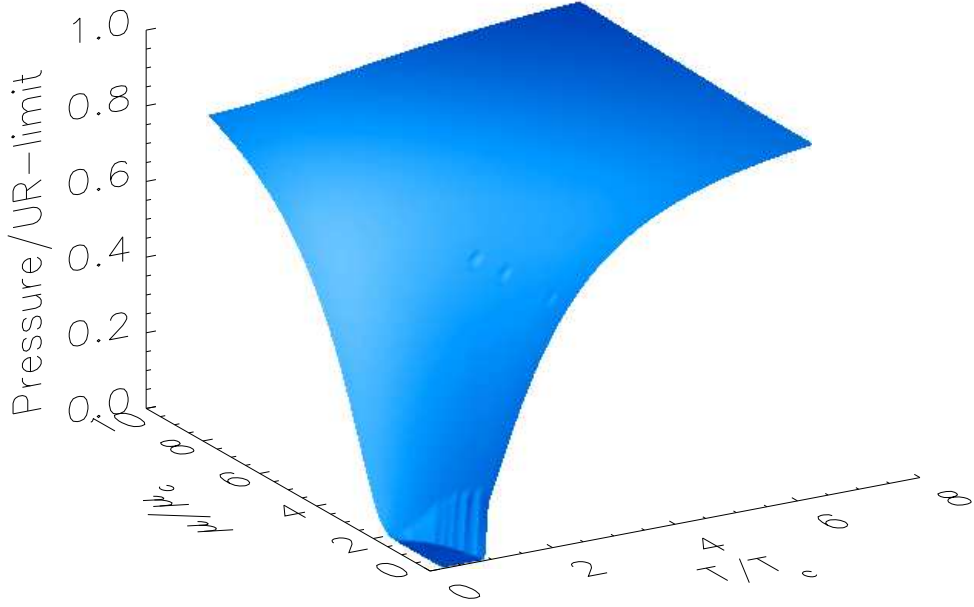,height=9.0cm}}
\caption{The quark pressure, $P_q(\bar T,\bar\mu)$, normalised to the free,
massless (or Ultra-Relativistic) result, Eq.~(\protect\ref{sbpres}).
\label{presfig}}
\end{figure}
It is plotted in Fig.~\ref{presfig}, which illustrates clearly that in this
model the free particle (Stefan-Boltzmann) limit is reached at large values
of $\bar T$ and $\bar\mu$.  The approach to this limit is slow, however.  For
example, at $\bar T \sim 0.3 \sim 2 \bar T_c^0$, or $\bar \mu \sim 1.0 \sim 3
\bar\mu_c^0$, Eq.~(\ref{qpres}) is only 50\% of the free particle pressure,
Eq.~(\ref{sbpres}).  A qualitatively similar result is observed in numerical
simulations of lattice-QCD actions at finite-$T$.\ucite{karsch97} This
feature results from the slow approach to zero with $y$ of
$\kappa(y,\bar\mu)$, illustrated in Fig.~\ref{figrho}, and emphasises the
persistence of the momentum dependent modifications of the quark propagator.

With the definition and calculation of the pressure, $P_q(T,\mu)$, all the
remaining bulk thermodynamic quantities that characterise the model can be
calculated.  As an example the ``interaction measure'': $\Delta:= \epsilon -
3 P$, where $\epsilon$ is the energy density, is plotted in
Fig.~\ref{energy}.  It is zero for an ideal gas, hence the name: $\Delta$
measures the interaction-induced deviation from ideal gas behaviour.
\begin{figure}[t]
\centering{\
\epsfig{figure=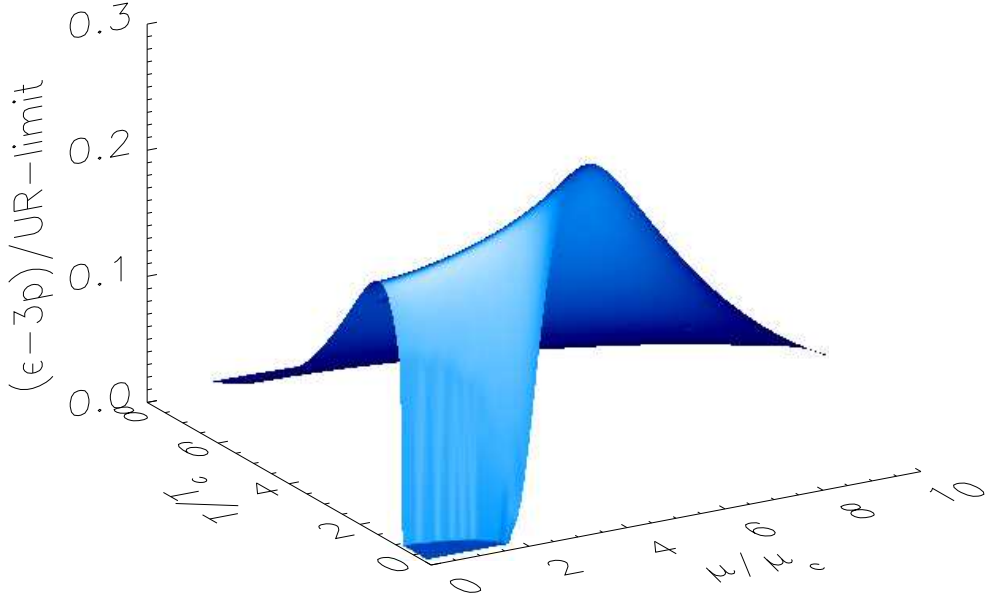,height=9.0cm}}
\caption{The ``interaction measure'', $\Delta(T,\mu)$, normalised to the
free, massless result for the pressure, Eq.~(\protect\ref{sbpres}).
\label{energy}}
\end{figure}
This figure provides a very clear indication of the persistence of
nonperturbative effects into the deconfinement domain, with a $\mu=0$ maximum
of $\Delta \approx 0.2\,P[S_0]$ at $T \approx 2 T_c$ and a $T=0$ maximum of
$\Delta \approx 0.3\,P[S_0]$ at $\mu \approx 3 \mu_c$.  Both
Figs.~\ref{presfig} and \ref{energy} indicate that there is a ``mirroring''
of finite-$T$ behaviour in the $\mu$-dependence of the bulk thermodynamic
quantities.
\vspace*{0.5\baselineskip}

{\it \arabic{section}.\arabic{subsection})~$\pi$ and $\rho$
properties.}\\[0.3\baselineskip] 
\addcontentsline{toc}{subsection}{\arabic{section}.\arabic{subsection})~$\pi$
and $\rho$ properties} 
\addtocounter{subsection}{1}
The model discussed in the last section has been used\ucite{schmidt98} to
study the $(T,\mu)$-dependence of $\pi$ and $\rho$ properties, and to
elucidate other features of the models described above that employ a more
sophisticated {\it Ansatz} for the dressed-gluon propagator.  In these
applications its simplicity is particularly helpful.

To begin, consider the vacuum quark condensate, which in this model is
\begin{equation}
\label{qbq}
-\langle \bar q q \rangle = 
\eta^3\,\frac{8 N_c}{\pi^2} \bar T\,\sum_{l=0}^{l_{\rm max}}\,
\int_0^{\bar\Lambda_l}\,dy\, y^2\,
{\sf Re}\left( \sqrt{\case{1}{4}- y^2 - \tilde\omega_{l}^2 }\right)\,:
\end{equation}
for $T=0=\mu$, $(-\langle \bar q q \rangle) = \eta^3 /(80\,\pi^2) = (0.11\,
\eta)^3$.  
\begin{figure}[t]
\centering{\
\epsfig{figure=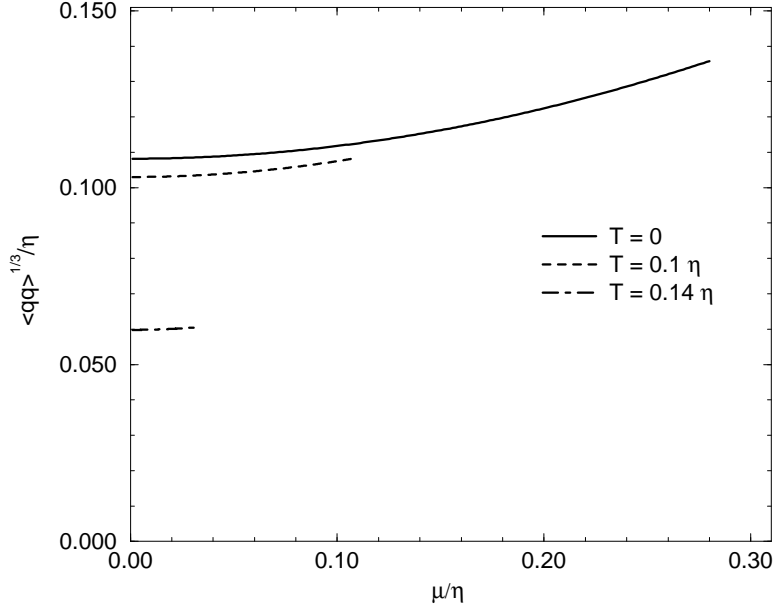,height=9.5cm}}\vspace*{-1.5\baselineskip}
\caption{\label{condensateB} The quark condensate, Eq.~(\protect\ref{qbq}),
as a function of $\mu$ for a range of values of $T$.  In all existing
studies, in which the quark mass function has a realistic momentum
dependence, it increases with $\mu$ and decreases with $T$.  At the critical
chemical potential, $\mu_c(T)$, $(-\langle \bar q q\rangle)$ drops
discontinuously to zero, as expected of a first-order transition.  For
$\mu=0$ it falls continuously to zero, exhibiting a second-order transition
at $T_c(\mu=0)= 0.16\,\eta$.}
\end{figure}
In Fig.~\ref{condensateB} one observes that $(-\langle \bar q q \rangle)$
decreases with $T$ but {\it increases} with increasing $\mu$, up to a
critical value of $\mu_c(T)$ when it drops discontinuously to zero.  These
results are in qualitative and semiquantitative agreement with the $(T=
0,\mu\neq 0)$ and $(T\neq 0,\mu = 0)$ studies described in Secs.~7.3)
and~7.4).  The increase with $\mu$ is also qualitatively identical to that
observed in a random matrix theory with the global symmetries of the QCD
partition function.\ucite{jackson} $(-\langle \bar q q \rangle)$ must
increase with $\mu$ in the confinement domain because confinement entails
that each additional quark must be locally-paired with an antiquark, thereby
increasing the density of condensate pairs.  This vacuum rearrangement is
manifest in the behaviour of the necessarily-momentum-dependent scalar part
of the quark self energy, $B(\tilde p_k)$.

In this model Eqs.~(\ref{pimass})-(\ref{fpi}) yield very simple expressions
in the chiral limit; for example,\footnote[2]{This is the expression for
$N_\pi^2$ from Eq.~(\protect\ref{pinormT}), which provides provides a better
approximation to the pion leptonic decay constant than
Eq.~(\protect\ref{fpi}) when one assumes $\Gamma_\pi(p;P) =
i\gamma_5\,B_0(p^2)$.}
\begin{eqnarray}
\label{npialg}
f_\pi^2 & = & \eta^2 \frac{16 N_c }{\pi^2} \bar T\,\sum_{l=0}^{l_{\rm max}}\,
\frac{\bar\Lambda_l^3}{3} \left( 1 + 4 \,\bar\mu^2 - 4 \,\bar\omega_l^2 -
\case{8}{5}\,\bar\Lambda_l^2 \right)\,.
\end{eqnarray}
Characteristic in Eq.~(\ref{npialg}) is the combination $\mu^2 - \omega_l^2$,
which entails that, whatever change $f_\pi$ undergoes as $T$ is increased,
the {\it opposite} occurs as $\mu$ is increased.  Without calculation,
Eq.~(\ref{npialg}) indicates that $f_\pi$ will {\it decrease} with $T$ and
{\it increase} with $\mu$.  This provides a simple elucidation of the results
described above.  Figure~\ref{fpimpi} illustrates this behaviour for $m\neq
0$.
\begin{figure}[t]
\centering{\ \epsfig{figure=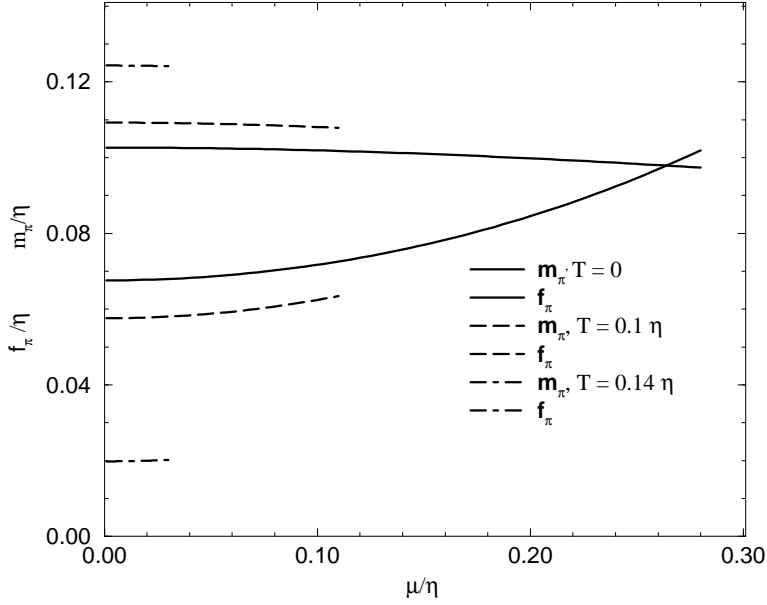,height=9.5cm}}\vspace*{-1.5\baselineskip}
\caption{\label{fpimpi} The pion mass, Eq.~(\protect\ref{pimass}), and weak
decay constant, Eq.~(\protect\ref{pinormT}), as a function of $\mu$ for a
range of values of $T$.  $m_\pi$ falls slowly and uniformly with $\mu$
[$m_\pi(T=0,\mu_c)= 0.95 \, m_\pi(T=0,\mu=0)$] but increases with $T$.  Such
a decrease is imperceptible if the ordinate has the range in
Fig.~\protect\ref{pirhomass}.  $f_\pi$ increases with $\mu$ and decreases
with $T$ [$f_\pi(T=0,\mu_c)= 1.51 \, f_\pi(T=0,\mu=0)$].}
\end{figure}
The $(T,\mu)$-dependence of $m_\pi$, from Eq.~(\ref{pimass}), is also
depicted in Fig.~\ref{fpimpi}.  It is {\it insensitive} to changes in $\mu$
and only increases slowly with $T$, until $T$ is very near the critical
temperature.  As in Sec.~7.4), this insensitivity is the result of mutually
cancelling increases in $\langle m\,\bar q q\rangle_\pi$ and $f_\pi$, and is
a feature of studies that preserve the momentum-dependence of the confined,
dressed-quark degrees of freedom in bound states.

With $\eta = 1.37\,$GeV and $m=30\,$MeV, one obtains $f_\pi=92\,$MeV and
$m_\pi= 140\,$MeV at $T=0=\mu$.  That large values of $\eta$ and $m$ are
required is a quantitative consequence of the inadequacy of
Eq.~(\protect\ref{mnprop}) in the ultraviolet: the large-$p^2$ behaviour of
the scalar part of the dressed-quark self-energy is incorrect.  This defect
is remedied easily\ucite{mr97} without qualitative changes to the results
presented here.\ucite{mPrivate}

$\rho$-meson properties are more difficult to study: one must solve the
vector-meson Bethe-Salpeter equation directly.  As described above, the
ladder truncation of the kernel in the inhomogeneous axial-vector vertex
equation and the rainbow truncation of the quark DSE form an AV-WTI identity
preserving pair.\ucite{bender96} It follows that the ladder BSE is accurate
for flavour-nonsinglet pseudoscalar and vector bound states of equal-mass
quarks because of a cancellation in these channels between diagrams of higher
order in the systematic expansion illustrated in Fig.~\ref{skeleton}.

A ladder BSE using the $T=0$ limit of Eq.~(\ref{mnprop}) was introduced in
Ref.~[\ref{mn83}].  It has one notable pathology: the bound state mass is
determined only upon the additional specification that the constituents have
zero relative momentum.  That specification leads to a conflict with
Eqs.~(\ref{bwti})-(\ref{gwti}), which follow from the AV-WTI, and is an
artefact of implementing the delta-function limit discontinuously; i.e.,
these identities are manifest for any finite-width representation of the
delta-function, as this width is reduced continuously to zero.  In other
respects this ladder BSE provides a useful qualitative and semi-quantitative
tool for analysing features of the pseudoscalar and vector meson masses.  For
example, Goldstone's theorem is manifest, in that the $\pi$ is massless in
the chiral limit, and also $m_\pi^2$ rises linearly with the current-quark
mass.  Further, there is a naturally large splitting between $m_\pi$ and
$m_\rho$, which decreases slowly with the current-quark mass.

To illustrate this and determine the response of $m_\rho$ to increasing $T$
and $\mu$, the BSE of Ref.~[\ref{mn83}] was generalised\ucite{schmidt98} to
finite-$(T,\mu)$ as
\begin{equation}
\label{bse}
\Gamma_M(\tilde p_k;\check P_\ell)= - \frac{\eta^2}{4}\,
{\sf Re}\left\{\gamma_\mu\,
S(\tilde p_i +\case{1}{2} \check P_\ell)\,
\Gamma_M(\tilde p_i;\check P_\ell)\,
S(\tilde p_i -\case{1}{2} \check P_\ell)\,\gamma_\mu\right\}\,,
\end{equation}
where $\check P_\ell := (\vec{P},\Omega_\ell)$.  The bound state mass is
obtained by considering $\check P_{\ell=0}$ and, in ladder truncation, the
$\rho$- and $\omega$-mesons are degenerate.

The $\pi$ equation admits the solution
\begin{equation} 
\Gamma_\pi(P_0) = \gamma_5 \left(i \theta_1 
        + \vec{\gamma}\cdot \vec{P} \,\theta_2 \right)
\end{equation}
and yields the mass plotted in Fig.~\ref{pirhomass}.  
\begin{figure}
\centering{\
\epsfig{figure=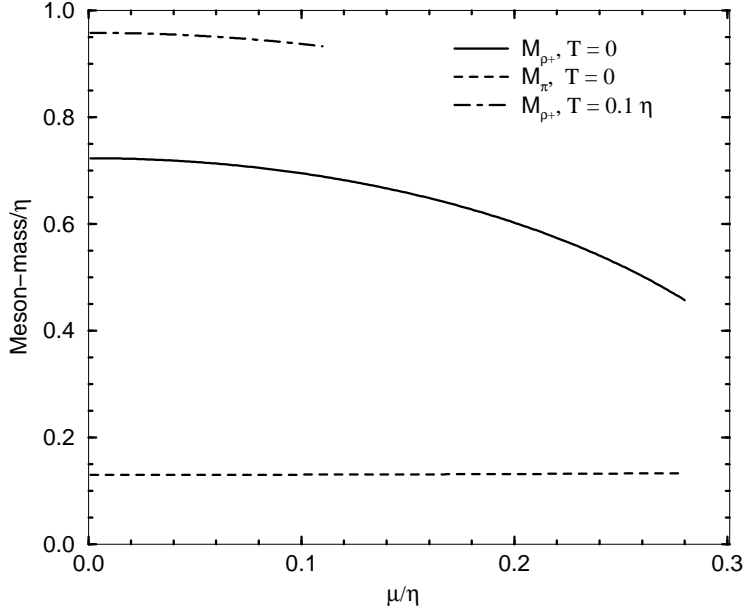,height=9.5cm}}\vspace*{-1.5\baselineskip}
\caption{\label{pirhomass} $M_{\rho+}$ and $m_\pi$ as a function of $\bar\mu$
for $\bar T = 0, 0.1$.  On the scale of this figure, $m_\pi$ is insensitive
to this variation of $T$.  The current-quark mass is $m= 0.011\,\eta$, which
for $\eta=1.06\,$GeV yields $M_{\rho+}= 770\,$MeV and $m_\pi=140\,$MeV at
$T=0=\mu$.}
\end{figure}
The mass behaves in qualitatively the same manner as $m_\pi$ in
Fig.~\ref{fpimpi}, from Eq.~(\ref{pimass}), as required if Eq.~(\ref{bse}) is
to provide a reliable guide.  In particular, it vanishes in the chiral limit.

For the $\rho$-meson there are two components: one longitudinal and one
transverse to $\vec{P}$.  The solution of the BSE has the form
\begin{equation}
\Gamma_\rho = \left\{
\begin{array}{l}
\gamma_4 \,\theta_{\rho+} \\
\left(
\vec{\gamma} - \case{1}{|\vec{P}|^2}\,\vec{P} \vec{\gamma}\cdot\vec{P}\right)\,
        \theta_{\rho-}
\end{array}
\right.\,,
\end{equation}
where $\theta_{\rho+}$ labels the longitudinal and $\theta_{\rho-}$ the
transverse solution.  The eigenvalue equation obtained from Eq.~(\ref{bse})
for the bound state mass, $M_{\rho\pm}$, is
\begin{equation}
\label{rhomass}
\frac{\eta^2}{2}\,{\sf Re}\left\{ \sigma_S(\omega_{0+}^2
        - \case{1}{4} M_{\rho\pm}^2)^2 
- \left[ \pm \,\omega_{0+}^2 - \case{1}{4} M_{\rho\pm}^2\right]
        \sigma_V(\omega_{0+}^2- \case{1}{4} M_{\rho\pm}^2)^2 \right\}
= 1\,.
\end{equation}

The equation for the transverse component is obtained with $[- \omega_{0+}^2
- \case{1}{4} M_{\rho-}^2]$ in (\ref{rhomass}).  Using the chiral-limit
solutions, Eq.~(\ref{ngsoln}), one obtains immediately that
\begin{equation}
M_{\rho-}^2 = \case{1}{2}\,\eta^2,\;\mbox{{\it independent} of $T$ and $\mu$.}
\end{equation}
This is the $T=0=\mu$ result of Ref.~[\ref{mn83}].  Even for nonzero
current-quark mass, $M_{\rho-}$ changes by less than 1\% as $T$ and $\mu$ are
increased from zero toward their critical values.  Its insensitivity is
consistent with the absence of a constant mass-shift in the transverse
polarisation tensor for a gauge-boson.

For the longitudinal component one obtains in the chiral limit:
\begin{equation}
\label{mplus}
M_{\rho+}^2 = \case{1}{2} \eta^2 - 4 (\mu^2 - \pi^2 T^2)\,.
\end{equation}
The characteristic combination $[\mu^2 - \pi^2 T^2]$ again indicates the
anticorrelation between the response of $M_{\rho+}$ to $T$ and its response
to $\mu$, and, like a gauge-boson Debye mass, that $M_{\rho+}^2$ rises
linearly with $T^2$ for $\mu=0$.  The $m\neq 0$ solution of
Eq.~(\ref{rhomass}) for the longitudinal component is plotted in
Fig.~\ref{pirhomass}.  As signalled by Eq.~(\ref{mplus}), $M_{\rho+}$ {\it
increases} with increasing $T$ and {\it decreases} as $\mu$
increases.\footnote[2]{There is a 25\% difference between the value of $\eta$
required to obtain the $T=0=\mu$ values of $m_\pi$ and $f_\pi$, from
Eq.~(\protect\ref{pimass}) and Eq.~(\protect\ref{pinormT}), and that required
to give $M_{\rho\pm}= 0.77\,$GeV.  This is a measure of the quantitative
accuracy of this algebraic model.}

I stated that contributions from skeleton diagrams not included in the ladder
truncation of the vector meson BSE do not alter the calculated mass
significantly because of cancellations between these higher order
terms.\ucite{bender96} This is illustrated explicitly in two calculations:
Ref.~[\ref{mitchell97}], which shows that the $\rho\to\pi\pi\to\rho$
contribution to the real part of the $\rho$ self-energy; i.e., the
$\pi$-$\pi$ induced mass-shift, is only $-3$\%; and
Ref.~[\ref{hollenberg92}], which shows, for example, that the contribution to
the $\omega$-meson mass of the $\omega\to 3\pi$-loop is negligible.
Therefore, ignoring such contributions does not introduce uncertainty into
estimates of the vector meson mass based on Eq.~(\ref{bse}).

Equation~(\ref{rhomass}) can also be applied to the $\phi$-meson.  The
transverse component is insensitive to $T$ and $\mu$, and the behaviour of
the longitudinal mass, $M_{\phi+}$, is qualitatively the same as that of the
$\rho$-meson: it increases with $T$ and decreases with $\mu$.  Using $\eta =
1.06\,$GeV, the model yields $M_{\phi\pm} = 1.02\,$GeV for $m_s = 180\,$MeV
at $T=0=\mu$.

In a 2-flavour, free-quark gas at $T=0$ the baryon number density is $\rho_B=
2 \mu^3/(3 \pi^2)\,$, by which gauge nuclear matter density,
$\rho_0=0.16\,$fm$^{-3}$, corresponds to $\mu= \mu_0 := 260\,$MeV$\,=
0.245\,\eta$.  At this chemical potential the algebraic model yields
\begin{eqnarray}
\label{mrhoa}
M_{\rho+}(\mu_0) & \approx & 0.75 M_{\rho+}(\mu=0)\,,\\
\label{mphia}
M_{\phi+}(\mu_0) & \approx & 0.85 M_{\phi+}(\mu=0)\,.
\end{eqnarray}
The study summarised in Sec.~7.4),\ucite{greg} indicates that a better
representation of the ultraviolet behaviour of the dressed-gluon propagator
expands the horizontal scale in Fig.~\ref{pirhomass}, with the critical
chemical potential increased by 25\%.  This suggests that a more realistic
estimate is obtained by evaluating the mass at $\mu_0^\prime=0.20\,\eta$,
which yields
\begin{eqnarray}
\label{mrhob}
M_{\rho+}(\mu_0^\prime) &\approx&  0.85 M_{\rho+}(\mu=0)\,,\\
\label{mphib}
M_{\phi+}(\mu_0^\prime) & \approx & 0.90 M_{\phi+}(\mu=0)\,;
\end{eqnarray}
a small, quantitative modification.  The difference between
Eqs.~(\ref{mrhoa}) and (\ref{mrhob}), and that between Eqs.~(\ref{mphia}) and
(\ref{mphib}), is a measure of the theoretical uncertainty in the estimates
in each case.  This reduction in the vector meson masses is quantitatively
consistent with that calculated in Ref.~[\ref{derek95}] and conjectured in
Ref.~[\ref{brown91}].  At the critical chemical potential for $T=0$,
$M_{\rho+} \approx 0.65\, M_{\rho+}(\mu=0)$ and $M_{\phi+} \approx 0.80\,
M_{\phi+}(\mu=0)$.

This simple model of \qcdtm\ preserves the momentum-dependence of gluon and
quark dressing, which is an important qualitative feature of more
sophisticated studies.  Its simplicity means that many of the consequences of
that dressing can be demonstrated algebraically.  For example, it elucidates
the origin of an anticorrelation, found for a range of quantities, between
their response to increasing $T$ and that to increasing $\mu$.

Both $(-\langle \bar q q)\rangle$ and $f_\pi$ decrease with $T$ and increase
with $\mu$, and this ensures that $m_\pi$ is insensitive to increasing $\mu$
and/or $T$ until very near the edge of the domain of confinement and DCSB.
The mass of the transverse component of the vector meson is insensitive to
$T$ and $\mu$ while the mass of the longitudinal component increases with
increasing $T$ but decreases with increasing $\mu$.  This behaviour is
opposite to that observed for $(-\langle \bar q q)\rangle$ and $f_\pi$, and
hence the scaling law conjectured in Ref.~[\ref{brown91}] is inconsistent
with this calculation, as it is with others of this type.

This study has two primary limitations.  First, the width of the vector
mesons cannot be calculated because the solution of Eq.~(\ref{bse}) does not
provide a realistic Bethe-Salpeter amplitude.  Second, the calculation of
meson-photon observables at $T=0=\mu$ only became possible with the
determination\ucite{ayse97} of the form of the dressed-quark-photon vertex.
Its generalisation to nonzero-$(T,\mu)$ is a necessary precursor to the study
of these processes.

\vspace*{\baselineskip}

\addtocounter{section}{1}
\setcounter{subsection}{1}
{\large\bf \arabic{section}.~Closing Remarks.}\\[0.7\baselineskip] 
\addcontentsline{toc}{section}{\arabic{section}.~Closing Remarks}
These lecture notes illustrate the contemporary application of
Dyson-Schwinger equations to the analysis of observable strong interaction
phenomena, highlighting the positive aspects and successes.  Many recent,
interesting studies have been neglected; a calculation of the electric dipole
moment of the $\rho$-meson\ucite{martin} and an exploration of
$\eta$-$\eta^\prime$ mixing\ucite{klabucar} among them.  However, a simple
enquiry of ``http://xxx.lanl.gov/find/hep-ph'' with the keywords:
``Dyson-Schwinger'' or ``Schwinger-Dyson'', will provide a guide to other
current research.

In all phenomenological applications, modelling is involved, in particular,
of the behaviour of the dressed Schwinger functions in the infrared.  [The
ultraviolet behaviour is fixed because of the connection with perturbation
theory.]  This is tied to the need to make truncations in order to define a
tractable problem.  Questions will always be asked regarding the fidelity of
the modelling.  The answers can only come slowly as, for example, more is
learnt about the constraints that Ward Identities and Slavnov-Taylor
identities in the theory can provide.  That approach has been particularly
fruitful in QED,\ucite{ayse97} and already in the development of a systematic
truncation procedure for the kernel of the quark DSE and meson
BSE.$\,^{\ref{bender96},\ref{QC96}}$ In the meantime, and as is common,
phenomenological applications provide a key to understanding which elements
of the approach need improvement: one must push and prod to find the weak
links.

{\bf Acknowledgments}.  I am grateful to the faculty and staff at JINR-Dubna
for their hospitality during this workshop, and especially to L.
Kalinovskaya and Yu. Kalinovsky for their particular care.  This work was
supported by the US Department of Energy, Nuclear Physics Division, under
contract number W-31-109-ENG-38, the National Science Foundation under grant
no. INT-9603385, and Deutscher Akademischer Austauschdienst.

\pagebreak
{\large\bf References.}\\[-1.6\baselineskip]
\addcontentsline{toc}{section}{References}
\addtocounter{section}{1}
%
\begin{flushleft}
\begin{enumerate}
\item \label{rw94} C. D. Roberts and A. G. Williams,
                Prog. Part. Nucl. Phys. {\bf 33} (1994) 477:\\
                hep-ph/9403224.
                \vspace*{-0.50\baselineskip}
\item \label{iitap} C. D. Roberts, ``Dyson Schwinger Equations in QCD'', in
                {\it Light-Front Quantization and Non-Perturbative QCD},
                edited by J. P. Vary and F. W\"olz (International Institute 
                of Theoretical and Applied Physics, Ames, 1997) pp.~212-239:\\
                http://www.iitap.iastate.edu/reports/lfw/contents.html.
                \vspace*{-0.50\baselineskip}
\item \label{bender96} A. Bender, C. D. Roberts and L. v. Smekal,
Phys. Lett. B {\bf 380} (1996) 7: \\ nucl-th/9602012.
                \vspace*{-0.50\baselineskip}
\item \label{QC96} C. D. Roberts in {\it Quark Confinement and the Hadron
Spectrum II}, edited by N. Brambilla and G. M. Prosperi (World Scientific,
Singapore, 1997), pp. 224-230: \\nucl-th/9609039.
                \vspace*{-0.50\baselineskip}
\item \label{ayse97} A. Bashir, A. Kizilersu and M.R. Pennington,
Phys. Rev. D {\bf 57}, 1242 (1998); and references therein: hep-ph/9707421.
\vspace*{-0.50\baselineskip}
\item \label{bp89} N. Brown and M. R. Pennington, Phys. Rev. D {\bf 39}
(1989) 2723.  \vspace*{-0.50\baselineskip}
\item \label{mr97} P. Maris and C. D. Roberts, Phys. Rev. C {\bf 56} (1997)
3369: nucl-th/9708029.  \vspace*{-0.50\baselineskip}
\item \label{mr98} P. Maris and C. D. Roberts, ``Differences between heavy-
and light-quarks'', in {\it Rostock 1997, Progress in heavy quark physics},
edited by M. Beyer, T. Mannel and H. Schr\"oder: nucl-th/9710062.
\vspace*{-0.50\baselineskip}
\item \label{misha} M. A. Ivanov, Yu. Kalinovsky, P. Maris and C. D. Roberts,
Phys. Rev. C {\bf 57} (1998) 1991:
nucl-th/9711023.\vspace*{-0.50\baselineskip}
\item \label{pichowsky} M. A. Pichowsky and T.-S. H. Lee, Phys. Rev. D {\bf
56} (1997) 1644: nucl-th/9612049; and M. A. Pichowsky, ``Nonperturbative
quark dynamics in diffractive processes'', PhD Thesis, University of
Pittsburgh.\vspace*{-0.50\baselineskip}
\item \label{cssm} P. Maris and C. D. Roberts, ``QCD bound states and their
response to extremes of temperature and density'', to appear in the
proceedings of the {\it Workshop on Nonperturbative Methods in Field Theory},
University of Adelaide, Adelaide, South Australia, Feb. 1998:
nucl-th/9806nnn.\vspace*{-0.50\baselineskip}
\item \label{mn83} H. J. Munczek and A. M. Nemirovsky, Phys. Rev. D {\bf 28}
(1983) 181.  \vspace*{-0.50\baselineskip}
\item \label{derek} D. B. Leinweber, Ann. Phys. {\bf 254} (1997) 328:
nucl-th/9510051.  \vspace*{-0.50\baselineskip}
\item \label{m90} V. Miransky, Mod. Phys. Lett. A {\bf 5} (1990) 1979.
\vspace*{-0.50\baselineskip}
\item \label{mrpion} P. Maris and C. D. Roberts, ``Pseudovector components of
the pion, $\pi^0\to \gamma\gamma$, and $F_\pi(q^2)$: nucl-th/9804062.
\vspace*{-0.50\baselineskip}
\item \label{munczek} H. Munczek, Phys. Lett. B {\bf 175} (1986)
215.\vspace*{-0.50\baselineskip} 
\item \label{burden} C. J. Burden, C. D. Roberts and A. G. Williams,
Phys. Lett. B {\bf 285} (1992) 347.\vspace*{-0.50\baselineskip}
\item \label{bc80} J. S. Ball and T.-W. Chiu, Phys. Rev. D {\bf 22} (1980)
2542.\vspace*{-0.50\baselineskip}
\item \label{cp92} D. C. Curtis and M. R. Pennington, Phys. Rev. D {\bf 46}
(1992) 2663.\vspace*{-0.50\baselineskip}
\item \label{hawes} F. T. Hawes, C. D. Roberts and A. G. Williams, Phys. Rev. D
{\bf 49} (1994) 4683.\vspace*{-0.50\baselineskip} 
\item \label{pdg96} Particle Data Group (R. M. Barnett {\it et al}.),
Phys. Rev. D {\bf 54} (1996) 1:
http://pdg.lbl.gov/.\vspace*{-0.50\baselineskip}
\item \label{amend} S. R. Amendolia, {\it et al}., Nucl. Phys. B {\bf 277}
(1986) 168.\vspace*{-0.50\baselineskip}
\item \label{cdrpion} C. D. Roberts, Nucl. Phys. A {\bf 605} (1996) 475:
hep-ph/9408233.\vspace*{-0.50\baselineskip}
\item \label{pctrev} P. C. Tandy, Prog. Part. Nucl. Phys. {\bf 39} (1997) 117:
nucl-th/9705018.\vspace*{-0.50\baselineskip}
\item \label{kaonFF} C.J. Burden, C.D. Roberts and M.J. Thomson
C. D. Roberts, Phys. Lett. B {\bf 371} (1996) 163:
nucl-th/9511012.\vspace*{-0.50\baselineskip}
\item \label{gppp} R. Alkofer and C. D. Roberts, Phys. Lett. B {\bf 369}
(1996) 101: hep-ph/9510284.\vspace*{-0.50\baselineskip}
\item \label{pipi} C. D. Roberts, R. T. Cahill, M. E. Sevior and N. Iannella,
Phys. Rev. D {\bf 49} (1994) 125: hep-ph/9304315.\vspace*{-0.50\baselineskip}
\item \label{pocanic} D. Po\v{c}ani\'c, ``Summary of $\pi$-$\pi$ scattering
experiments'', in {\it Chiral Dynamics: Theory and Experiment}, edited by
A. M. Bernstein and B. R. Holstein, Lecture Notes in Physics, Vol.~452
(Springer, Berlin, 1995), 95: hep-ph/9412339.\vspace*{-0.50\baselineskip}
\item \label{ewff} Yu. Kalinovsky, K. L. Mitchell and C. D. Roberts,
Phys. Lett. B {\bf 399} (1997) 22:
nucl-th/9610047.\vspace*{-0.50\baselineskip}
\item \label{bebek} C. J. Bebek, {\it et al}., Phys. Rev. D {\bf 13}, 25
(1976).\vspace*{-0.50\baselineskip} 
\item \label{bebekB} C. J. Bebek, {\it et al}., Phys. Rev. D {\bf 17}, 1693
(1978).\vspace*{-0.50\baselineskip}  
\item \label{FJ} G. R. Farrar and D. R. Jackson, Phys. Rev. Lett. {\bf 43}, 246
(1979).\vspace*{-0.50\baselineskip}  
\item \label{sham} W. D. Shambroom, {\it et al}., Phys. Rev. D {\bf 26}
(1982) 1.\vspace*{-0.50\baselineskip}  
\item \label{aubert} J. J. Aubert, {\it et al}., Phys. Lett. B {\bf 161}
(1985) 203.\vspace*{-0.50\baselineskip}
\item \label{arneodo} M. Arneodo, {\it et al}., Nucl. Phys. B {\bf 429}
(1994) 503.\vspace*{-0.50\baselineskip}
\item \label{derrick} M. Derrick, {\it et al}., Phys. Lett. B {\bf 377}
(1996) 259: hep-ex/9601009.\vspace*{-0.50\baselineskip}
\item \label{derrickb} M. Derrick, {\it et al}., Phys. Lett. B {\bf 380}
(1996) 220: hep-ex/9604008.\vspace*{-0.50\baselineskip}
\item \label{derrickc} M. Derrick, {\it et al}., Phys. Lett. B {\bf 350}
(1995) 120: hep-ex/9503015.\vspace*{-0.50\baselineskip}
\item \label{aid} S. Aid, {\it et al}., Nucl. Phys. B {\bf 468}
(1996) 3: hep-ex/9602007.\vspace*{-0.50\baselineskip}
\item \label{collinsperry} J. C. Collins and M. J. Perry,
Phys. Rev. Lett. {\bf 34} (1975) 1353.\vspace*{-0.50\baselineskip}
\item \label{krishna} K. Rajagopal, ``The Chiral Phase Transition in QCD:
Critical Phenomena and Long-wavelength Pion Oscillations'', in {\it
Quark-gluon plasma}, edited by R. C. Hwa (World Scientific, New York, 1995),
484: hep-ph/9504310. \vspace*{-0.5\baselineskip}
\item \label{karsch97} J. Engels {\it et al}., Phys. Lett. B {\bf 396} (1997)
210: hep-lat/9612018; and references therein.\vspace*{-0.50\baselineskip}
\item \label{wiringa} R. B. Wiringa, V. Fiks and A. Fabrocini, Phys. Rev. C
{\bf 38} (1988) 1010.\vspace*{-0.50\baselineskip} 
\item \label{drees97} A. Drees, ``Dilepton enhancement at the CERN SpS'',
{\it in} Proceedings of the {\it XXVth International Workshop on Gross
Properties of Nuclei and Nuclear Excitations}, Hirschegg 1997, edited by
H. Feldmeier, J. Knoll, W. N\"orenberg and J. Wambach (GSI-Darmstadt, 1997),
178.\vspace*{-0.50\baselineskip} 
\item \label{na45home} This figure can be found at:
http://ceres6.physi.uni-heidelberg.de/ceres/referenceData/Index.html.
\vspace*{-0.50\baselineskip} 
\item \label{brown95} G. Q. Li, C. M. Ko and G. E. Brown,
Phys. Rev. Lett. {\bf 75} (1995) 4007:
nucl-th/9504025.\vspace*{-0.50\baselineskip}
\item \label{rapp97} R. Rapp, G. Chanfray and J. Wambach, Nucl. Phys. A {\bf
617} (1997) 472: hep-ph/9702210.\vspace*{-0.50\baselineskip}
\item \label{derek95} X. Jin and D. B. Leinweber, Phys. Rev. C {\bf 52}
(1995) 3344: nucl-th/9510064.\vspace*{-0.50\baselineskip}
\item \label{klingl97} F. Klingl, N. Kaiser and W. Weise, Nucl. Phys. A {\bf
624} (1997) 527: hep-ph/9704398.\vspace*{-0.50\baselineskip}
\item \label{schmidt98} P. Maris, C. D. Roberts and S. Schmidt, Phys. Rev. C
{\bf 57} (1998) R2821: nucl-th/9801059 .\vspace*{-0.50\baselineskip}
\item \label{kapusta} J. I. Kapusta, ``Finite-Temperature Field Theory''
(Cambridge University Press, Cambridge UK, 1989).\vspace*{-0.50\baselineskip}
\item \label{htl} R. D. Pisarski, Phys. Rev. Lett. {\bf 63} (1989) 1129;
E. Braaten and R. D. Pisarski, Nucl. Phys. B {\bf 337} (1990) 569; {\it ibid}
B {\bf 339} (1990) 310; Phys. Rev. Lett. {\bf 64} (1990) 1338; J. Frenkel and
J. C. Taylor, Nucl. Phys. B {\bf 334} (1990) 199.\vspace*{-0.50\baselineskip}
\item \label{detar} C. De Tar, ``Quark Gluon Plasma in Numerical Simulations
of Lattice QCD'', in {\it Quark-gluon plasma}, ed. R. C. Hwa (World
Scientific, New York, 1995), 1: hep-ph/9504325. \vspace*{-0.5\baselineskip}
\item \label{ukawa90} A. Ukawa, Nucl. Phys. B {\bf 17} (Proc. Suppl.) (1990)
118; and references therein.\vspace*{-0.50\baselineskip}
\item \label{dks96} M.-P Lombardo, J. B. Kogut and D. K. Sinclair,
Phys. Rev. D {\bf 54} (1996) 2303:
hep-lat/9511026.\vspace*{-0.50\baselineskip}
\item \label{adam} M. A. Halasz, A. D. Jackson and J. J. M. Verbaarschot,
Phys. Rev. D {\bf 56} (1997) 5140:
hep-lat/9703006.\vspace*{-0.50\baselineskip} 
\item \label{karsch95} F. Karsch, Nucl. Phys. A {\bf 590} (1995)
367c: hep-lat/9503010.\vspace*{-0.50\baselineskip}
\item \label{blum} T. Blum, {\it et al}., Phys. Rev. D {\bf 51} (1995) 5153:
hep-lat/9410014.\vspace*{-0.50\baselineskip}
\item \label{laermann} E. Laermann, ``Thermodynamics using Wilson and
Staggered Quarks'': hep-lat/9802030.\vspace*{-0.50\baselineskip}
\item \label{prl} A. Bender, D. Blaschke, Yu. Kalinovsky and C.D. Roberts,
Phys. Rev. Lett. {\bf 77} (1996) 3724:
nucl-th/9606006.\vspace*{-0.50\baselineskip} 
\item \label{thermo} D. Blaschke, C.D. Roberts and S. Schmidt, Phys. Lett. B
{\bf 425} (1998) 232: nucl-th/9706070.\vspace*{-0.50\baselineskip}
\item \label{greg} A. Bender, {\it et al}., ``Deconfinement at finite
chemical potential'', Phys. Lett. B, in press:
nucl-th/9710069.\vspace*{-0.50\baselineskip}
\item \label{jackson96} A. D. Jackson and J. J. M. Verbaarschot, Phys. Rev. D
{\bf 53} (1996) 7223: hep-ph/9509324.\vspace*{-0.50\baselineskip} 
\item \label{stingl} U.~H\"{a}bel, R.~K\"{o}nning, H.-G.~Reusch, M.~Stingl
  and S.~Wigard, Z.~Phys.~A {\bf 336} (1990) 423; U.~H\"{a}bel,
  R.~K\"{o}nning, H.-G.~Reusch, M.~Stingl and S.~Wigard, Z.~Phys.~A {\bf 336}
  (1990) 435.\vspace*{-0.50\baselineskip}
\item \label{hrw94} F. T. Hawes, C. D. Roberts and A. G. Williams,
Phys. Rev. D {\bf 49} (1994) 4683:
hep-ph/9309263.\vspace*{-0.50\baselineskip}
\item \label{m95} P. Maris, Phys. Rev. D {\bf 52} (1995) 6087:
hep-ph/9508323.\vspace*{-0.50\baselineskip}
\item \label{fr96} M. R. Frank and C. D. Roberts, Phys. Rev. C {\bf 53}
(1996) 390: hep-ph/9508225.\vspace*{-0.50\baselineskip}
\item \label{sevior92} M. E. Sevior, Nucl. Phys. A {\bf 543} (1992)
275c.\vspace*{-0.50\baselineskip}
\item \label{kl94} F. Karsch and E. Laermann, Phys. Rev. D {\bf 50} (1994) 6954:
hep-lat/9406008.\vspace*{-0.50\baselineskip}
\item \label{arne} D. Blaschke, A. Hoell, C. D. Roberts and S. Schmidt,
``Analysis of chiral and thermal susceptibilities'', Phys. Rev. C, in press:
nucl-th/9803030.\vspace*{-0.50\baselineskip}
\item \label{mPrivate} P. Maris, private
communication.\vspace*{-0.50\baselineskip}
\item \label{Betal95} D. Blaschke, {\it et al}., Nucl. Phys. A {\bf 592}
(1995) 561.\vspace*{-0.50\baselineskip}
\item \label{haymaker} R. W. Haymaker, Riv. Nuovo Cim. {\bf 14} (1991) series
3, no. 8.\vspace*{-0.50\baselineskip}
\item \label{cahill} R. T. Cahill, Aust. J. Phys. {\bf 42} (1989)
171.\vspace*{-0.50\baselineskip}
\item \label{reg85} R. T. Cahill and C. D. Roberts, Phys. Rev. D {\bf 32}
(1985) 2419.\vspace*{-0.50\baselineskip}
\item \label{kubodera} H. Yabu, F. Myhrer and K. Kubodera, Phys. Rev. D {\bf
50} (1994) 3549: nucl-th/9402014.\vspace*{-0.50\baselineskip}
\item \label{brown} G. Brown, Nucl. Phys. A {\bf 488} (1988)
689c.\vspace*{-0.50\baselineskip}
\item \label{jackson} M. A. Halasz, private
communication.\vspace*{-0.50\baselineskip}
\item \label{mitchell97} K. L. Mitchell and P. C. Tandy, Phys. Rev. C {\bf
55}, 1477 (1997): nucl-th/9607025.\vspace*{-0.50\baselineskip}
\item \label{hollenberg92} L. C. L. Hollenberg, C. D. Roberts and
B. H. J. McKellar, Phys. Rev. C {\bf 46}, 2057
(1992).\vspace*{-0.50\baselineskip}
\item \label{brown91} G. E. Brown and M. Rho, Phys. Rev. Lett. {\bf 66}, 2720
(1991).\vspace*{-0.50\baselineskip}
\item \label{martin} M. B. Hecht and B. H. J. McKellar, Phys. Rev. C {\bf 57}
(1998) 2638: hep-ph/9704326.\vspace*{-0.50\baselineskip}
\item \label{klabucar} D. Klabucar and D. Kekez, ``$\eta$ and $\eta^\prime$
     at the limits of applicability of a coupled Schwinger-Dyson and
     Bethe-Salpeter approach in the ladder approximation'':
     hep-ph/9710206.\vspace*{-0.50\baselineskip}
\end{enumerate}
\end{flushleft}
\end{document}